\documentclass[fleqn,usenatbib,useAMS]{mnras}

\usepackage[T1]{fontenc}
\usepackage{ae,aecompl}
\usepackage{graphicx}	% Including figure files
\usepackage{amsmath}	% Advanced maths commands
\usepackage{amssymb}	% Extra maths symbols
\usepackage{natbib}
\usepackage{subfigure}  
\usepackage{multirow}
\usepackage{hhline} 
\usepackage{comment}
\usepackage[table,xcdraw]{xcolor}
\usepackage{hyperref}
\newcommand{\vir}[1]{``#1''}
\newcommand{\te}[1]{\text{#1}}

\newcommand{\su}{_\odot}
\newcommand{\gaia}{\textit{Gaia} }
\newcommand{\gaiap}{\textit{Gaia}}

\title[First all-sky view of the Galactic halo]{The first all-sky view
  of the Milky Way stellar halo with {\it Gaia}+2MASS RR Lyrae}

\author[G. Iorio et al.]{
G. Iorio,$^{1,2,3}$\thanks{giuliano.iorio@unibo.it}
V. Belokurov,$^{3}$
D. Erkal,$^{3,4}$
S. E. Koposov,$^{3,5}$
C. Nipoti,$^{1}$
F. Fraternali$^{1,6}$
\\
$^{1}$Dipartimento di Fisica e Astronomia, Universit\`a di Bologna, via Gobetti 93/2, I-40129, Bologna, Italy\\
$^{2}$INAF -  Osservatorio Astronomico di Bologna, via Gobetti 93/3, I-40129, Bologna, Italy\\
$^{3}$Institute of Astronomy, University of Cambridge, Madingley Road, Cambridge CB3 0HA, UK\\
$^{4}$Department of Physics, University of Surrey, Guildford, GU2 7XH, UK \\
$^{5}$McWilliams Center for Cosmology, Department of Physics, Carnegie
Mellon University, 5000 Forbes Avenue, Pittsburgh, PA 15213, USA\\
$^{6}$Kapteyn Astronomical Institute, University of Groningen, Landleven 12, 9747 AD Groningen, The Netherlands\\
}
\date{Accepted XXX. Received YYY; in original form ZZZ}

\pubyear{2017}
\hypersetup{draft}
\begin{document}
\label{firstpage}
\pagerange{\pageref{firstpage}--\pageref{lastpage}}
\maketitle
% * <giuliano.iorio89@gmail.com> 2016-11-27T14:54:39.043Z:
\begin{abstract}
We exploit the first \gaia data release to study the properties of the
Galactic stellar halo as traced by RR Lyrae. We demonstrate that it is
possible to select a pure sample of RR Lyrae using only photometric
information available in the \gaiap+2MASS catalogue. The final sample
contains about 21600 RR Lyrae covering an unprecedented fraction
($\sim60\%$) of the volume of the Galactic inner halo ($\te{R}<28$
kpc).  We study the morphology of the stellar halo by analysing the RR
Lyrae distribution with parametric and non-parametric
techniques. Taking advantage of the uniform all-sky coverage, we test
halo models more sophisticated than usually considered in the
literature, such as those with varying flattening, tilt and/or offset
of the halo with respect to the Galactic disc. A consistent picture
emerges: the inner halo is well reproduced by a smooth distribution of
stars settled on triaxial density ellipsoids. The shortest axis is perpendicular
to the Milky Way's disc, while the longest axis
forms an angle of $\sim70^{\circ}$ with the axis connecting the Sun and the Galactic centre.
The elongation along the
major axis is mild ($\te{p}=1.27$), and the vertical flattening is
shown to evolve from a squashed state with $\te{q}\approx0.57$ in the
centre to a more spherical $\te{q}\approx0.75$ at the outer edge of
our dataset. 
Within the radial range probed,
the density profile of the stellar halo is well approximated by a single
power-law with exponent $\alpha=-2.96$. We do not find
evidence of  tilt or  offset of the halo with respect to the Galaxy's disc.

\end{abstract}

%%%%%%%%%%%%%%%%%%%%%%%%%%%%%%%%%%%%%%%%%%%%%%%%%%

%%%%%%%%%%%%%%%%% BODY OF PAPER %%%%%%%%%%%%%%%%%%

\section{Introduction}

The diffuse cloud of stars observed around the Milky Way and known as
the stellar halo is the {\it alter ego} of the much more massive structure,
whose presence is inferred indirectly: the dark matter (DM) halo.
The
  DM halo dominates the Galactic mass budget and, according to the
  currently favoured theories of structure formation, holds clues to
  a number of fundamental questions in astrophysics. These include, amongst
  others, the properties of the DM particles \citep[see
    e.g][]{dave2001,governato2004,lovell2014}, the nature of gravity itself
  \citep[see e.g.][]{milgrom1983,screen2012}, as well as the coupling
  between the DM and the baryons
  \citep[e.g.][]{kauffmann1993,sommer2003,fire2015}.
The two halos emerge alongside each other, sharing the formation mechanism, i.e.\ a
combination of the accretion onto the Galaxy and the subsequent
relaxation and phase mixing. Thus, there ought to exist a bond between
them, which can be exploited to reveal the properties of the dark halo
through the study of the luminous one. For example, by positing the
continuity of the phase-space flow, the DM halo can be mapped
out if the stellar halo spatial shape is known and complemented by
stellar kinematics \citep[see
  e.g.][]{jeans1915,Helmi_review,posti2015,williams_evans_2015_actions}.

Leaving the ideas of James Jeans aside, could the structural
parameters of the stellar halo alone inform our understanding of the
mass assembly of the Galaxy? At our disposal are the numbers
pertaining to the slope of the stellar halo's radial density profile
and its vertical flattening \citep[see
  e.g.][]{juric,bellhalo,deasonhalo,xuehalo}. The radial profile has
so far been measured with a variety of stellar tracers. Studies based
on Main Sequence Turn-Off stars \citep[e.g.][]{sesarmsto,pilahalo},
Blue Straggler and Horizontal Branch stars \citep[e.g.][]{deasonhalo}
and RR Lyrae \citep[e.g.][]{sesarhalo,watkins} seem to favour a
``broken'' profile. According to these datasets, somewhere between 20
and 30 kpc from the MW centre, the density slope changes from a
relatively shallow one, as described by power-law index of
approximately $-2.5$, to a much steeper one, consistent with a
power-law index of $\approx-4$.

In an attempt to interpret the observed radial density profile,
\citet{deasonbreak} conjecture that the presence or absence of a break
is linked to the details of the stellar halo accretion history. In
their exposition, a prominent break can arise if the stellar halo is
dominated by the debris from a i) single, ii) early and iii) massive
accretion event. This hypothesis appears to be supported by
semi-analytic MW stellar halo models \citep[see
  e.g.][]{bj2005,amoriscoatlas} but is yet to be fully tested with
Cosmological zoom-in simulations \citep[see,
    however][]{pillepich2014}. Nonetheless, a consistent picture is
now emerging: in line with other pieces of evidence, the ``broken'' MW
stellar halo appears to be the tell-tale sign of an early-peaked and
subsequently quiescent accretion history. Note also that such
destitute state of the stellar halo is not permanent but rather
transient \citep[see e.g.][]{deasonhabits,amoriscowindow}. In
agreement with simulations, the MW stellar halo is destined to
transform dramatically with the dissolution of the debris from the Sgr
dwarf and the Magellanic Clouds.

While the radial density profile can be gauged based on the data from
a limited number of sight-lines through the Galaxy, the shape of the
stellar halo requires a much more complete coverage of the sky. So
far, much of the halo modelling has relied on the Sloan Digital Sky
Survey data, which is biased towards the Northern celestial
hemisphere. It is therefore possible that the incomplete view has
troubled the efforts to simultaneously infer the details of the radial
density evolution and the shape of the halo. For example, using
A-coloured stars, \citet{deasonhalo} measure substantial flattening of
the stellar halo in the direction perpendicular to the Galactic disc
plane, but no evidence for the change of the shape with radius. On the
other hand, \citet{xuehalo} use a sample of
spectroscopically-confirmed K giants to detect a noticeable change of
flattening with radius. Furthermore, they argue that if the halo shape
is allowed to vary with radius, then a break in the radial density
profile is not required. Finally, to add to the puzzle, based on a set
of BHB stars with spectra, \citet{das} report both evolving halo shape
and the break in the radial density.  
Pinning down the shape of
the stellar halo is important both for the dynamical inference of
the shape of the DM halo \citep[see e.g.][]{williams_evans_2015_haloes,bowden2016} and for our
understanding of the response of the DM distribution to the presence
of baryons \citep[see][]{kaza2004,gnedin2004,duffy2010,abadi2010}.

Looking at some of the earliest halo studies, which inevitably had to
rely on much more limited samples of tracers, it is worth pointing out
that, strikingly, glimpses of the variation of the halo shape were
already caught by \citet{kinmanhalo}. This pioneering work took
advantage of perhaps the most reliable halo tracer, the RR Lyrae stars
(RRLs, hereafter). These old and metal-poor pulsating stars suffer
virtually no contamination from other populations of the Milky Way and
have been used to describe the Galactic halo with unwavering success
over the last 50 years \citep[see
  e.g.][]{hawkins1984,saha1984,wm1996,ivezic2000,Vivas,catelan2009,watkins,S82,rrl3,oglerrl2014,gabriel2015,oglerrl2016}.

While deep, wide-area samples of RRLs now exist, for example
provided by the Catalina Sky Survey \citep[CSS,][]{CSS}, Palomar
Transient Factory \citep[PTF,][]{ptfbrani} and Pan-STARRS1
\citep[PS1,][]{ps1nina,ps1brani}, they have yet to be used to model
the Galactic halo globally. In case of CSS, this might be due to the
varying completeness of the sample. For PTF and PS1, this is sadly due
to the public unavailability of the data. To remedy this, here we
attempt to extract an all-sky sample of RRLs from the \gaia Data 
Release 1 
\citep[GDR1,][]{gaiaDR1} data. Our primary goal is to use the thus
procured RRL candidates to model the global properties of the MW
stellar halo. Therefore, we are not concerned with maximizing the
completeness but instead strive to achieve homogeneous selection
efficiency and reasonably high purity. While GDR1 does not contain any
explicit variability information for stars across the sky,
\citet{bel16} and \citet{deasonMC} show that likely variable objects
can be extracted from the \texttt{GaiaSource} table available as part
of GDR1. We build on these ideas and combine \gaia and Two Micron All
Sky Survey (2MASS) photometry (and astrometry) to produce a sample of
$\approx21,600$ RRLs out to $\approx20$ kpc from the Sun, with constant
completeness of $\approx20\%$ and purity $\approx90\%$.

Armed with this unprecedented dataset, we simultaneously extract the
radial density profile as well as the shape of the Galactic stellar
halo. Furthermore, taking advantage of the stable completeness and the
all-sky view provided by {\it Gaia}+2MASS, we explore whether the
density slope and the shape evolve with radius out to $\approx30$
kpc. Finally, we also allow the halo to be i) arbitrary oriented ii)
triaxial and iii) off-set from the nominal MW centre.

The analysis of the density distribution of the stars in our sample is based on the fit of density models, rather than on the fit of full dynamical models (see e.g. \citealt{DasK,das}). The main reason behind this choice is that we do not have any kinematic information, so the use of self-consistent dynamical models does not add any significant improvement to our study. Moreover, the knowledge of the spatial density distribution of the stellar halo is a useful piece of information not only if the halo is stationary, but also if it is not \vir{phase-mixed}, as suggested by cosmological $N$-body simulations \citep{HelmiAcq}.

The paper is
organized as follows.  In Section~\ref{sec:sample} we describe the
\gaia data as well as the method used to select an all-sky sample of
RRL candidates from a cross-match between \gaia and the 2MASS.  Here,
we also give the estimates of the purity and completeness of the
resulting sample. In Section~\ref{sec:halodist} we show and discuss
the spatial distribution of the selected RRLs. Section~\ref{sec:bay}
presents the details of the maximum likelihood approach employed to
fit the data with different halo density models and the final results
of this analysis. In Section~\ref{sec:ref} the best-fit halo model is
discussed together with the possible biases that can affect our
results.  The summary of the results can be found in
Section~\ref{sec:summary}.

\begin{comment}

The purpose of this work is to present a first study of the Galactic
halo exploiting the properties of the first data release of \gaia.
The results of this work are a little \vir{taste} of what will be
possible to achieve with the future \gaia data in the study of the
properties of the Galactic halo.

The Galactic halo is a relic of the Galactic formation and it is
composed mostly by old and metal-poor stars \citep{eggen}, therefore
to reveal its properties we need a sample of stars that can be found
in the halo, with known distances and bright enough to be seen up to
the extreme border of the Galaxy. In this context the RRLyrae stars
(RRLs, hereafter) can be considered one of the best halo tracers: they
are old and metal-poor variable stars, they lies on the
horizontal-branch and as a consequence they can be considered almost
standard-candles.  Moreover they are relatively bright so that can be
seen up to 100 kpc by the current surveys (e.g. SDSS, \citealt{S82}).

In this work we selected a sample of RRLs stars from the \gaia data
release 1 (GDR1, hereafter) and then we used them to study the
properties of the halo both making star density-maps and fitting
different models to the data with a Bayesian approach.

\end{comment}

%\cite{deasonhalo} 

\section{The RR Lyrae Sample} \label{sec:sample}
In this Section we describe the method used to select a sample of RRLs
from GDR1.

\subsection{\gaia Data release 1} \label{sec:dr1}

\gaia is an all-sky scanning space observatory, currently collecting
multi-epoch photometric and astrometric measurements of about a
billion stars in the Galaxy. More details on the \gaia mission and on
GDR1 can be found in \cite{gaiaMission}. In the first data release,
the information available for most faint sources is limited to
basic properties, such as positions on the sky and fluxes in the  broad \gaia $G$ band, which covers most of the visible spectra from
approximately 400\,nm to 10000\,nm \citep{gaiaG}.

In this work, we used the table \texttt{GaiaSource} released as part of
the GDR1 \citep{gaiaDR1}. \texttt{GaiaSource} contains a
number of auxiliary pieces of information, which provide plenty of
added value to the GDR1. For example, the errors on the mean flux
measurements can be used to separate constant and variable sources, and
even gauge the amplitude of the variability \citep[see,
  e.g.][]{bel16,deasonMC}. Moreover, the quality of the
astrometric fit, encapsulated by the so-called astrometric excess
noise, contains information regarding the morphology of the source,
and can be used to separate stars from galaxies
\cite[see][]{gaiasat}. The relevant \texttt{GaiaSource}
quantities used here (other than the sky coordinates RA, Dec), are:

\begin{itemize}
\item{$N_\text{obs}$}, the number of times a
  source has crossed a CCD in the {\it Gaia}'s focal plane;
\item{$\te{F}_{G}$}, the flux (electron per
  second) measured in the $G$ band averaging over
  $N_\text{obs}$ single flux measurements;
\item{$\sigma_{\te{F}_{G}}$}, the standard
  deviation of the $N_\text{obs}$ flux measurements;
\item{$G$}, the mean magnitude in the \gaia $G$ band
  \citep{gaiaphot} calculated from $\text{F}_G$;
\item{AEN}, the astrometric excess noise, which measures strong
  deviations from the best astrometric solution. The AEN should be large
  for objects whose behaviour deviates from that of point-like
  sources, as, for example, unresolved stellar binaries or
  galaxies (see \cite{aen} for  details).
\end{itemize}

Additionally, relying on the cross-match between \gaia and 2MASS, we calculated:

\begin{itemize}
\item{PM}, the total proper motion of each object.
  PM$=\sqrt{\mu^2_\alpha \cos^2{\delta} + \mu^2_\delta}$, where
  $\mu_\alpha$ and $\mu_\delta$ are the proper motions measured along RA and Dec, respectively.
\end{itemize}

\subsection{RR Lyrae in \gaia DR1}

%Measuring the  flux  of  stars several times with different time baselines, \gaia represents a perfect variable star machine.
%The final sample of variable objects is expected to be released  toward the end on the  emission,  however 

We use the parameters provided in \texttt{GaiaSource} to select RRLs from the GDR1.
Following the method outlined in \cite{bel16} and \cite{deasonMC}, we defined the quantity
\begin{equation}
\te{AMP}\equiv \log_{10}\left(\sqrt{N_\te{obs}} \frac{\sigma_{\te{F}_{G}}}{\te{F}_{G}} \right), 
\label{eq:amp}
\end{equation}
which can be used as a proxy for the amplitude of the stellar variability.
Indeed, for variable stars with well sampled light curves, $\sigma_{\te{F}_G}$ is proportional to the amplitude of the flux oscillation, while for non-variable stars it is just a measure of photon-count Poisson errors.
Thus, it is possible to set a threshold value for AMP that will select only variable candidates.
For instance, \cite{bel16} showed that most variable stars like Cepheids
and RRLs have $\te{AMP}>-1.3$.

The selection of variables through the AMP parameter suffers from the limitations that AMP is a time-averaged information and does not allow to distinguish between various types of variable objects: RRLs, Cepheids or Mira variables.
However, these different classes of pulsating stars populate a well-defined strip in the colour-magnitude diagram. 
Therefore, as we show below, one can overcome this problem by applying selection cuts both in AMP and in colour obtaining a fairly clean sample of RRLs.
Note that, high values of AMP are also expected for contaminants (e.g.\ eclipsing binaries) and artefacts (e.g.\ spurious variations related to \gaia cross-match failures). 
%that cannot be singled out without the light curves.
We discuss their importance in Sect.\ \ref{sec:cont}.

\subsubsection{The \gaia + 2MASS sample}

\begin{figure*}
\centering \includegraphics[width=1.0\textwidth]{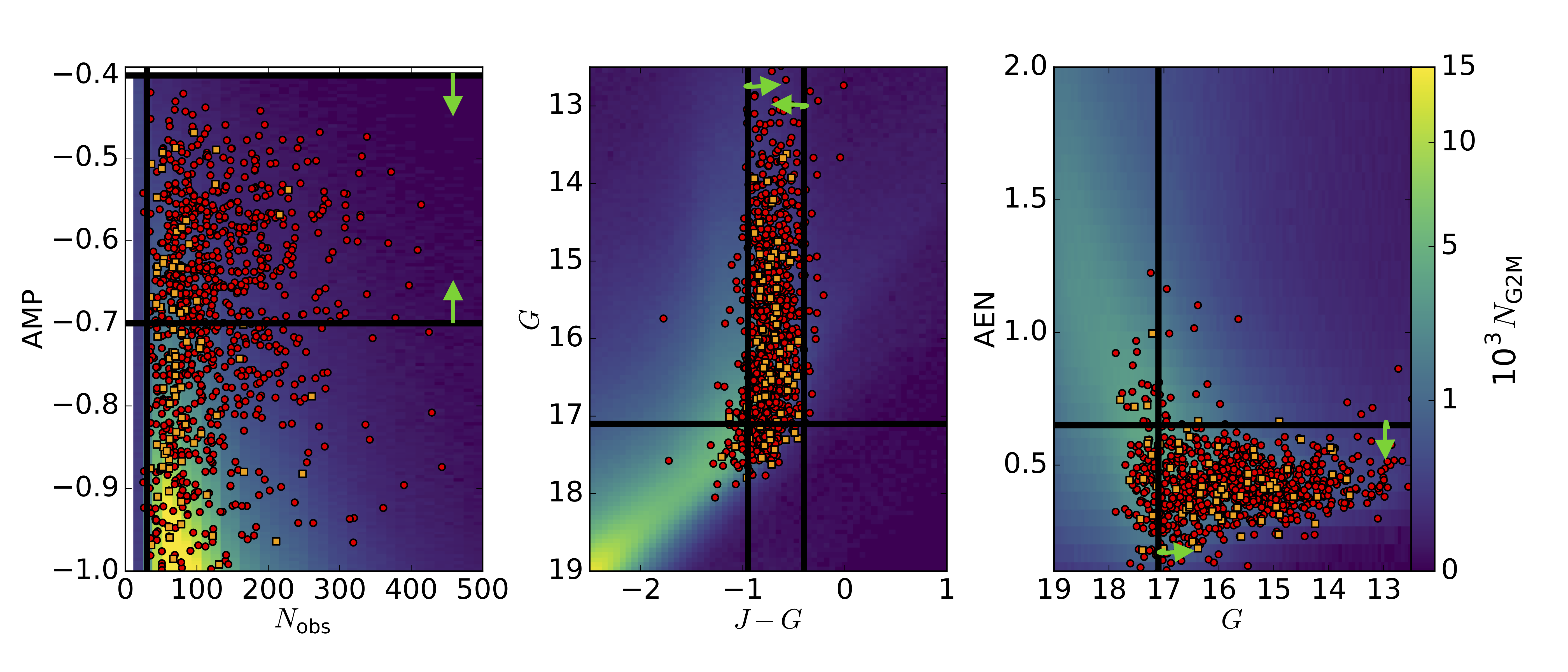}
\caption{Diagnostic diagrams for the selection criteria of our RRL
  sample.  Left-hand panel: distribution of stars in the
  AMP-$N_\te{obs}$ space; middle panel: distribution of stars in the
  colour-magnitude diagram ($G$ magnitude from \gaiap, $J$ magnitude
  from 2MASS); right-hand panel: distribution of stars in the AEN-$G$
  magnitude space.  The colour-maps show the distribution of objects
  in the G2M catalogue, while points show a a randomly selected subsample of bona fide
  RRLs from GCSS (red circles, 5\% of the original sample) and GS82 (orange squares, 35\% of the original sample) catalogues
  (see Sec. \ref{sec:aux}). The horizontal-black and vertical-black
  lines show the selection cuts used to obtain the final sample of
  RRLs (see Tab. \ref{tab:tab_cut}), while the green arrows indicate
  the regions used to obtain the final sample.  }
\label{fig:cut_sel}
\end{figure*}

As mentioned, GDR1 reports photometric information only in the $G$ band.
We derived a colour ($J-G$) for each source by cross-matching \texttt{GaiaSource} with the 2MASS survey data \citep{2mass} using the nearest-neighbour method with an aperture of 10 arcsec obtaining the final  \texttt{GaiaSource} + 2MASS sample of stars (G2M, hereafter).
We chose 2MASS mainly due to its uninterrupted all-sky coverage. 
%The colour index $J-G$ has been calculated using the $G$ magnitude from \gaia and the $J$ magnitude from 2MASS. 
The observed magnitudes have been corrected for extinction due to interstellar dust using the maps of \cite{dustext} and the transformation $A_G=2.55 E(B-V)$ for
the $G$ band \citep[see][]{bel16} and $A_J=0.86 E(B-V)$ for
the $J$ band \citep{2massext}.

\subsubsection{Auxiliary RR Lyrae datasets}\label{sec:aux}

In order to extract a reliable sample of RRL stars from the G2M catalogue, we must apply ad-hoc selection criteria.
To this aim, we used two samples of bona-fide RRLs: the CSS \citep{CSS,drake2} and the Stripe 82 (S82, \citealt{S82}) catalogues.
These samples allowed us to identify the optimal
selection criteria, analyse the completeness and the
contamination of the catalogue\footnote{The completeness indicates the fraction of
  recovered true RRLs as a function of the apparent magnitude, while the
  contamination is an estimate of the fraction of spurious objects
  (non RRLs) that \vir{pollute} our sample. \label{fn:comp}} and 
 estimate the RRL absolute magnitude in the $G$ band.
The CSS contains about 22700 type-ab RRL stars distributed over a large area of the sky (about 33,000 deg$^2$ between $0^\circ<\alpha<360^\circ$ and $-75^\circ<\delta<65^\circ$) and extended up to a distance of 70 kpc. 
The completeness of this sample is constant (at $\sim 65\%$) only for $13<{V}<15$, while it quickly decreases outside this range. 
Most importantly, as shown in Fig.~13 
of \citet{CSS}, for objects fainter than $V\sim15$, the
completeness is a strong function of the number of observations and
thus varies appreciably across the sky.
SDSS's Stripe 82 covers a 2.5$^{\circ}$-wide and $100^{\circ}$-long
patch of sky aligned with the celestial equator and contains \vir{only} 483 RRLs. 
However, the sample is very pure (with less than $<1\%$ of contaminants) and complete up to a distance of 100 kpc. 

The large number of stars in CSS is useful to
define the selection criteria (see Sec. \ref{sec:selection}) and to
estimate the absolute magnitude in the $G$ band (see Sec.
\ref{sec:amag}), while the high quality of S82 sample is ideal to
analyse the completeness and contamination of our final sample (see
Sec. \ref{sec:comcont} and Sec. \ref{sec:cont}). 
A cross-matching of the CSS and S82
catalogues with G2M using an aperture of 1 arcsec led to the two samples
CSS+G2M (GCSS hereafter) and S82+G2M (GS82 hereafter).
%CSS+G2M (11700 objects, GCSS hereafter) and S82+G2M (233
%objects, GS82 hereafter).

\subsubsection{RR Lyrae selection cuts} \label{sec:selection}

In this section we describe how the final sample of RRLs was obtained
from the G2M catalogue. The selection was driven by the properties of
the bona fide RRLs in the GCSS and GS82 catalogues (see
Sec. \ref{sec:aux}) in order to maximise completeness of the sample
and its spatial uniformity, while keeping the level of contamination
low (see Sec. \ref{sec:comcont}).

First of all,  in order to exclude a region likely dominated by the Galactic disc, we removed all the stars in the G2M catalogue located between the Galactic latitudes $\te{b}=-10^\circ$ and $\te{b}=10^\circ$.
Our limit in latitude ($|\te{b}|=10^\circ$) is lower with respect to other works in literature (e.g. \citealt{deasonhalo,das}), in which, however, the choice was mainly motivated by the limited sky-coverage of the used survey. Given the unprecedented sky-coverage of our data sample, we decided to push forward the study of the halo structure exploring also the region at low Galactic latitude that is usually not well sampled by other surveys.
We are aware that our final sample could be polluted by Galactic disc contaminants, but in the following sections we carefully analyse the level of contamination and all our results are obtained taking into account  the possible biases due to stars of the Galactic disc.

Figure~\ref{fig:cut_sel} shows the distribution of the G2M stellar
density (yellow-blue-purple colour-maps), a randomly selected subsample  of  RRLs in GCSS (red points) and  in GS82 (orange squares) in the $N_\te{obs}$-AMP (left-hand panel), color-magnitude (middle panel) and AEN-$G$ (right-hand panel) planes. 
The bona fide RRLs occupy a
well defined strip in colour, thus we excluded all the stars with the
$J-G$ colour index greater than $-0.4$ and lower than $-0.95$ as shown by
the vertical black lines in the middle panel of
Fig.~\ref{fig:cut_sel}. 
It is worth noting that most of the \vir{normal} stars occupy this colour interval, therefore this cut mostly eliminates artefacts.
The left-hand panel shows that the genuine
RRLs are almost uniformly distributed in the AMP range of the G2M
sample, however the contamination by spurious
objects increases rapidly for AMP$<-0.7$ (see
Sec.\ \ref{sec:comcont} and Fig.~\ref{fig:cont}), thus we only retained stars with variability amplitudes above this value. 
With regards to completeness, the faint magnitude limit plays an important role. According to our analysis, $G=17.1$ is the faintest magnitude that we can reach to obtain a sample
with spatially uniform completeness (see Sec.\ \ref{sec:comcont} for further
details).  
The number of  bright stars with $G<10$, corresponding to RR Lyrae with distances less than 1 kpc from the Sun, is very small compared to the number of objects in our final catalogue.  Therefore, instead of extending our completeness/contamination analysis at the very bright magnitudes (see Sec. \ref{sec:comcont} and \ref{sec:cont}), we decided to put the bright magnitude limit at $G=10$.

The selection criteria described above involving colour, AMP and magnitude have
the largest impact on the definition of our sample of RR Lyrae. 
However, we also applied a few minor refinements. 
The right-hand panel of Fig.~\ref{fig:cut_sel} shows that most of the bona fide
RRLs have a very small value of AEN, so we excluded all sources with
$\te{AEN}>0.65$ as shown by the horizontal-black line. 
This cut likely removes
contaminant extragalactic objects since they typically have
$\te{AEN}\approx2$ \citep{bel16} and some of the eclipsing binaries
that survive the colour selection. Additionally, to further clean
the sample from possible nearby Galactic disc contaminants, we cull
all the stars with a total proper motion, PM, greater than $50$\,mas/yr. Given differences in the light-curves and its sampling, the significance
of AMP (Eq. \ref{eq:amp}) might depend on the number of photometric measurements
$N_\te{obs}$. With this in mind, we impose $N_\te{obs}>30$: the focal plane of \gaia has an array of 9$\times$7 CCDs, so
all objects with less than 3 complete \gaia transits are excluded. 

It would be useful to have an estimate of the photometric metallicity to retain only genuine  metal poor stars from the halo and effectively exclude metal rich contaminants from the Galactic disc. However, the photometric metallicity estimate requires a basic knowledge of the shapes of the RRL light curves which is not available in our dataset (see e.g. \citealt{jurcsik}).

Finally, we masked a few regions of the sky. 
First we removed the area near the Magellanic Clouds using two circular apertures: one centred on $(\te{l}, \te{b})=(280.47^\circ, -32.89^\circ)$ with an angular radius of $9^\circ$  for the Large Magellanic Cloud (LMC) and the other centred on $(\te{l}, \te{b})=(302.80^\circ, -44.30^\circ)$ with a radius of $7^\circ$ for the Small Magellanic Cloud (SMC). 
By inspecting the sky distribution of the stars in our RRL sample, we noticed the presence of two extended structures (S1 and S2, hereafter), that were not connected to any known halo substructures, but are likely objects instead produced by \gaia cross-match failures (see Sec.~\ref{sec:comcont}) that \vir{survived} our selection cuts. We decided to mask these sky regions as well by removing all the stars in the following boxes $\te{l}=[167^\circ, 189^\circ]$ $\te{b}=[16^\circ, 22^\circ]$  for S1 and $\te{l}=[160^\circ, 190^\circ]$ $\te{b}=[63^\circ, 73^\circ]$  for S2. 
The final sample has been further cleaned to exclude \vir{hot pixels} using  a simple median-filter method. 
We first built a sky-map using pixels of $30\arcmin$, then we replaced the number of stars in each pixel by the median of the star-counts calculated in a squared window of four pixels. Finally, we calculated the ratio between the original sky-map and the one processed with the median filter and all the objects in pixels with a ratio larger than 10 were removed. The properties of the median-filter has been gauged to reveal small-scale features, since the most evident large-scale structures have been already removed (LMC, SMC, S1 and S2). The \vir{spotted} hot-pixels correspond to known globular clusters (e.g.\ M3 and M5) or are connected to \vir{remnants} of cross-match failure structures (see Sec.~\ref{sec:comcont}).
In order to fully exploit the all-sky capacity of our sample, we decided not to exclude {\it a priori} any portion of the sky containing known substructures (e.g. unlike \citealt{deasonhalo} which masked out the Sagittarius stream). The analysis of the most significant substructures found in this work can be found in Sec. \ref{sec:halodist} and \ref{sec:ref}.

The final sample contains about 21600 RRLs  that can be used to have a direct look at the distribution of stars in the Galactic halo (Sec. \ref{sec:halodist}). 
The final number of object is similar to the one in the CSS catalogue, however  we cover a larger area of the sky (almost all-sky).
Our sample populates 58\% of the halo spherical volume within the Galactocentric distance of about 28\,kpc which represents a significant improvement in volume fraction as compared to previous works (e.g.\ 20\% in \citealt{deasonhalo}).
%A supplementary cut on the Galactocentric latitude $\theta$ (defined in Eq.~\ref{eq:freftheta}) has been applied to obtain a cleaner sample of halo RRLs (about 13700 stars with halo volume coverage of 44\%) that has been analysed with a likelihood sampling approach in Sec.~\ref{sec:bay}. 
A summary of the applied selection cuts can be found in Tab.~\ref{tab:tab_cut}.

\begin{table}
\centering
\begin{tabular}{lc}
\hline
\multicolumn{2}{c}{Selection cuts} \\ \hline
|$\te{b}$| {[}deg{]} & $>10$ \\ 
$G$ {[}mag{]} & (10, 17.1) \\
AMP & (-0.7, -0.4) \\
$J-G$ & (-0.95, -0.4) \\
$N_\te{obs}$ & $>30$ \\
AEN & $<0.65$ \\
PM {[}mas/yr{]} & $<50$\\
|$\theta$| {[}deg{]} $^\dagger$ & $>20$ \\ 
\hline
\multicolumn{2}{c}{Structure cuts} \\ \hline
LMC&  $\te{D}_\te{LMC}>9^\circ$ \\  
SMC&  $\te{D}_\te{SMC}>7^\circ$\\  
S1&  $\te{l} \notin [167^\circ,189^\circ] \vee \te{b} \notin [16^\circ,22^\circ]$\\  
S2&  $\te{l} \notin [160^\circ,190^\circ]  \vee \te{b} \notin [63^\circ,73^\circ]$\\  
\hline
$N_\te{stars}$ & 21643 (13713$^\dagger$) \\ 
$f_\te{V}$ {[}\%{]} & 58 (44$^\dagger$) \\ \hline
\end{tabular}
\caption{Summary of the selection cuts used to obtain the final sample of RRLs from the G2M catalogue. The description of parameters used in the sample selection and the details on the cut substructures can be found in Sec.~\ref{sec:dr1}. $\te{D}_\te{LMC}$ and $\te{D}_\te{SMC}$
are the sky angular distances from the LMC and SMC, respectively.
$\theta$ is the Galactocentric latitude (defined in Eq.~\ref{eq:freftheta}) and was  estimated by assuming an RRL absolute magnitude of $M_\te{RLL}=0.525$. The $\dagger$ refers to the subsample used in the likelihood analysis in Sec.~\ref{sec:bay}. The bottom part of the table gives a summary of the whole sample. $N_\te{stars}$ is the number of stars in the sample and $f_\te{V}$ is the fraction of the spherical volume of the halo sampled by our stars between Galactocentric distance 0\,kpc and 28\,kpc. }
\label{tab:tab_cut}
\end{table}

\subsubsection{Absolute magnitude and distance estimate} \label{sec:amag}

\begin{figure}
\centering
\includegraphics[width=1.0\columnwidth]{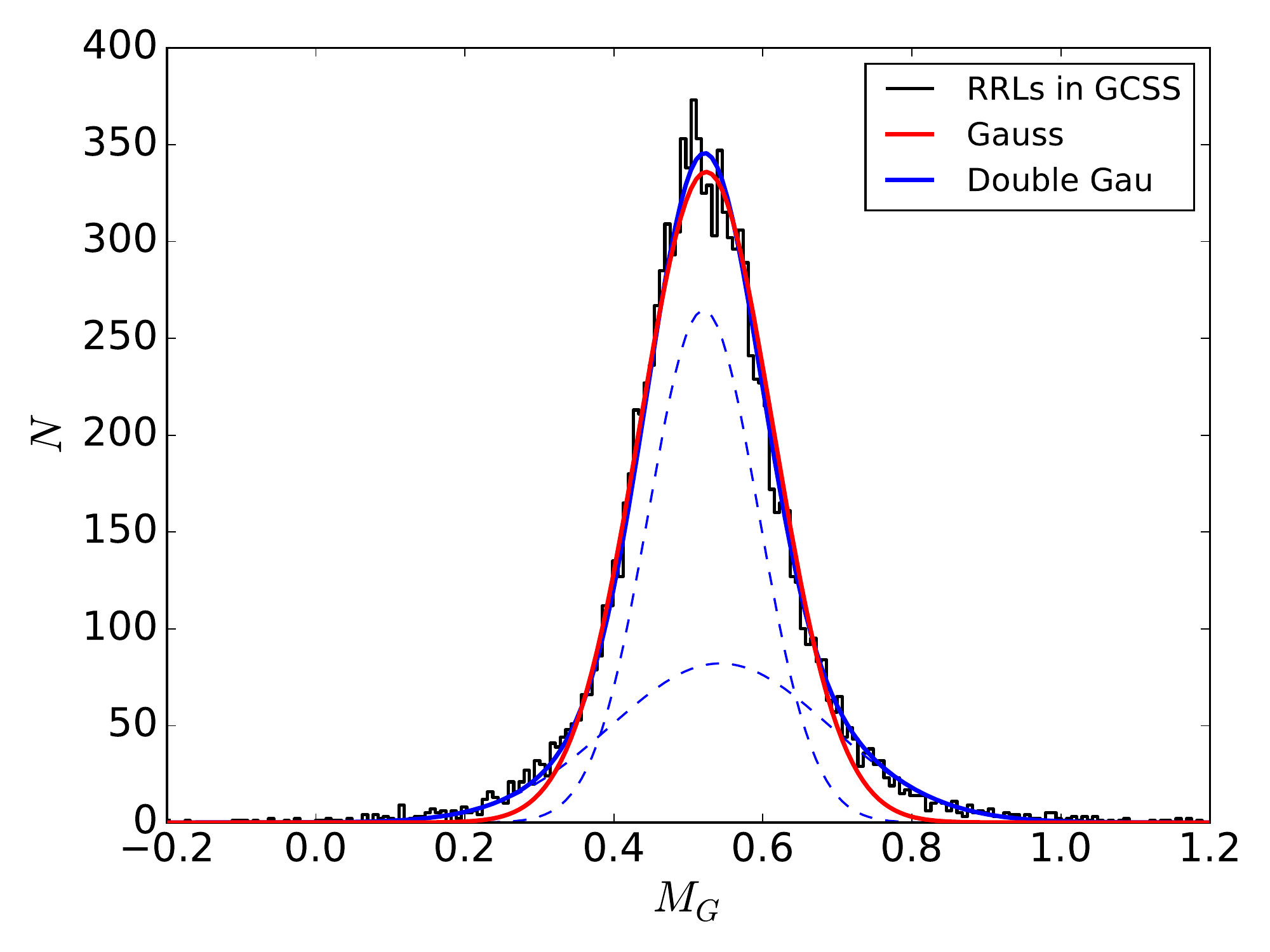}
\caption{Distribution of $G$ band absolute magnitudes for bona fide RRLs  in the GCSS catalogue (\gaiap+CSS, see Sec.~\ref{sec:aux}). The coloured lines show the fit with different functions: single Gaussian (red) and  double Gaussian (blue, with blue dashed lines showing individual Gaussian components of the fit).}
\label{fig:Mg}
\end{figure}

Despite photometric variability, RRLs have an almost constant
absolute magnitude and, having the apparent magnitudes $G$, we can
directly estimate the heliocentric distances through
\begin{equation}
\log\left(\frac{\te{D}\su}{\te{kpc}}\right)=\frac{G-M_G}{5}-2,
\label{eq:amag}
\end{equation}
where $M_G$ is the absolute magnitude in the \gaia $G$ band.
To estimate $M_G$ we used the RRLs in GCSS, as they have heliocentric distances estimated from the period-luminosity relation. 
%and we can obtain their absolute Magnitude in the $G$ band using Eq.~\ref{eq:amag}. 
The resulting $M_G$ distribution is shown in Fig.~\ref{fig:Mg}: it has a well defined peak at $M_G\approx0.5$ and a small dispersion. We fit this distribution with a Gaussian function obtaining a mean of $0.525$ and a dispersion of $0.090$, comparable to the uncertainties of the V-band absolute magnitude of RRLs (see e.g. \citealt{Vivas}).
The above Gaussian function perfectly describes the data in the central parts, but the distribution shows broader wings for $M_G<0.3$ and $M_G>0.7$.  
The objects that populate the wings could be Oosteroff type II RRLs and objects influenced by the Blazhko effect (\citealt{CSS} and references therein).  
A better fit can be obtained using a double Gaussian model, where two Gaussian components peak at about the same absolute magnitude of the single Gaussian fit (see Fig. \ref{eq:amag}). Given the small dispersion around the mean, we decided to set the absolute magnitude for all the RRLs in our sample to a single value $M_\te{RRL}=0.525$. An analysis of the effect of this approximation on the results can be found in Sec.~\ref{sec:biasamag}.

\subsubsection{Completeness} \label{sec:comcont}

\begin{figure}
\centering
\includegraphics[width=1.0\columnwidth]{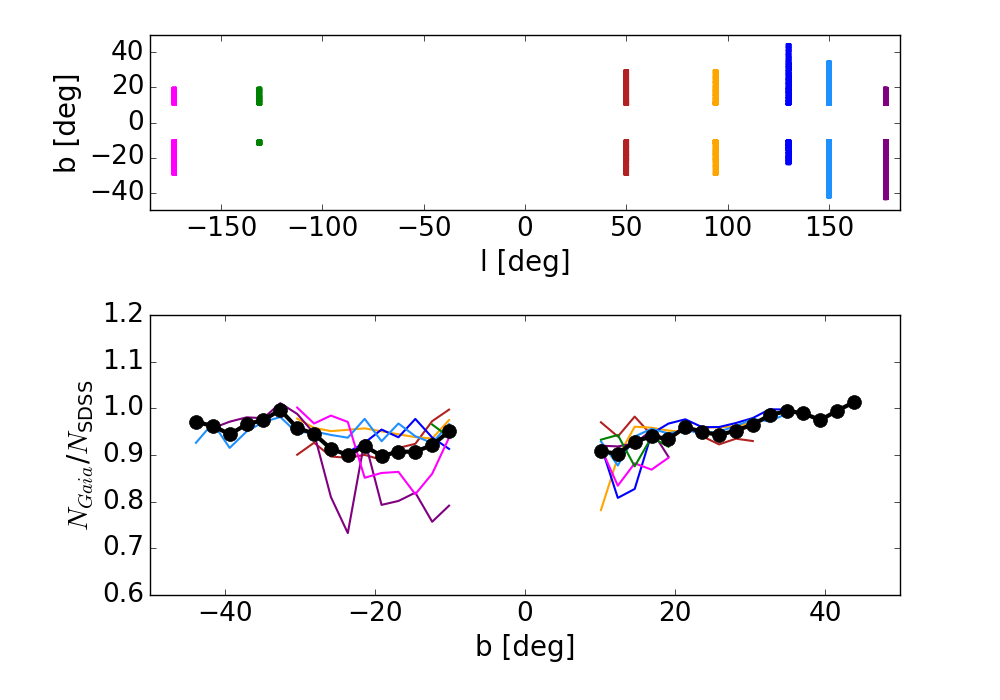}
\caption{Completeness analysis of the sources in \texttt{GaiaSource} (Sec. \ref{sec:dr1}) as a function of Galactic latitude. 
\textbf{Top panel}: regions of the sky considered in this analysis, each stripe has a  constant Galactic longitude.
\textbf{Bottom panel}: ratio between the number of stars in the \texttt{GaiaSource} ($N_\gaia$) and stellar sources in the SDSS DR7 ($N_\te{SDSS}$, \citealt{sdss}) in bins of Galactic latitude. The lines refer to the ratio obtained for stars located in regions of the same colour shown in the top panel. The dots and the black line indicate the ratio obtained considering the stars in all the \vir{stripes}. The stars in \gaia have been selected using the $16<G<18$ cut, and the SDSS sources using the $16<r<18$ cut (further details on the text).}
\label{fig:comptot}
\end{figure}

Before studying the  properties of the stellar halo (as traced by RRLs), it is fundamental to consider the completeness of our sample of RRLs.  

First of all, we checked that the scanning law of \gaia does not cause
an intrinsic decrease of the completeness at low Galactic latitude. We
compared the number of objects in \texttt{GaiaSource} with the number
of stellar sources in the Data Release 7 of the Sloan Digital Sky
Survey (SDR7, \citealt{sdss}) in a series of stripes at fixed Galactic
longitudes, selected using the footprint of SDR7. The top panel of
Fig. \ref{fig:comptot} shows the position of the stripes in Galactic
coordinates. We selected all stellar sources in the SDR7 with $16<r<18$, where $r$ is the $r$-band magnitude. In this
magnitude range the SDR7 can be considered 100\%
complete\footnote{\url{http://classic.sdss.org/dr7/products/general/completeness.html}}.
We chose the $r$ band because the peak of the filter response is
almost coincident in wavelength with the one of the \gaia $G$ band
(see Sec.~\ref{sec:dr1}), therefore the two magnitudes are
directly comparable.
The bottom panel of Fig. \ref{fig:comptot} shows the ratio between the
number of stars in \gaia and SDR7 in bins of Galactic latitude for
individual SDSS stripes and considering all the stripes together. The
ratio does not show significant variations as a function of $\te{b}$ with values
between 0.9 and 1.0. We conclude that there is no evidence for strong
intrinsic completeness variations in the \gaia catalogue for
$\te{b}>10^\circ$.

\begin{figure}
\centering
\includegraphics[width=1.0\columnwidth]{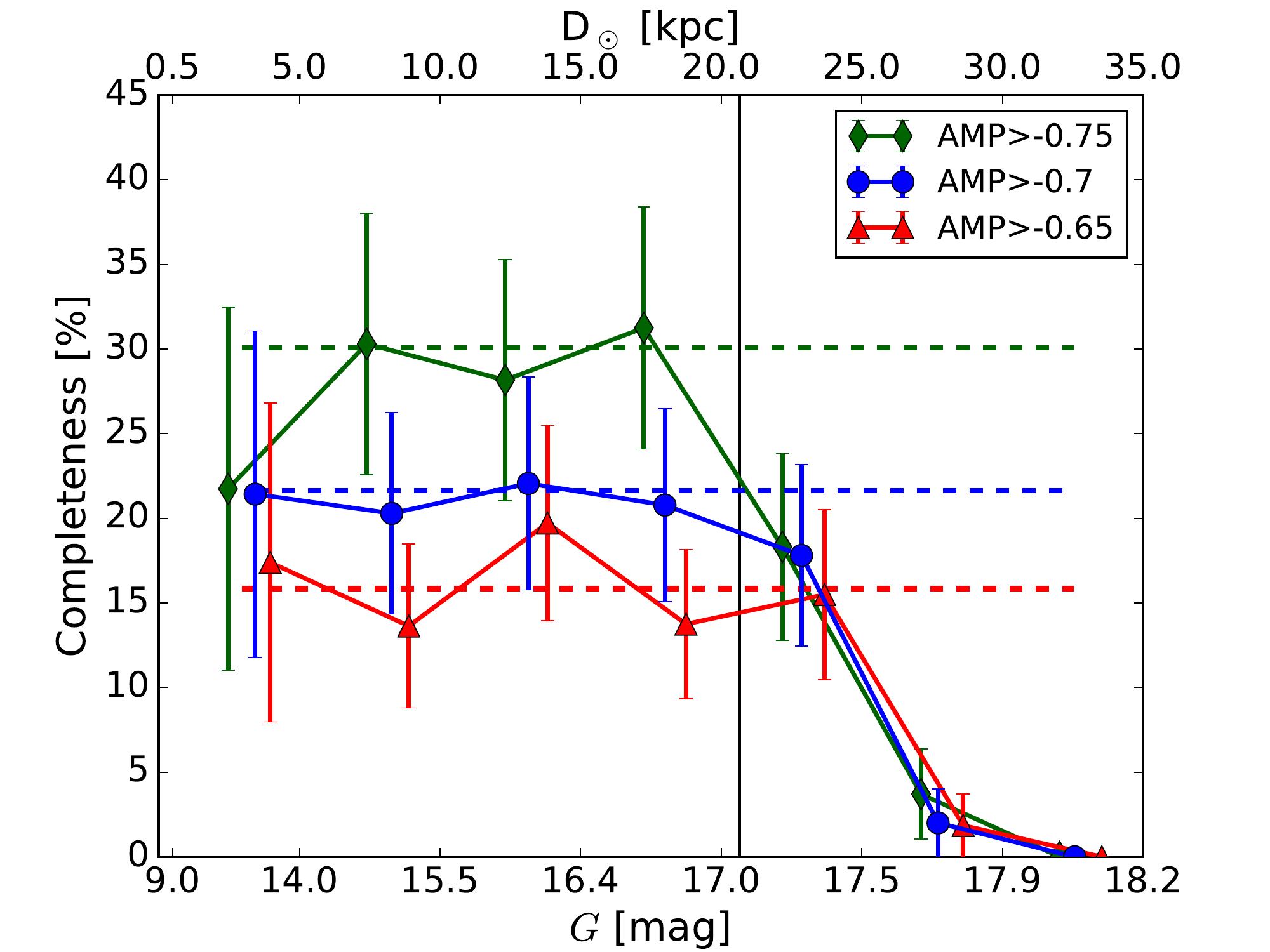}
\caption{Completeness of our samples of RRLs as a function of distance from the Sun ($\te{D}_\odot$) or $G$ magnitude. The conversion from $\te{D}_\odot$ to $G$ has been obtained from Eq.~\ref{fig:Mg} assuming a constant absolute magnitude $M_\te{RRL}=0.525$.
Different symbols indicate completeness for samples obtained using different AMP cuts: red-triangles for $\te{AMP}>-0.65$, blue circles for $\te{AMP}>-0.70$ and green diamonds for $\te{AMP}>-0.75$
The error bars were calculated using the number of stars in each magnitude range and Poisson statistics. The dashed lines indicate the completeness estimated in the $G$ magnitude range of $15<G<17$, while  the vertical black line marks the $G=17.1$ faint magnitude limit of our final sample (see Sec.~\ref{sec:comcont}).}
\label{fig:comp}
\end{figure}

We estimated the completeness (and the contamination) using the RRLs in our auxiliary catalogues: CSS and S82 (Sec. \ref{sec:aux}).
In particular, S82 represents a complete ($\sim$100 \%) and pure ($\sim$99 \%)  catalogue of RRLs located up to 100\,kpc from the Sun, so it is perfect to test both the contamination and the completeness of our sample.  
We compared the number of stars in the original S82 sample with the ones contained in the GS82 after the selection cuts described in Sec.~\ref{sec:selection}, in bins of heliocentric distance.   
Fig.~\ref{fig:comp} shows the level of completeness as a function of magnitude/distance assuming different lower limits in AMP ($-0.65$  red triangles,  $-0.70$  blue circles, $-0.75$ green diamonds) in the range of magnitudes $15<G<17$. 
The results are in agreement with the distance-based estimate of completeness as shown by the dashed lines. 
The level of completeness is relatively low, ranging from about 15\% for $\te{AMP}>-0.65$ to about 30\% for $\te{AMP}>-0.75,$ but it is reasonably constant up to about $G=17.5$ then it abruptly decreases to 0 at $G\approx18,$ so we decided to conservatively cut our sample at $G=17.1$ (vertical black line in Fig.~\ref{fig:comp}, see Sec.~\ref{sec:selection}).

\begin{figure}
\centering
\includegraphics[width=1.0\columnwidth]{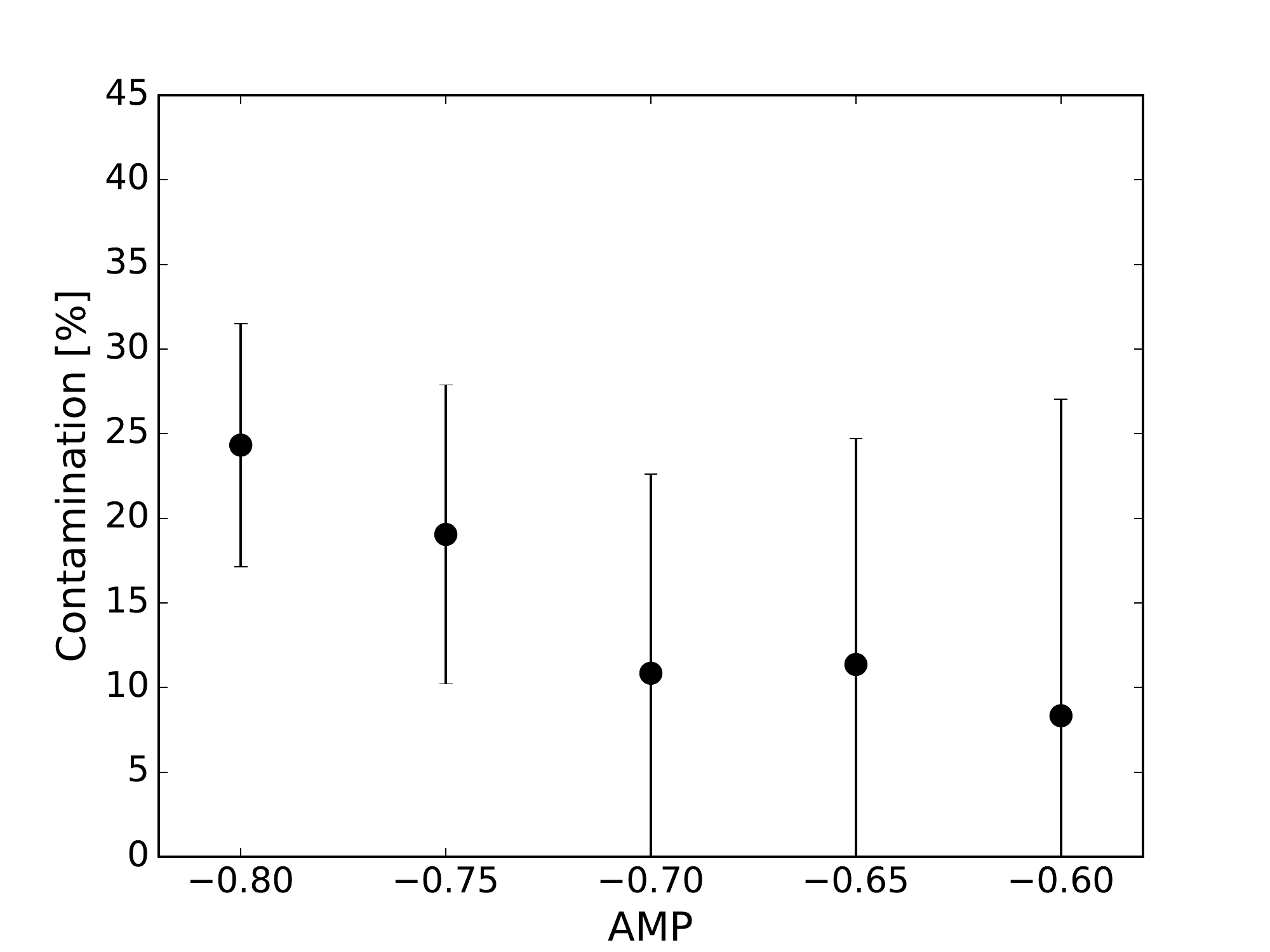}
\caption{Contamination of the RRLs sample  as a function of the AMP cut. 
The contamination is estimated using the S82 catalogue (see Sec. \ref{sec:aux}) in the range of $G$ magnitudes $15<G<17$. The error-bars have been calculated by propagating Poisson uncertainties from the number of stars in AMP bins.}
\label{fig:cont}
\end{figure}

We also used the GCSS catalogue and found no significant variation of the completeness as a function of the Galactic sky coordinates $\te{l}$ and $\te{b}$, although we found a mild trend for increasing $N_\te{obs}$. 
This is expected since the larger the number of flux measurements the better the sampling of the light curves and, as a consequence, the AMP cut (Eq. \ref{eq:amp}) improves its effectiveness in selecting RRLs. 
However, the increase is not dramatic as all the differences are within 10\%. In conclusion, for the purpose of this work, we considered the completeness of our catalogue of RRLs uniform across our sky coverage and the considered magnitude range (see Tab. \ref{tab:tab_cut}).

\subsubsection{Contamination} \label{sec:cont}

We estimated the contamination of spurious sources as
\begin{equation}
\te{Contamination}=\frac{N_{\rm S} - N_\te{GS82}}{N_S},
\end{equation}
where $N_{\rm S}$ and $N_\te{GS82}$ indicate the number of stars in our
samples and in GS82 catalogue after the selection cuts in
Sec.~\ref{sec:selection}. 
The S82 sample is pure \citep{S82}, so all
the stars in our sample that are not present in the GS82 catalogue are likely contaminants. 
As before, we considered the
magnitude range $15<G<17$ for different lower limits of AMP.
For AMP threshold lager than
$-0.7$ the level of contamination is lower than 10\% (Fig.\ \ref{fig:cont}) with a mild
increase toward low Galactic latitudes, where the contribution
of the Galactic disc is larger. 
For AMP cut lower than $-0.7$ the
contamination fraction rapidly increases (about 25\% for
$\te{AMP}=-0.8$). We expect that the contaminants of our sample are 
eclipsing-binaries in the Galactic disc and possible instrumental
artefacts. As shown and discussed in detail in \cite{bel16} some
regions of the \gaia all-sky map are crossed by sharp strips with a
large excess of objects. These strips are similar in
width to the field of view of \gaia and are regularly spaced on
the sky. \cite{bel16} propose that these are spurious features due to
failures in the cross-match procedure of \gaiap, so that at some
epochs the flux of a star comes from a different object. The spurious
measurement of the flux increases $\sigma_{\te{F}_G}$ and moves the
stars in the region of high AMP \vir{polluting} our samples. We found
that for $\te{AMP}>-0.8$ the contaminants are mostly related to
cross-match failures. 
Unlike the disc contaminants, the cross-match
failures have a complicated and poorly understood spatial
distribution, so these structures are difficult to be taken in account
in the study of the properties of the Galactic halo. For this reason,
we decided to use $\te{AMP}=-0.7$ as the lower limit in our selection cut (see Sec.~\ref{sec:selection}).

%CN 8 July 17

\section{Distribution of the RRL\lowercase{s} in the inner stellar Halo} \label{sec:halodist}

\subsection{A first Gaia look at the stellar halo} \label{sec:look}

\begin{figure*}
\centering
\includegraphics[width=0.9\textwidth]{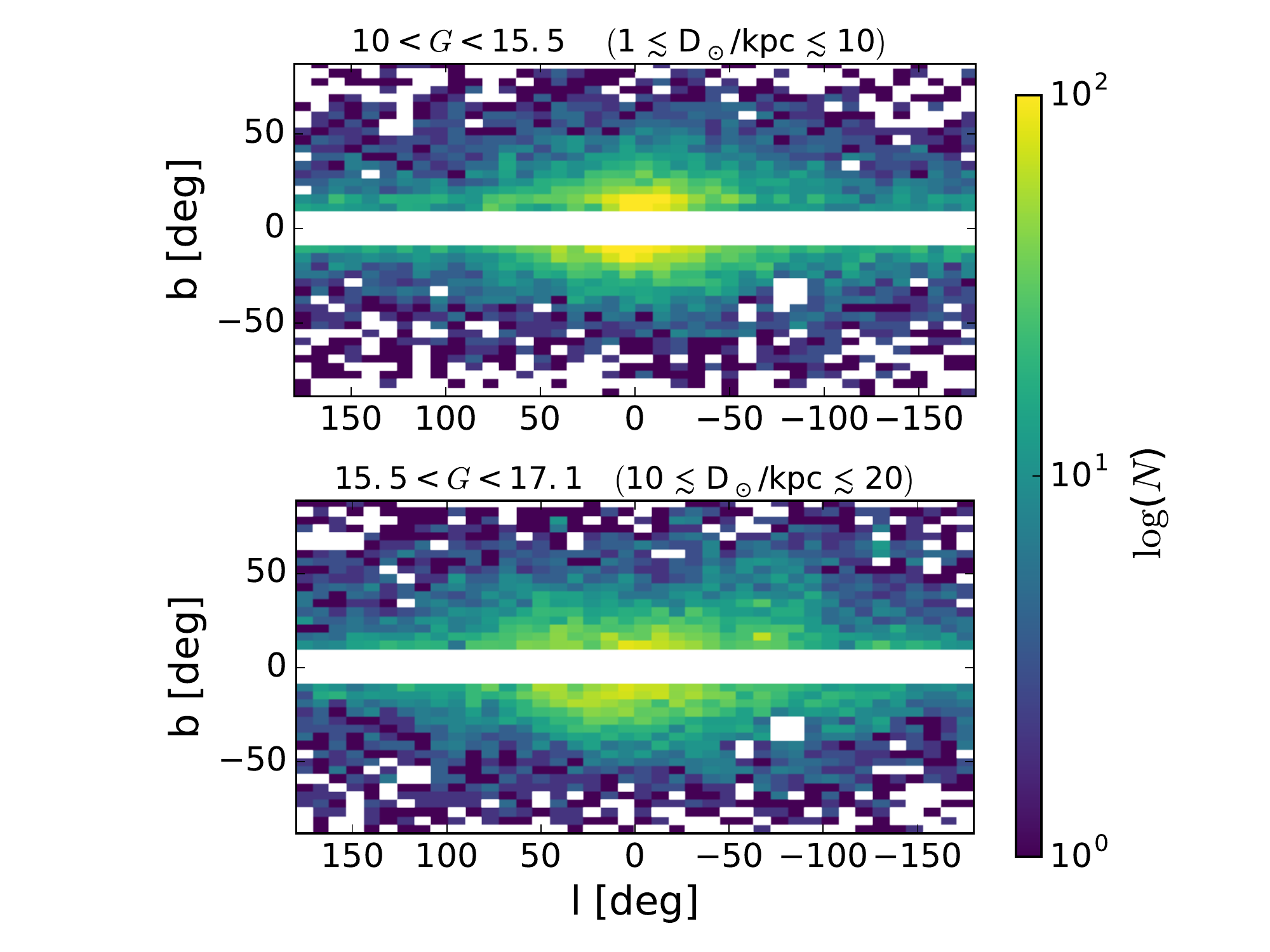}
\caption{Star-count all-sky maps in the Galactic coordinates (l, b)
  for the RRLs in our sample (Sec. \ref{sec:selection}) in the
  magnitude intervals $10<G<15.5$ (upper panel) and $15.5<G<17.1$
  (lower panel). The \vir{holes} between $\te{l}=-50^\circ$ and
  $\te{l}=-100^\circ$ are due to the mask used to eliminate the
  contribution of the Large and Small Magellanic Clouds (see
  Sec. \ref{sec:selection}).}
\label{fig:skymap}
\end{figure*}

Fig. \ref{fig:skymap} shows the distribution of the RRLs of our final
sample as a function of the Galactic coordinates (l, b). We split the
stars in the sample into $G$ magnitude intervals: $10<G<15.5$,
corresponding to heliocentric distances between $\sim$1 kpc and 10 kpc
(Eq. \ref{eq:amag}), and $15.5<G<17.1$, i.e.  with heliocentric
distances between 10 kpc and 20 kpc (assuming that all RRLs have
absolute magnitude $M_G=0.525$). We stress that Fig. \ref{fig:skymap}
represents the first all-sky view of the distribution of the RRLs in
the inner halo.

The first magnitude range covers a portion of the Galaxy mostly
located in the side of the Galaxy containing the Sun between the
Galactic radii 1.4 kpc and 18 kpc.  The distribution of stars in this
region is quite regular: as expected most of the stars are in the
direction of the Galactic centre ($\te{l}\approx 0$) and there are not
evident asymmetries with respect to the Galactic plane.  The second
magnitude range covers the Galactic distances between 18 kpc and 28
kpc in the side of the Galaxy containing the Sun and between 3 kpc and
12 kpc in the other side. In this distance range the distribution of
the RRLs is less regular with clear asymmetries: in particular, an
excess of stars is evident at high Galactic latitude around
$\te{l}=-50^\circ$.  In both magnitude intervals an over-dense band of
stars at low Galactic latitude ($\te{|b|}<15^\circ-20^\circ$) can be
seen running all around the Galaxy. A discussion of the nature of
these structures can be found in Sec. \ref{sec:struct}.

\subsection{Setting the frame of reference} \label{sec:fref}

We set a left-handed Cartesian reference frame
$(\te{X}_\te{g},\te{Y}_\te{g},\te{Z}_\te{g})$ centred in the Galactic
centre and such that the Galactic disc lies in the plane
$(\te{X}_\te{g},\te{Y}_\te{g})$, the Sun lies on the positive
$\te{X}$-axis and the Sun rotation velocity is
$\dot{\te{Y}}_\te{g}>0$.  The actual vertical position of the Sun with
respect to the galactic disc is uncertain, but it is estimated to be
be smaller than 50 pc (\citealt{zsun} and reference therein), and thus
negligible for the purpose of this work.  In this Cartesian reference
frame, the coordinates of an object with Galactic longitude $\te{l}$,
Galactic latitude $\te{b}$ and at a distance $\te{D}_{\odot}$ from the
Sun are
\begin{equation}
\renewcommand{\arraystretch}{1.5}
\left\{ \begin{array}{lll}
	\te{X}_\te{g}=\te{X}_{\te{g},\odot} - \te{D}_{\odot} \cos \te{b} \cos \te{l},\\
	\te{Y}_\te{g}= \te{D}_{\odot} \cos \te{b} \sin \te{l},\\
	\te{Z}_\te{g}= \te{D}_{\odot} \sin \te{b}, \end{array} \right.
 \label{eq:fref}
\end{equation}
where $\te{X}_{\te{g},\odot}$ is the distance of the Sun from the
Galactic centre.  In this work we assume $\te{X}_{\te{g},\odot}=8
\ \te{kpc}$ \citep{rsunbovy}, but the main results of our work are
unchanged for other values of $\te{X}_{\te{g},\odot}$, within the
observational uncertainties (between 7.5 kpc, e.g. \citealt{rsunfra},
and 8.5 kpc, e.g. \citealt{rsunsch}).

For a star with Galactocentric Cartesian coordinates
$(\te{X}_\te{g},\te{Y}_\te{g},\te{Z}_\te{g})$ we define the distance
from the Galactic centre
\begin{equation}
\te{D}_\te{g}=\sqrt{\te{X}^2_\te{g} + \te{Y}^2_\te{g}+\te{Z}^2_\te{g}},
\label{eq:frefDg}
\end{equation}
the Galactocentric cylindrical radius
\begin{equation}
\te{R}=\sqrt{\te{X}^2_\te{g} + \te{Y}^2_\te{g}},
\label{eq:frefR}
\end{equation}
the Galactocentric latitude
\begin{equation}
\theta= \arctan \frac{\te{Z}_\te{g}}{\te{R}},
\label{eq:freftheta}
\end{equation}
and the Galactocentric longitude
\begin{equation}
\phi= \arctan{\frac{\te{Y}_\te{g}}{\te{X}_\te{g}}}.
\label{eq:frefphi}
\end{equation}

Assuming that the stellar halo is stratified on concentric ellipsoids,
it is useful to introduce another Cartesian frame
$(\te{X},\te{Y},\te{Z})$, aligned with the ellipsoid principal axes,
and to define the elliptical radius
\begin{equation}
\te{r}_\te{e}=\sqrt{\te{X}^2 + \te{Y}^2\te{p}^{-2} + \te{Z}^2\te{q}^{-2} },
\label{eq:ellrad}
\end{equation}
where $\te{p}$ and $\te{q}$ are, respectively, the
$\te{Y}$-to-$\te{X}$ and $\te{Z}$-to-$\te{X}$ ellipsoid axial
ratios. In general we will allow the $(\te{X},\te{Y},\te{Z})$ to
differ from $(\te{X}_\te{g},\te{Y}_\te{g},\te{Z}_\te{g})$ in both
orientation and position of the origin. When the origin of
$(\te{X},\te{Y},\te{Z})$ is the Galactic centre and $\te{p}=1$, as we
will assume in this section, the system $(\te{X},\te{Y},\te{Z})$ can
be identified with $(\te{X}_\te{g},\te{Y}_\te{g},\te{Z}_\te{g})$
without loss of generality.

\subsection{Density distribution of the halo RRLs} \label{sec:rrldens}

Given the all-sky RRL sample illustrated in Fig. \ref{fig:skymap}, we
can compute their volume number density distribution $\rho$. In
particular we define the number density of halo RRLs in a cell $\Delta
\vec{\epsilon}$ centred at $\vec{\epsilon}$, where $\vec{\epsilon}$ is
a coordinate vector, such as for instance $(\te{R},\te{Z}_{\te{g}})$,
as
\begin{equation}
\rho(\vec{\epsilon},\Delta \vec{\epsilon}) \approx  N(\vec{\epsilon}, \Delta \vec{\epsilon})  f^{-1}_\te{V}(\vec{\epsilon}, \Delta \vec{\epsilon})  \te{V}^{-1} (\vec{\epsilon},\Delta \vec{\epsilon}),
\label{eq:rho}
\end{equation}
where $N(\vec{\epsilon}, \Delta \vec{\epsilon})$ is the number of
stars observed in the cell, V is the volume of the cell and
$f_\te{V}$
is the fraction of this volume accessible to our analysis,
which depends on the sky coverage, on the selection cuts and the mask that we applied (see
Sec.  \ref{sec:selection}), and on the off-set between the Sun and the
Galactic centre. We estimate $f_\te{V}$ numerically as follows.
First, we define a Cartesian Galactic grid sampling each of the axes
$(\te{X}_\te{g},\te{Y}_\te{g},\te{Z}_\te{g})$ with 300 points linearly
spaced between $-\te{D}_\te{g,max}$ and $\te{D}_\te{g,max}$
($\te{D}_\te{g,max}\simeq 28$ kpc is the maximum value of
$\te{D}_\te{g}$, given our magnitude limit at $G=17.1$), so the
reference Galactic volume (a cube of 56 kpc side centred on the
Galactic centre) is sampled with 27 million points with a density of
about 125 points per kpc$^3$. 
Each of the points of the grid is a \vir{probe} of the Galactic volume. We assign to each of them Galactic coordinates, heliocentric coordinates  and observational coordinates (l, b, $\alpha$, $\delta$).
A secondary (non-uniform) grid is built by applying the selection cuts in Tab.\ 1 to the points on the primary grid, so we end up with a grid sampling the complete Galactic volume and a grid sampling the portion of the volume accessible to our analysis. Given a cell in a certain coordinates space (e.g. R, $\te{Z}_\te{g}$),  we define $f_\te{V}$ as the ratio between the number of points of the secondary grid and the number of points of the primary grid contained in the cell.
In summary, when we estimate the number density in a 1D or 2D space we build three binned maps: the first contains the number of stars in each bin/cell ($N$), the second the total volume of each bin/cell  ($\text{V}$)  and the third the fraction of this volume that we are actually sampling ($f_\te{V}$). These three quantities are inserted in Eq. 10 to estimate the stellar number density.
We tested this method with
both simple analytic distributions and mock catalogues (see
Sec. \ref{sec:mock}).

\subsubsection{Meridional plane}

\begin{figure}
\centering
\includegraphics[width=1\columnwidth]{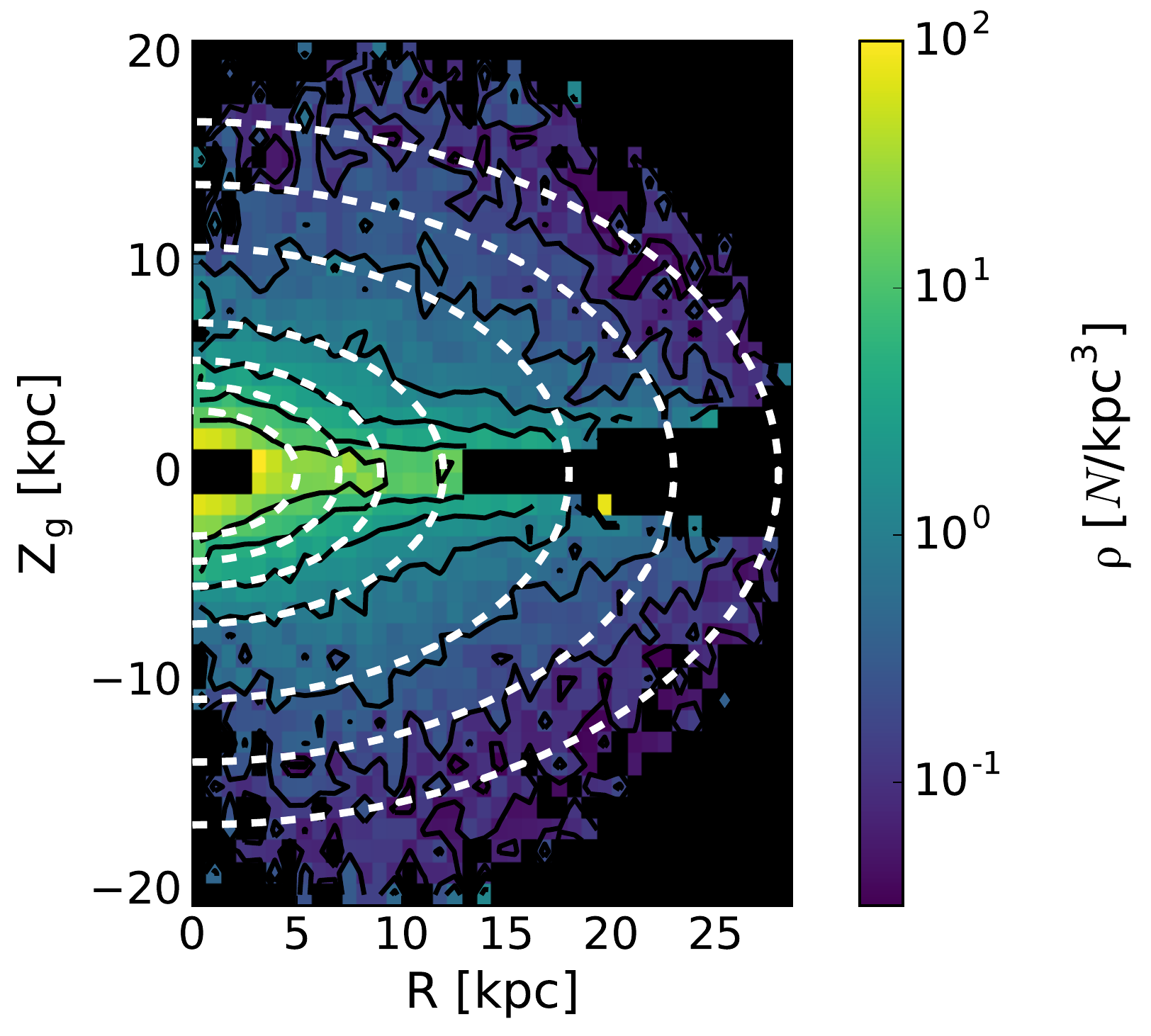}
\caption{Number density in the Galactic $\te{R}-\te{Z}_\te{g}$ plane
  for the RRLs in our sample.  The number density is calculated
  dividing the number of stars found in a cylindrical ring by the
  volume of the ring corrected for the non-complete volume sampling
  of the data (Eq. \ref{eq:rho}). The black contours are plotted with
  spacing $\Delta \log \rho=0.4$ from $\log \rho=-1.2$ to $\log
  \rho=1.6$, where $\rho$ is in units of $\te{kpc}^{-3}$; the
  white dashed lines represent elliptical contours with an axial ratio
  $\te{q}=0.6$.}
\label{fig:Rzdens}
\end{figure}

Fig. \ref{fig:Rzdens} shows the number density of the RRLs of our
sample in the Galactic meridional plane $\te{R}-\te{Z}_\te{g}$.  The
shape of the iso-density contours clearly shows the presence of two
components: a spheroidal component and a discy component. The discy
component causes the flattening of the contours at low
$\te{Z}_\te{g}$.  The nature of this component is uncertain: it could
represent the disc RRL stars, it could be a low-latitude substructure
of the MW halo or, finally, it could be due to non-RRL contaminants
from the Galactic disc (both genuine variables such as eclipsing
binaries as well as artefacts, e.g. due to cross-match failures). A
more detailed analysis of this low-latitude substructure can be found
in Sec. \ref{sec:struct}.  The density at higher $\te{Z}_\te{g}$ is
less contaminated by the discy component and it represents more
directly the density behaviour of the RRLs in the halo. The
iso-density contours nicely follow the the $\te{q}=0.6$ elliptical
contours (overplotted in Fig. \ref{fig:Rzdens}) out to
$\te{R}\approx15-20$ kpc.  At larger radii the contours tend to be
rounder.  The overall density distribution looks reasonably symmetric
with respect to the Galactic plane, although there is an over-dense
region at $\te{Z}_\te{g}>10$ kpc, which does not seem to have a
counterpart below the Galactic plane (see Sec. \ref{sec:vasy} for
further details).

\subsubsection{Density profile}  \label{sec:dprof}

\begin{figure}
\centering
\includegraphics[width=0.9\columnwidth]{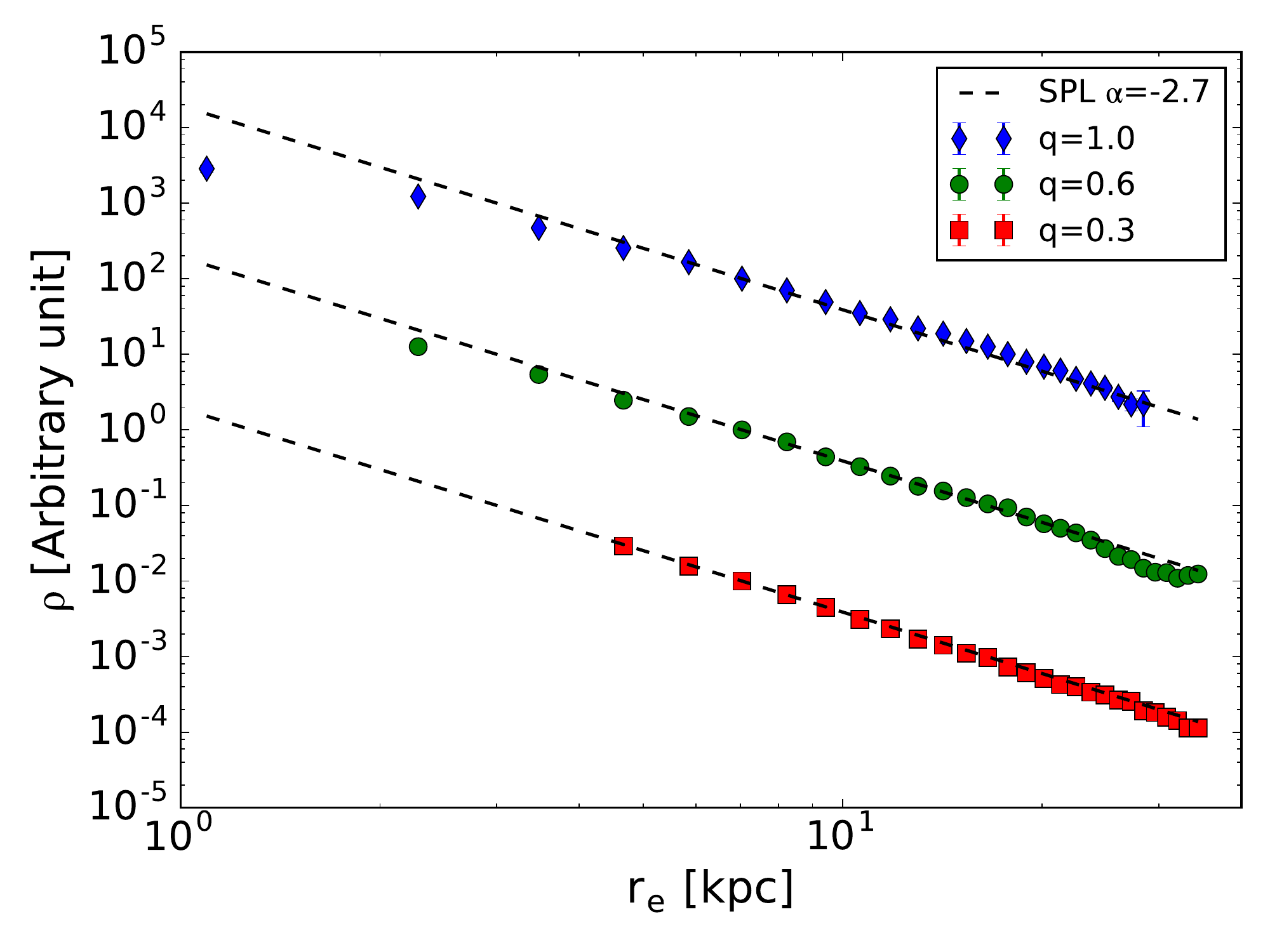}
\caption{1D number density profiles of the halo RRLs as functions of
  the elliptical radius $\te{r}_\te{e}$ (Eq. \ref{eq:ellrad}) assuming
  that the halo is stratified on spheroids with $\te{p}=1$ and
  different values of $\te{q}$: $\te{q}=1.0$ (blue diamonds),
  $\te{q}=0.6$ (green dots) and $\te{q}=0.3$ (red squares).  The
  black-dashed line shows, for comparison, a single power law with
  index $\alpha=-2.7$.  The Poissonian errors are smaller than the
  size of the symbols. The density normalisation is arbitrary and
  different for each profile.}
\label{fig:mdens}
\end{figure}

Fig. \ref{fig:Rzdens} demonstrates that the RRLs in the inner halo are
consistent with being stratified on spheroids (except in the region close to the Galactic plane), thus we estimate the 1D
RRL density profile by counting the stars in spheroidal ($\te{p}=1$,
$\te{q}\neq 1$) shells and dividing this number by the shell volume
corrected by the coverage of our sample (Eq. \ref{eq:rho}).  The
profiles are shown in Fig. \ref{fig:mdens} for $\te{q}=1.0$,
$\te{q}=0.6$ and $\te{q}=0.3$. 
Independently of the assumed value of q,
the density of RRLs follows a single power law with no significant evidence of
change of slope out to an elliptical radius of 35 kpc. It must be noted that the most distant region is only sampled by a low number of stars located in a small portion of the total volume, so it is possible that a change of slope starting near the edge of our elliptical radial range could be missed with this analysis. As the elliptical radius depends on the assumed axial ratios, the results of this analysis are not straightforwardly comparable with previous works (see Sec.\ \ref{sec:comparison} for a detailed comparison with results obtained in the literature).

The slope of this power law is similar
in the three cases and very close to $\alpha=-2.7$ (see
Fig. \ref{fig:mdens}), but, based on this simple analysis, all values
of $\alpha$ in the range $-3<\alpha<-2.5$ are consistent with the
data. 
%It is difficult to assess the compatibility of our profiles with
%the significant change of slope (from $\approx -2.5$ to $\approx -4$)
%claimed in many previous works (e.g. \citealt{deasonhalo}) as our
%sample barely reaches the so-called ``break-radius''. 

\subsubsection{Halo flattening} \label{sec:hflat}

\begin{figure*}
\centering
\includegraphics[width=0.9\textwidth]{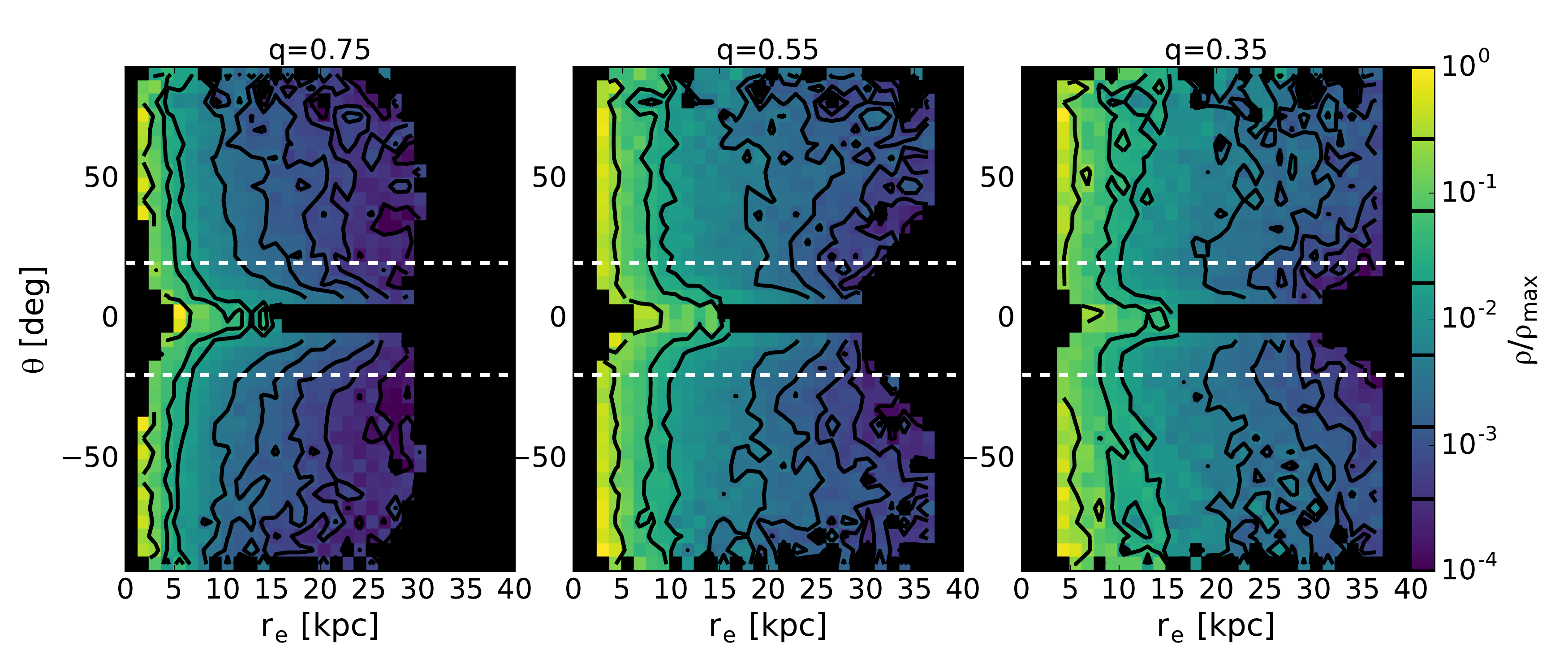}
\caption{Number density of RRLs in the elliptical radius
  ($\te{r}_\te{e}$) - Galactocentric latitude ($\theta$) space for
  $\te{q}=0.75$ (left-hand panel), $\te{q}=0.55$ (middle panel) and
  $\te{q}=0.35$ (right-hand panel), assuming $\te{p}=1$.  The density
  is normalized to the maximum value of each panel. The contours show
  the normalised density levels
  $(0.0002,0.00055,0.001,0.002,0.004,0.02,0.04,0.2)$, while the dashed
  lines indicate $|\theta|=20^\circ$.}
\label{fig:mb10}
\end{figure*}

The distribution of RRLs in the meridional Galactic plane
(Fig. \ref{fig:Rzdens}) suggests that the inner halo should be
reasonably well represented by a spheroidal stratification with
$\te{q}\approx0.6$. 
However, in order to more rigorously study the halo flattening, we
estimate the density in the $\te{r}_\te{e}-\theta$ plane (see
equations \ref{eq:fref} and \ref{eq:ellrad}). 
In practice, for a given
value of $q$, we scan the density as a function of $\te{r}_\te{e}$ at
fixed $\theta$. If the RRLs are truly stratified on similar spheroids
with axial ratio $q$, the density is independent of $\theta$, so the
isodensity contours in the $\te{r}_\te{e}-\theta$ plane are vertical
stripes.  If the assumed value of $\te{q}$ is smaller than the true
value $\te{q}_\te{true}$, $\te{r}_\te{e}$ is underestimated, the
estimated density is a monotonic decreasing function of $\theta$ and
the isodensity contours in the $\te{r}_\te{e}-\theta$ plane are bent
in the direction of $|\theta|$ increasing with $\te{r}_\te{e}$
(provided that the density is a decreasing function of
$\te{r}_\te{e}$).  The isodenisty contours are bent in the opposite
direction, if one assumes $\te{q}>\te{q}_\te{true}$.  The shape of the
iso-density contours in the $\te{r}_\te{e}-\theta$ plane is a very
efficient and direct diagnostic of the evolution of q as function of
the elliptical radius.

The number density maps in the $\te{r}_\te{e}-\theta$ plane of the
RRLs in our sample are shown in Fig. \ref{fig:mb10} for $\te{q}=0.75$,
$\te{q}=0.55$ and $\te{q}=0.35$, assuming $\te{p}=1$. Below
$|\theta|=20^\circ$ (indicated by the white-dashed lines) the contours
are nearly horizontal, because the density is dominated by the highly
flattened discy component (see Fig. \ref{fig:Rzdens}) for which q is
over-estimated in all the panels.  At higher latitudes
($|\theta|>20^\circ$) the contours give a direct indication on the
flattening of the halo: in the right-hand panel of Fig. \ref{fig:mb10}
($\te{q}=0.35$) the iso-density contours are significantly inclined in
a way that implies $\te{q}_\te{true}>0.35$.  For $\te{r}_\te{e}<20$
kpc a flattening of about 0.55 gives a good description of the data as
shown by the vertical iso-density contours in the middle panel of
Fig. \ref{fig:mb10} ($\te{q}=0.55$), but beyond 20 kpc the contours
start to bend so that $\te{q}_\te{true}>0.55$.  The last two
iso-density contours in the left-hand panel ($\te{q}\approx0.75$) look
vertical enough to assert that at the outer radii the halo becomes
more spherical.  This analysis, together with the recent works of
\cite{lamost}, shows the first direct evidence of a change of shape of
the stellar halo going from the inner to the outer halo. Moreover the
unique all-sky view of our sample allows us to confirm that this trend
is symmetric with respect to the Galactic plane. A variation of the
halo flattening was also proposed in previous works
(e.g. \citealt{xuehalo} and \citealt{das}). In Sec. \ref{sec:results}
we perform a comprehensive model fitting analysis of the RRL dataset
and compare our results to those found in the literature.

\subsubsection{Vertical asymmetries} \label{sec:vasy}

\begin{figure*}
\centering
\includegraphics[width=0.9\textwidth]{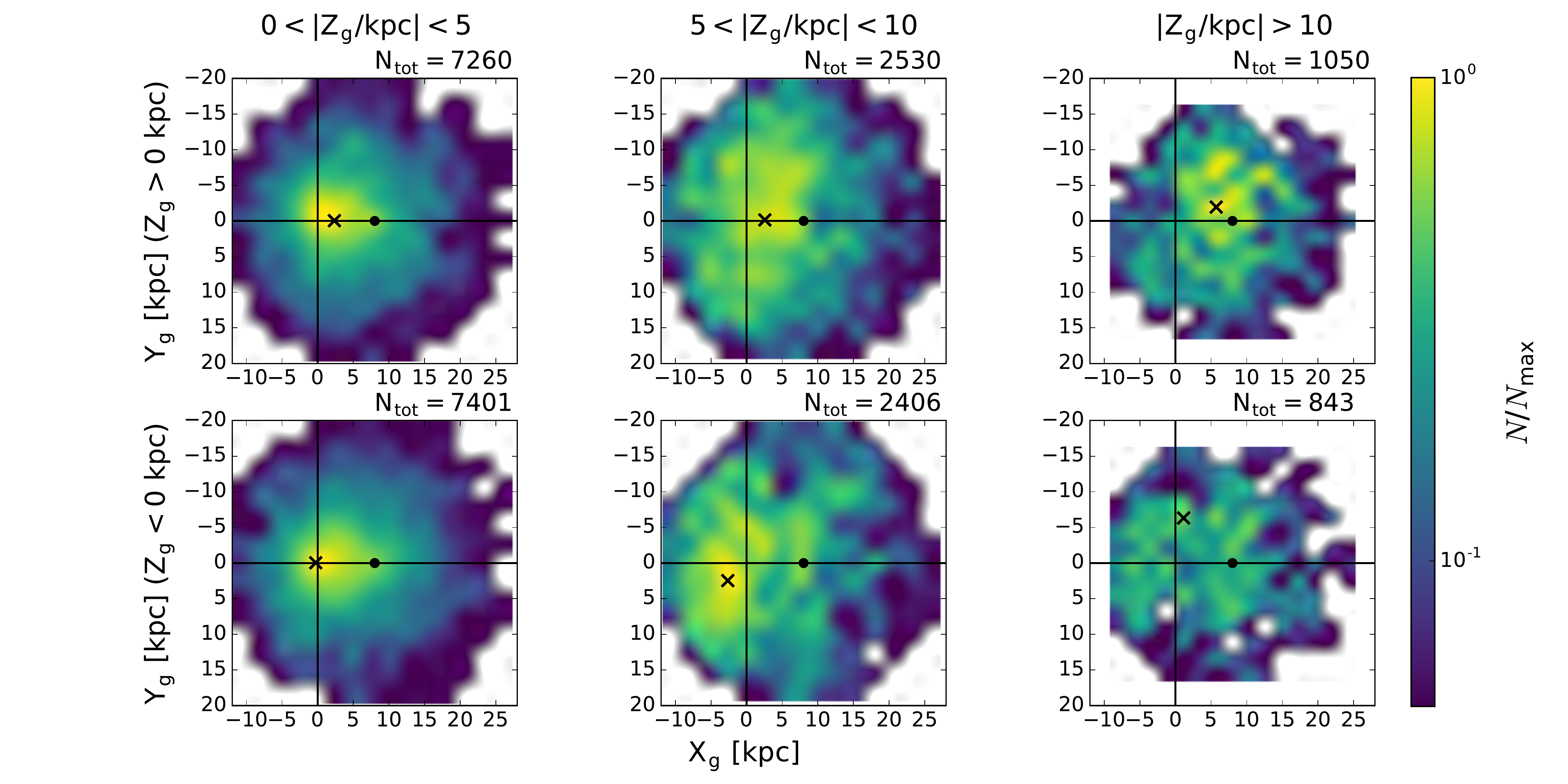}
\caption{Number counts of stars in the $\te{X}_\te{g}-\te{Y}_\te{g}$
  plane for different $\te{Z}_\te{g}$-slabs: left-hand panels with
  $0<|\te{Z}_\te{g}/\te{kpc}|<5$, middle panels with
  $5<|\te{Z}_\te{g}/\te{kpc}|<10$ and right-hand panels with
  $|\te{Z}_\te{g}/\te{kpc}|>10$. The top and bottom panels show the
  layers above ($\te{Z}_\te{g}>0$) and below ($\te{Z}_\te{g}<0$) the
  Galactic plane, respectively.  The colour maps are normalised to the
  maximum number in each column of plots. $N_\te{tot}$ is the total
  number of stars found on each $\te{Z}_\te{g}$-slab. The black dots
  show the position of the Sun (at $\te{Z}_\te{g}=0$), while the
  crosses indicate the maximum of the star counts. The maps have been
  smoothed with a spline kernel.}
\label{fig:zslab}
\end{figure*}

Contours of the RRL density shown in Fig. \ref{fig:Rzdens} and
Fig. \ref{fig:mb10} display an overall symmetry between the Northern
and Southern Galactic hemispheres, however at $\te{Z}_\te{g}>10$ kpc
and around $\theta\approx 60^\circ$ an over-density of RRLs above the
disc plane is evident. Thanks to the unprecedented sky coverage of our
sample we can directly analyse the distribution of stars in different
$\te{Z}_\te{g}$-slabs to understand whether the over-density is
compatible with the Poisson star-count fluctuations or it is due to a
genuine halo substructure.

Fig. \ref{fig:zslab} shows star-count maps in the
$\te{X}_\te{g}-\te{Y}_\te{g}$ plane integrated along three different
$\te{Z}_\te{g}$-slabs.  Close to the plane (left-hand panels) the two
maps look very similar and the difference between the total number of
stars is small (less than 2\%) and compatible within the Poisson
errors.  In these maps, an elongated component (also visible in the
other density maps, Figs. \ref{fig:Rzdens} and \ref{fig:mb10}) can be
seen stretching from the Galactic centre to the Sun and beyond: this
component appears symmetric with respect to the Galactic plane but is
present only in the side of the Galaxy containing the Sun.  In the
intermediate $\te{Z}_\te{g}$-slabs (middle panels) the difference
between the number of stars in the region above and below the Galactic
plane is small (less than 5\%) as in the previous case, but the maps
look less symmetric. In particular the peaks of the star counts are
now apart by about 10 kpc. Finally, the slabs at the highest $\te{Z}_\te{g}$
(right-hand panels) show a significant difference in the total number
of stars (about 20\%) that can not be explained by Poisson
fluctuations only. Moreover the excess of stars above the Galactic
plane is strongly clustered in the regions between
$\te{X}_\te{g}\approx(5,15)$ kpc and $\te{Y}_\te{g}\approx(0,10)$ kpc.

Note that some of the differences between the star counts in the
regions above and below the Galactic plane could be due to the mask
used to eliminate the contribution of the Magellanic Clouds (see
Sec. \ref{sec:selection}). Indeed, some mismatch is expected given
that we have excluded a relatively large region of the halo volume
below the Galactic plane. In order to quantify the differences
introduced by the mask, we produce mock catalogues (see
Sec. \ref{sec:mock}) of different halo models and find that the
Magellanic Clouds mask introduces a difference of about the 7\% in the
number of objects above and below the Galactic plane for
$|\te{Z}_\te{g}|>10$ kpc, significantly less than the value of 20\%
obtained here. Therefore, the structure seen at high positive
$\te{Z}_\te{g}$ appears to be genuine, and is most likely related to
the \vir{Virgo over-density} \citep{juric, Vivas2}. Across all
$\te{Z}_\te{g}>0$ slabs, portions of the Virgo Cloud are visible as an
excess of stars at positive $\te{X}_\te{g}$. Virgo's counterpart
underneath the disc is the Hercules-Aquilla Cloud \citep[see][]{hac,
  hac_simion}, discernible in middle bottom panel as strong
over-density at negative $\te{X}_\te{g}$ (and $\te{Z}_\te{g}$).

%We also note that in our halo models, the peak of the star-count in
%the intermediate Z-slabs (5 kpc $<|\te{Z}_\te{g}|<10$ kpc) is
%qualitatively more compatible with the distribution of stars seen in
%the data in the regions below the Galactic plane (middle-bottom panel
%in Fig. \ref{fig:zslab}) than in the one above (middle-top panel in
%Fig. \ref{fig:zslab}). In the latter case, the region containing the
%peak is roughly coincident with the substructure seen at higher
%$\te{Z}_\te{g}$, so it may represent a tail of this substructure
%toward the Galactic plane.

\section{Model fitting}\ \label{sec:bay}
In this section we present the stellar halo models and we
compare them with the observed sample of RRLs.

\subsection{Clean sample} \label{sec:csample}

Fig.  \ref{fig:Rzdens} and Fig. \ref{fig:mb10} show  the  presence of  a highly \vir{flattened} structure close to the Galactic plane. The properties of this structure are  clearly at  odds  with the distribution  of stars  at  high  Galactic latitude that are more  likely  a  \vir{genuine} tracer of the  halo population. 
The \vir{flattened} component  contains about 35\% of the RRLs in our sample, so it must be taken into account to infer the properties of the stellar halo.
Therefore, we built a  \vir{clean} sample of RRLs eliminating all the stars belonging to the substructure from our original catalogue (Sec. \ref{sec:selection}). 
Fig. \ref{fig:mb10} suggests that the substructure can be effectively eliminated with a selection in angle $\theta$ (see Eq. \ref{eq:freftheta}).

In particular, at $|\theta|=20^\circ$ there is a transition between a very flattened component and a more spheroidal structure. Therefore, we define our clean sample of halo RRLs as the stars with $|\theta|>20^\circ$: this subsample contains  13713  objects and covers approximately 44\% of the halo volume within a sphere of radius 28 kpc (see Tab. \ref{tab:tab_cut}). We stress that this cut is based on values of $\theta$ obtained from Eq. \ref{eq:amag} and Eq. \ref{eq:freftheta}  assuming the same value of the absolute magnitude, $M=M_\te{RRL}=0.525$, for all the stars (see Sec. \ref{sec:amag}).

We also tried to exclude the flattened structure with alternative cuts, e.g. using a higher cut on the Galactic latitude b or a direct cut on  $\te{Z}_\te{g}$. We found that the results obtained for the sample with  $|\theta|>20^\circ$  (see Sec. \ref{sec:results}) are qualitatively similar to the results obtained using a sample with $|\te{b}|>30^\circ$ (the lowest Galactic latitude used in most of the previous works, e.g. \citealt{deasonhalo}) or a sample with $|\text{Z}_\te{g}|>6$ kpc (\citealt{fermani} uses $|\text{Z}_\te{g}|>4$ kpc to cut disc stars). The cuts on b and $\text{Z}_\te{g}$ significantly reduce the number of tracers in the inner part of the halo and the final number of stars is smaller, in both cases, with respect to the one obtained cutting the original sample at $|\theta|=20^\circ$.

In the following subsections we  present the method  used to fit  halo models to  this subsample of RRLs and  the  final results.

\subsection{Halo models} \label{sec:hmodel}
As in Sec. \ref{sec:look}, we assume here that the number density of the halo RRLs is stratified on ellipsoids with axial ratios p and q.
Therefore, a  halo model  is defined  by a  functional form for the density profile (number density as a function of the elliptical radius)  and  a \vir{geometrical} model for the ellipsoidal iso-density surfaces (in practice, characterized by values of p and q, and the orientation of the principal axes with respect to the Galactic disc).

\subsubsection{Density profiles} \label{sec:densmod}
We consider five families of number density profiles: double power-law (DPL), single power-law (SPL), cored power-law (CPL), broken power-law (BPL) and Einasto profiles (EIN).
The DPL profile has a number density
\begin{equation}
\rho(\te{r}_\te{e})\propto\left(  \frac{\te{r}_\te{e}}{\te{r}_\te{eb}}  \right)^{-\alpha_\te{inn}}  \left( 1+ \frac{\te{r}_\te{e}}{\te{r}_\te{eb}}  \right)^{-(\alpha_\te{out}-\alpha_\te{inn})}, 
\label{eq:plgeneral}
\end{equation}
where $\alpha_\te{inn}$ and  $\alpha_\te{out}$ indicate the inner and outer power-law slopes and $\te{r}_\te{eb}$ is the scale length. 
The number density of the SPL is given by Eq. \ref{eq:plgeneral} when 
$\alpha_\te{inn}=\alpha_\te{out}$, while the number density of the CPL has $\alpha_\te{inn}=0$, in which case $\te{r}_\te{eb}$ represents the length of the inner core.
The BPL number density profile is given by a piecewise function:
\begin{equation}
\rho(\te{r}_\te{e})\propto
\begin{cases}
\te{r}_\te{e}^{-\alpha_\te{inn}} & \te{r}_\te{e}\leq\te{r}_\te{eb}  \\
\te{r}_\te{eb}^{\alpha_\te{out}-\alpha_\te{inn}} \te{r}_\te{e}^{-\alpha_\te{out}} & \te{r}_\te{e}\geq\te{r}_\te{eb}
\end{cases}.
\label{eq:bpl}
\end{equation}
The EIN profile \citep{einastor} is given by
\begin{equation}
\rho(\te{r}_\te{e}) \propto \text{exp} \left[ -d_\te{n} \left( \left( \frac{\te{r}_\te{e}}{\te{r}_\te{eb}} \right)^{\frac{1}{\te{n}}} -1 \right) \right],
\label{eq:einasto}
\end{equation}
where  $d_\te{n}=3\te{n} -0.3333 +0.0079/\te{n}$ for  $\te{n}\geq0.5$ \citep{einasto}.
The steepness of the EIN  profile, $\alpha_\te{EIN}$, changes continuously as a function of $\te{r}_\te{e}$ tuned by the  parameter $\te{n}$, 
\begin{equation}
\alpha_\te{EIN}=-\frac{d_\te{n}}{\te{n}} \left(   \frac{\te{r}_\te{e}}{\te{r}_\te{eb}} \right)^{\frac{1}{\te{n}}}.
\label{eq:einastostep}
\end{equation}

The EIN profile is the only density law among those considered here that assures a halo model with a finite total mass for any choice of the parameters. 
The power-law profiles with $\alpha_\te{inn}<-3$ or with $\alpha_\te{out}>-3$ imply halos with infinite total mass, however our study focuses only on a limited radial range and we do not exclude \textit{a priori} any solution. 
We anticipate that our best density model is a SPL with $\alpha_\te{inn}<-3$ (see Sec. \ref{sec:results}), therefore there should be a physical radius, outside our radial range, beyond which the profile becomes, either abruptly (e.g. BPL or an exponential truncation) or gently (e.g. DPL), steeper.

\subsubsection{Iso-density ellipsoidal surfaces} \label{sec:ellsurf}

Concerning the iso-density ellipsoidal surfaces we define four different models:
\begin{itemize}
\item{spherical (SH)}: we set $\te{p}=1$ and $\te{q}=1$ in  Eq. \ref{eq:ellrad},  so that  $\te{r}_\te{e}$ is  just  the  spherical  radius $\te{D}_\te{g}$ (see definition \ref{eq:frefDg})   in the  Galactic  frame  of reference;
\item{disc-normal axisymmetric (DN)}: we set $\te{p}=1$, the axis of symmetry is normal to the Galactic disc, and q is a free-parameter;
\item{disc-plane axisymmetric (DP)}: we set $\te{q}=1$, the axis of symmetry is within the Galactic plane making an (anticlockwise) angle $\gamma$ with respect to the Galactic  Y-axis, and p is a free-parameter,. 
\item{triaxial (TR)}: both p and q  are considered free-parameters, the Z-axis of symmetry is coincident with the normal to the Galactic plane and the X and Y axes are within the plane  making an (anticlockwise) angle $\gamma$ with respect to the Galactic X and Y axes, respectively.
\end{itemize}
Given the  unprecedented sky coverage of our sample, we also tested more complex models for the iso-density ellipsoidal surfaces:
\begin{itemize}
\item{q-varying (\textit{qv})}: q depends on the elliptical radius as
\begin{equation}
\te{q}(\te{r}_\te{e})=\te{q}_\infty - (\te{q}_\infty-\te{q}_0) \te{exp}\left[ 1- \frac{\sqrt{\te{r}_\te{e}^2+\te{r}_\te{eq}^2}}{\te{r}_\te{eq}} \right],
\label{eq:qvar}
\end{equation} 
so q varies from  $\te{q}_0$ at the centre  to  the asymptotic  value  $\te{q}_\infty$ at large radii  and the variation is tuned by the  exponential scale length $\te{r}_\te{eq}$;
\item{tilt (\textit{tl}),} in this model  we assume that the principal axes  of the ellipsoids are tilted with respect to the Galactic plane, in practice, before calculating the elliptical radii (Eq. \ref{eq:ellrad}), we transform the  Galactic coordinates (see Eq. \ref{eq:fref}) of each star by applying a  rotation matrix $R(\gamma,\beta,\eta)$ following a  ZYX formalism so that $\gamma$ is the   rotation angle  around the  original Z-axis, $\beta$ the one  around the new  Y-axis and  finally $\eta$ is the  rotation angle  around the final X-axis, all the rotation are defined in the anticlockwise direction;
\item{offset (\textit{off}),} in this model the  elliptical radius of each star is estimated  with respect to a point  offset  by  ($\te{X}_\te{off}$, $\te{Y}_\te{off}$, $\te{Z}_\te{off}$) with respect to the Galactic centre.
\end{itemize}

The specific functional form of  Eq. \ref{eq:qvar} is empirical: the same expression was adopted by in \cite{xuehalo} and \cite{das}, where, however, q is a function of the  spherical radius ($\te{D}_\te{g}$). We decided to maintain the dependence on $\te{r}_\te{e}$  given that this approach is self-consistent  with  our assumption that the RRLs are stratified on  ellipsoidal surfaces.
If follows that for our models with a varying q, the elliptical radius $\te{r}_\te{e}$  of a star with coordinates $(\te{X}_\te{g}, \te{Y}_\te{g}, \te{Z}_\te{g})$ is the root of
\begin{equation}
\te{r}^2_\te{e} - \te{X}^2_\te{g} - \te{Y}^2_\te{g} \te{p}^{-2} - \te{Z}^2_\te{g} \te{q}(\te{r}_\te{e})^{-2}=0,
\end{equation}
where $\te{q}(\te{r}_\te{e})$ is defined in Eq. \ref{eq:qvar}. In our fitting code we solve this equation numerically with a Newton-Raphson root finder.

\subsubsection{Complete halo model} \label{sec:label}

Each complete halo model is defined by a model for the density law (SPL, DPL, CPL, BPL, EIN) plus a model for the shape of the iso-density surfaces (SH, DN, DP, TR) and any combination of geometrical variants ({\it qv, tl, off}).
For instance, a SPL-SH model has a spherical distribution of stars with a single power-law density profile, while a
$\text{EIN-TR}^\textit{qv,tl}$ model is a triaxial tilted model with varying flattening along the Z-axis of symmetry and an Einasto density profile (see Tab. \ref{tab:prior} for a reference on the various model labels).
In the next sections when we discuss the properties or the results of density models  (e.g. SPL models), unless otherwise stated, we are implicitly referring to all the models that share the same density model whatever the geometrical and geometrical variants properties are. The same applies when we focus on geometrical models only.
We fit our data with all possible combinations of density laws (SPL, DPL, CPL, BPL, EIN), geometrical models (SH, DN, DP, TR) and model variants ({\it qv, tl, off}).
See Tab. \ref{tab:fit} for a summary of results obtained with a sample of such models.

\subsection{Comparing models with observations} \label{sec:method}

\subsubsection{Density of stars in the observed volume} \label{sec:dstars}

Given  a certain halo model (Sec. \ref{sec:hmodel}), the normalised number density of stars in an infinitesimal volume $\te{dV}_\te{g}=\te{d}\te{X}_\te{g} \te{d}\te{Y}_\te{g} \te{d}\te{Z}_\te{g}$ is given by
\begin{equation}
\tilde{\rho}(\te{X}_\te{g}, \te{Y}_\te{g}, \te{Z}_\te{g}| \vec{\mu})\propto \frac{\te{d}N}{\te{dV}_\te{g}},
\label{eq:drho}
\end{equation}
where $\vec{\mu}$ is the vector containing all the  model parameters  (e.g. $\vec{\mu}=(\alpha_\te{inn}, \te{q})$ for an SPL+disc-normal axisymmetric model, see Sec. \ref{sec:hmodel}) and $\tilde{\rho}$ is defined such that
\begin{equation}
\int \tilde{\rho}(\te{X}_\te{g}, \te{Y}_\te{g}, \te{Z}_\te{g}| \vec{\mu}) \te{dV} =1.
\label{eq:drhonorm}
\end{equation}
It is useful to define the normalised star number density
\begin{equation}
\tilde{\nu}(m,\te{l},\te{b}|M,\vec{\mu})= |\te{J}| \tilde{\rho}(\te{X}_\te{g}, \te{Y}_\te{g}, \te{Z}_\te{g}| m, \te{l}, \te{b}, M, \vec{\mu})
\label{eq:nurho}
\end{equation}
within the infinitesimal projected (onto the sky) volume element  $\te{d}S=\te{d}m \te{dl} \te{db}$ 
centred at ($m$,l,b), where $m$ is the observed magnitude, l and b are the Galactic longitude and latitude, and
\begin{equation}
|\te{J}|= \frac{\ln 10}{5} \te{D}^3_\odot(m,M) \cos \te{b} 
\label{eq:finjac}
\end{equation}
is the determinant of the Jacobian matrix
$J=\left[\frac{\te{dV}_\te{g}}{\te{d}S}\right]$ (Appendix \ref{app:jaco}).
The  number density $\tilde{\nu}$ depends also on the  additional  parameter $M$, which is  the  absolute magnitude  needed to pass from the  observable variables  $(m, \te{l}, \te{b})$ to  the  Cartesian  variables (see Eqs. \ref{eq:amag} and \ref{eq:fref}).
Substituting Eq. \ref{eq:finjac} in Eq. \ref{eq:nurho}  we obtain  
\begin{equation}
\tilde{\nu}(m,\te{l},\te{b}|M,\vec{\mu})= \frac{\ln 10}{5} \tilde{\rho}(\te{X}_\te{g}, \te{Y}_\te{g}, \te{Z}_\te{g}| m, \te{l}, \te{b}, M, \vec{\mu}) \te{D}^3_\odot(m,M) \cos \te{b}.
\label{eq:nurho2}
\end{equation}

\subsubsection{Likelihood of a single star} \label{sec:ratef}

From the normalised density function (Eq. \ref{eq:nurho2})  we can define  
the expected rate function for finding a star with ($m$, l, b) given  the absolute magnitude $M$ and a halo model with  parameters  $\vec{\mu}$
\begin{equation}
\lambda(m,\te{l},\te{b},M|\vec{\mu})= A \tilde{\nu}(m,\te{l},\te{b}|M,\vec{\mu})W(m,\te{l},\te{b}|M_\te{RRL})P(M).
\label{eq:denslamb_bn}
\end{equation}
In Eq. \ref{eq:denslamb_bn}, $A$ is a constant,  $P(M)$ represents  the probability density function (pdf) of  the absolute magnitude of the stars, while the  function $W$ is  the selection function that takes into account the incomplete coverage of the Galactic volume. It is  a function of l, b and $m$ and returns a result in the Boolean domain  $\te{B}=(0, 1)$.  
In particular, $W$ is always equal to 1 except for the points ($m$, l, b) that are outside the volume covered by our clean sample (see Tab. \ref{tab:tab_cut}).

\begin{table*}
\centering
\tabcolsep=0.08cm
\begin{tabular}{cc||ccc|ccccc}
\hline
\multirow{2}{*}{\begin{tabular}[c]{@{}c@{}}Model\\ component\end{tabular}} & \multirow{2}{*}{\begin{tabular}[c]{@{}c@{}}Model\\ label\end{tabular}} & \multicolumn{3}{c|}{Density parameters} & \multicolumn{5}{c}{Iso-density surfaces parameters} \\ \cline{3-10} 
& & $\alpha_\te{inn}$/n & $\alpha_\te{out}$ & $\te{r}_\te{eb}$ (kpc) & p & q($\te{q}_0 / \te{q}_\infty$) & $\te{r}_\te{q}$ (kpc) & $\te{X}_\te{off} / \te{Y}_\te{off} / \te{Z}_\te{off}$ (kpc) & $\gamma / \beta / \eta$ (degree) \\ \hline
\rowcolor[HTML]{C0C0C0} \hline
Single power-law & SPL  & U{[}0,20{]} & $\delta(\alpha_\te{inn})$ & $\delta(1)$ &  &  &  &  &  \\
Cored power-law & CPL & $\delta(0)$ & U{[}0,20{]} & U{[}0.01,10{]} &  &  &  &  &  \\
\rowcolor[HTML]{C0C0C0} 
Double power-law &DPL & U{[}0,20{]} & U{[}0,20{]} & U{[}$\te{r}^{\te{min}}_\te{e}$,$\te{r}^{\te{max}}_\te{e}${]} $\dagger$  &  &  &  &  &  \\
Broken power-law &BPL & U{[}0,20{]} & U{[}0,20{]} & U{[}$\te{r}^{\te{min}}_\te{e}$,$\te{r}^{\te{max}}_\te{e}${]} $\dagger$ &  &  &  &  &  \\
\rowcolor[HTML]{C0C0C0} 
Einasto & EIN & U{[}0,100{]} &  & U{[}0.1,500{]} &  &  &  &  &  \\
Spherical & SH &  &  &  & $\delta(1)$ & $\delta(1)$ &  &  &  \\
\rowcolor[HTML]{C0C0C0} 
Disc-normal axsym & DN &  &  &  & $\delta(1)$ & U{[}0.1,10{]} &  &  &  \\
Disc-plane axsym & DP &  &  &  & U{[}0.1,10{]} & $\delta(1)$ &  &  & U{[}-80,80{]}  \\
\rowcolor[HTML]{C0C0C0} 
Triaxial & TR &  &  &  & U{[}0.1,10{]} & U{[}0.1,10{]} &  &  & U{[}-80,80{]}  \\
q-var & {\it qv} &  &  &  &  &  & U{[}0.1,100{]} &  &  \\
\rowcolor[HTML]{C0C0C0} 
Offset & {\it off} &  &  &  &  &  &  & U{[}0,10{]} &  \\
Tilted & {\it tl} &  &  &  &  &  &  &  & U{[}-80,80{]} \\ \hline
\end{tabular}
\caption{Prior distribution $\Pi$ (Eq. \ref{eq:likel}) of the halo model parameters. Each row  refers to a given component of the halo models, each column indicates a  single  parameter  or a group of parameters if they  share  the same  prior distribution.  
The second column indicates the labels where we indicate the halo models through the text (Sec. \ref{sec:label}).
The third column refers to the parameter n for the Einasto profile and to the parameter $\alpha_\te{inn}$ for the other density models.
The U indicates  a uniform  distribution within the values shown inside the square brackets; a $\delta$ indicates a Dirac delta distribution.
$\dagger$ The  prior range of the parameter $\te{r}_\te{eb}$  for the BPL and DPL density models  is not the same in all models, depending on the minimum ($\te{r}^{\te{min}}_\te{e}$) and maximum ($\te{r}^{\te{max}}_\te{e}$) elliptical radius of the stars in the sample.}
\label{tab:prior}
\end{table*}

In this work,  we decided to set the absolute magnitude of the RRLs to a single value, $M_\te{RRL}=0.525$ (Sec. \ref{sec:amag}), so we are assuming that  $P(M)$ in  Eq.   \ref{eq:denslamb_bn} is a Dirac delta.  As a consequence, we can marginalize  Eq. \ref{eq:denslamb_bn} over $M$ to obtain
\begin{equation}
\lambda(m,\te{l},\te{b}|M_\te{RRL}, \vec{\mu})=  A \tilde{\nu}(m,\te{l},\te{b}|M_\te{RRL},\vec{\mu})W(m,\te{l},\te{b}|M_\te{RRL}),
\label{eq:denslamb_b}
\end{equation}
and we define the normalised  rate function as 
\begin{equation}
\tilde{\lambda}(m,\te{l},\te{b}|M_\te{RRL}, \vec{\mu}) = \frac{\lambda(m,\te{l},\te{b}|M_\te{RRL}, \vec{\mu})}{\int \lambda(m,\te{l},\te{b}|M_\te{RRL}, \vec{\mu})  \te{d}m \ \te{dl} \ \te{db}}.
\label{eq:denslamb2}
\end{equation}
The  normalised rate function in Eq. \ref{eq:denslamb2} is the  pdf of stars, i.e. the likelihood per star, at a certain position ($m$, l, b) for halo model parameters $\vec{\mu}$.

\subsubsection{The total likelihood } \label{sec:lkdata}

Consider  a sample of stars  $D$  with  coordinates  ($m$, l, b) and a halo model with parameters  $\vec{\mu}$. Given the pdf of the stellar distribution  $\tilde{\lambda}(\te{m}, \te{l}, \te{b})$ (Eq. \ref{eq:denslamb2})  the logarithmic likelihood  is
\begin{equation}
\ln\mathcal{L}=\sum^{N_\text{s}}_{i=1} \ln \tilde{\lambda}(m_i,\te{l}_i,\te{b}_i|\vec{\mu}),
\label{eq:lnnp}
\end{equation}
where $N_\text{s}$ is the number of stars in our sample.
Plugging Eq. \ref{eq:nurho2} and Eq. \ref{eq:denslamb_b} into Eq. \ref{eq:denslamb2} we can write the logarithmic   likelihood as  
\begin{equation}
\ln\mathcal{L}=   \sum^{N_\text{s}}_{i=1} \ln \frac{\tilde{\rho}(\te{X}_\te{g}, \te{Y}_\te{g}, \te{Z}_\te{g}| m_i, \te{l}_i, \te{b}_i, M_\te{RRL}, \vec{\mu}) \te{D}^3_\odot(m_i,M_\te{RRL}) \cos \te{b}}{\te{V}_c(\vec{\mu})}
\label{eq:lnn2}
\end{equation}
where 
\begin{equation}
\te{V}_c(\vec{\mu})=\int^{90^\circ}_{-90^\circ} \cos \te{b} \ \text{db}   \int^{360^\circ}_{0^\circ}   \text{dl} \int^{G_\te{max}}_{G_\te{min}} \tilde{\rho} \text{D}^3_\odot W \text{d}m
\label{eq:vc2a}
\end{equation}
is the normalisation integral.
Notice that the numerator of Eq. \ref{eq:lnn2} is evaluated only in regions of the sky where $W=1$ (i.e. where we observe stars).

\subsubsection{Sampling of the parameter space} \label{sec:mcmc}

Considering the Bayes's law, the logarithmic posterior probability  of  the  parameters $\vec{\mu}$ of an halo  model  given the data is 
\begin{equation}
\ln P\left(\vec{\mu}|D\right)= \ln P\left(D|\vec{\mu}\right) + \ln \Pi \left(\vec{\mu}\right)=\ln\mathcal{L} + \ln \Pi \left(\vec{\mu}\right),
\label{eq:likel}
\end{equation}
where $P\left(D|\vec{\mu}\right)=\mathcal{L}$ is the probability of the data, $D$, given the parameters $\vec{\mu}$  (See sec. \ref{sec:lkdata}) and $\Pi \left(\vec{\mu}\right)$ represents the prior probability  of $\vec{\mu}$ (Tab. \ref{tab:prior}).
In Eq. \ref{eq:likel} we omit the Bayesian evidence term $P(D)$ that  is defined as the integral of the likelihood over the whole parameter space. This term is negligible in the determination of the best set of parameters for a given model (see below).

Using Eq. \ref{eq:lnn2} we can  write the  posterior  probability as
\begin{equation}
\begin{split}
\ln & P\left(\vec{\mu}|D\right) = \\
&\sum^{N_\text{s}}_{i=1} \ln \frac{\tilde{\rho}(\te{X}_\te{g}, \te{Y}_\te{g}, \te{Z}_\te{g}|m_i, \te{l}_i, \te{b}_i, M_\te{RRL}, \vec{\mu}) \te{D}^3_\odot(m_i,M_\te{RRL})  \cos \te{b}}{\te{V}_c(\vec{\mu})}\\
&+ \ln \Pi\left(\vec{\mu}\right).
\end{split}
\label{eq:likel3}
\end{equation}
In order to explore the parameter space without biases, we decided to use very large priors for most of our parameters (see Tab. \ref{tab:prior}).
In particular, the very large ranges for $\alpha_\text{inn}$ and $\alpha_\text{out}$ approximate
uniform infinite priors.

Concerning the parameter  $\te{r}_\te{eb}$, we assume, for both the BPL and DPL density models, that the prior distribution is uniform between $\te{r}^{\te{min}}_\te{e}$ and  $\te{r}^{\te{max}}_\te{e}$ which  represent  the  minimum  and maximum elliptical  radii (Eq. \ref{eq:ellrad}) of stars in our sample. For most of our models,  $\te{r}^{\te{min}}_\te{e}$ remains constant at about 2.4 kpc, while  $\te{r}^{\te{max}}_\te{e}$ ranges between roughly 31 kpc and 38 kpc.

We analysed the posterior probability of the parameters, $\vec{\mu}$, for a given halo model  using the  Eq. \ref{eq:likel3}  and sampled the parameters  space with  Goodman \& Weare's Affine Invariant Markov chain Monte Carlo (mcmc, \citealt{mcmc}), making use of the \texttt{Python}   module  emcee\footnote{\url{https://github.com/dfm/emcee}} \citep{emcee}.
The technical details of our approach can be found in Appendix \ref{app:tec}.

The  final best-fit values of the model parameters have been estimated using  the 50th percentile of the posterior distributions and  the  16th and 84th percentiles have been used to estimate the 1-$\sigma$ uncertainties.

\subsubsection{Tests on mock catalogues} \label{sec:mock}

In order to test our fitting method, we developed a simple
  \texttt{Python} script to generate mock catalogues with the same
  properties (e.g. number of stars, magnitude limits) as our clean
  sample of RRLs (Sec. \ref{sec:csample}). This algorithm distributes
  the stars using a combination of the density laws
  (Sec. \ref{sec:densmod}) and according to the assumptions on the
  properties of the iso-density ellipsoidal surfaces
  (Sec. \ref{sec:ellsurf}).  The absolute magnitude of the stars in the mock catalogues are distributed using the best-fit double Gaussian functional form shown in Fig. \ref{fig:Mg}. The final mock catalogues do not  include  \gaia and 2MASS photometric and sky-coordinates errors because the uncertainties they cause on the estimate of the distance  are negligible.
 
 We applied our fitting method to mock
catalogues finding that we are able to recover the input parameters
for all possible model combinations, including in the sample a
fraction of Galactic disc contaminants (see Sec. \ref{sec:biasdisc}).

Analysing the mock catalogues, we can estimate what properties of the
halo we are able to measure and constrain using the stars in our
catalogue. We found that we are able to detect a core in the density
profile (CPL density model) all the way down to 100 pc. However for a
very small core size most of the information comes from a
relative small region where the transition between the inner core and
the outer density profile takes place.  The detection of a change of
slope depends on the halo flattening: for $\te{p}=1$ and $\te{q}=1$ we
can detect a break in the density profile within about
$\te{r}_\te{eb}=28$ kpc (Eq. 21). For more flattened models ($\te{q}$
around $0.5/0.6$) this range extends up to about $\te{r}_\te{eb}=35$
kpc. However, in this latter case most of the information for regions
where $\te{r}_\te{e}>28$ kpc comes from a small number of stars at
very high Galactic latitude, therefore the fit can be easily biased by
the presence of substructures such as the one highlighted in
Sec. \ref{sec:vasy} and Fig. 9.  In conclusion, we are confident that
we are able to robustly detect significant deviations from a SPL
within a range of elliptical radii from less than 1 kpc to about 30
kpc. Finally, we verified that the method described above can recover a
variation of the halo flattening given a realistic mock dataset.

\subsection{Results}  \label{sec:results}
In this section we  present the  main results obtained  applying the method described in Sec. \ref{sec:method}  to  the  RRLs in our clean sample (Sec. \ref{sec:csample}).

% \usepackage{multirow}
\begin{table*}
\centering
\tabcolsep=0.11cm
\begin{tabular}{cc||lcc}
\hline
\multicolumn{2}{c||}{Model} & \multicolumn{1}{c}{} &  &    \\
Density law & Surface & \multicolumn{1}{c}{\multirow{-2}{*}{Parameters}}  & \multirow{-2}{*}{$\Delta \ln (\mathcal{L}_\te{max})$} & \multirow{-2}{*}{$\Delta \te{BIC}$} \\ \hline
\rowcolor[HTML]{C0C0C0} 
SPL & SH & $\alpha_\te{inn}=2.61\pm0.01 \ (2.61)\dagger$ &  -868 & 1688 \\
SPL & DP & \begin{tabular}[c]{@{}l@{}}$\alpha_\te{inn}=2.64\pm0.01 \ (2.64)\dagger$,  \\ 
$\te{p}=1.59\pm0.03 \ (1.59)$,  $\gamma=-22.0\pm1.7 \ \te{deg} \ (-22.3 \ \te{deg})\ddagger$\end{tabular}   & -405 & 779 \\
\rowcolor[HTML]{C0C0C0} 
SPL & DN & \begin{tabular}[c]{@{}l@{}}$\alpha_\te{inn}=2.71\pm0.01 \  (2.71)\dagger$, \\ $\te{q}=0.58\pm0.01 \ (0.58)$\end{tabular}  &  -81 & 124 \\
BPL & DN & \begin{tabular}[c]{@{}l@{}}$\alpha_\te{inn}=2.70\pm0.01 \ (2.70)$, $\alpha_\te{out}=2.90\pm0.11 \ (2.87)$, \\ $\te{r}_\te{eb}=22.3^{+3.3}_{-2.8} \ (22.2)$ kpc,  $\te{q}=0.58\pm0.01 \ (0.58)$\end{tabular} &  -79 & 141 \\
\rowcolor[HTML]{C0C0C0} 
DPL & DN & \begin{tabular}[c]{@{}l@{}}$\alpha_\te{inn}=2.70\pm0.03 \ (2.71)$, $\alpha_\te{out}=2.74^{+0.04}_{-0.02} \ (2.72)$,\\ $\te{r}_\te{eb}=21.1^{+5.7}_{-9.9} \  (23.1)$ kpc, $\te{q}=0.58\pm0.01 \ (0.58)$\end{tabular} & -81 & 143 \\
CPL & DN& \begin{tabular}[c]{@{}l@{}}$\alpha_\te{out}=2.72\pm0.02 \ (2.71)$, $\te{r}_\te{eb}=0.03^{+0.03}_{-0.01} \ (0.03)$ kpc,  \\ $\te{q}=0.58\pm0.01 \ (0.58)$\end{tabular} &  -81 & 134 \\
\rowcolor[HTML]{C0C0C0} 
EIN & DN & \begin{tabular}[c]{@{}l@{}}$\te{n}=37.5^{+3.5}_{-4.6} \  (40)$, $\te{r}_\te{eb}=391.1^{+84.6}_{-169.8} \ (474)$ kpc, \\ $\te{q}=0.58\pm0.01 \ (0.58)$\end{tabular} & -87 & 145 \\
SPL & $\text{DN}^{\it qv}$ & \begin{tabular}[c]{@{}l@{}}$\alpha_\te{inn}=2.93\pm0.05 \ (2.93)\dagger$, \\ $\te{q}_0=0.52\pm0.02 \ (0.52)$, $\te{q}_\infty=0.74\pm0.05 \ (0.75)$, \\ $\te{r}_\te{eq}=14.8\pm1.9 \ (15.0)$ kpc\end{tabular}  & -66 & 113 \\
\rowcolor[HTML]{C0C0C0} 
SPL & TR & \begin{tabular}[c]{@{}l@{}}$\alpha_\te{inn}=2.71\pm0.01 \ (2.71)\dagger$, \\  $\te{p}=1.27\pm0.03 \ (1.27)$, $\gamma=-21.1\pm2.6 \ \te{deg} \ (-21.1 \ \te{deg})\ddagger$ \\ $\te{q}=0.65\pm0.01 \ (0.65)	\ddagger$\end{tabular}  & -22 & 23 \\
SPL & $\text{TR}^{\it qv}$ & \begin{tabular}[c]{@{}l@{}}$\alpha_\te{inn}=2.96\pm0.05 \ (2.96)\dagger$, \\ $\te{p}=1.27\pm0.03 \ (1.27)$, $\gamma=-21.3\pm2.6 \ \te{deg} \ (-21.3 \ \te{deg})\ddagger$ \\ $\te{q}_0=0.57\pm0.02 \ (0.57)$, $\te{q}_\infty=0.84\pm0.06 \ (0.84)$,\\  $\te{r}_\te{eq}=12.2^{+2.4}_{-1.8} \ (12.2)$ kpc\end{tabular}  & 0 & 0 \\ \hline
\end{tabular}
\caption{Summary of properties and results for a sample of families of  halo models. For each family we report the assumed density law (Sec. \ref{sec:densmod}) and geometry of the  iso-density surfaces (Sec. \ref{sec:ellsurf}). 
See Tab. \ref{tab:prior} for a reference on the model labels.
For each fitted parameter  we report  the  median and  the uncertainties estimated as the 16th and 84th percentiles  of the posterior distribution, the values in parentheses indicate the value for which we obtain the maximum value of the likelihood $\mathcal{L}_\te{max}$ (Eq. \ref{eq:lnn2}).
The last two columns indicate the logarithmic likelihood and BIC differences  between  the best model of the family and the best of of all the presented models. $\dagger$ In the SPL models $\alpha_\te{inn}=\alpha_\te{out}$, see Sec. \ref{sec:densmod}. $\ddagger$ The (anticlockwise) angle $\gamma$ indicates the  tilt of the X and Y axes of symmetry of the halo with respect to the  Galactic X and Y  axes, see Sec. \ref{sec:ellsurf}.}
\label{tab:fit}
\end{table*}

\subsubsection{Model comparison} \label{sec:compare}

In addition to estimating the  parameters  for a single model,  it is important to compare the results  of different models to determine which of them gives the best description of the data. The  most  robust  way to  perform a model comparison  is  through  the  ratio of the  Bayesian evidences.
Under  the assumption that the posterior distributions are almost Gaussian, the Bayesian evidence can be approximated by the  Bayesian information criterion  (BIC, \citealt{bic_def}) defined as
\begin{equation}
\text{BIC}=\dim(\vec{\mu})\ln\left(N_\text{S}\right) -  2 \ln\left(\mathcal{L}_\te{max}\right),
\label{eq:BIC}
\end{equation}
where  $N_\te{S}$ is the number of objects  in the sample  and $\mathcal{L}_\te{max}$ is the maximum value  of the likelihood  (Eq. \ref{eq:lnn2}).

The BIC is often used to compare different models with different dimensions in parameter  space and the model with the lowest BIC is preferred. The BIC is similar to the maximum likelihood  criterion, but  it includes a  penalty depending on the number of free parameters, such that for two models with the same likelihood, the one with more parameters is penalised.  
The best-fit parameters together with a comparison of the maximum  likelihood and the BIC  for a representative sample  of halo models  can be found  in  Tab. \ref{tab:fit}. These quantities have been calculated  taking the maximum  of the  $14400$ likelihood estimates  obtained  with the final mcmc sampling (see Sec. \ref{sec:mcmc}).

Other than the BIC,  we also compared the ability of the different models to describe the observed properties of the RRLs (e.g. distribution of the stars in the sky).

\subsubsection{RRLs density law} \label{sec:rrlsdl}

\begin{figure}
\centering
\includegraphics[width=1.0\columnwidth]{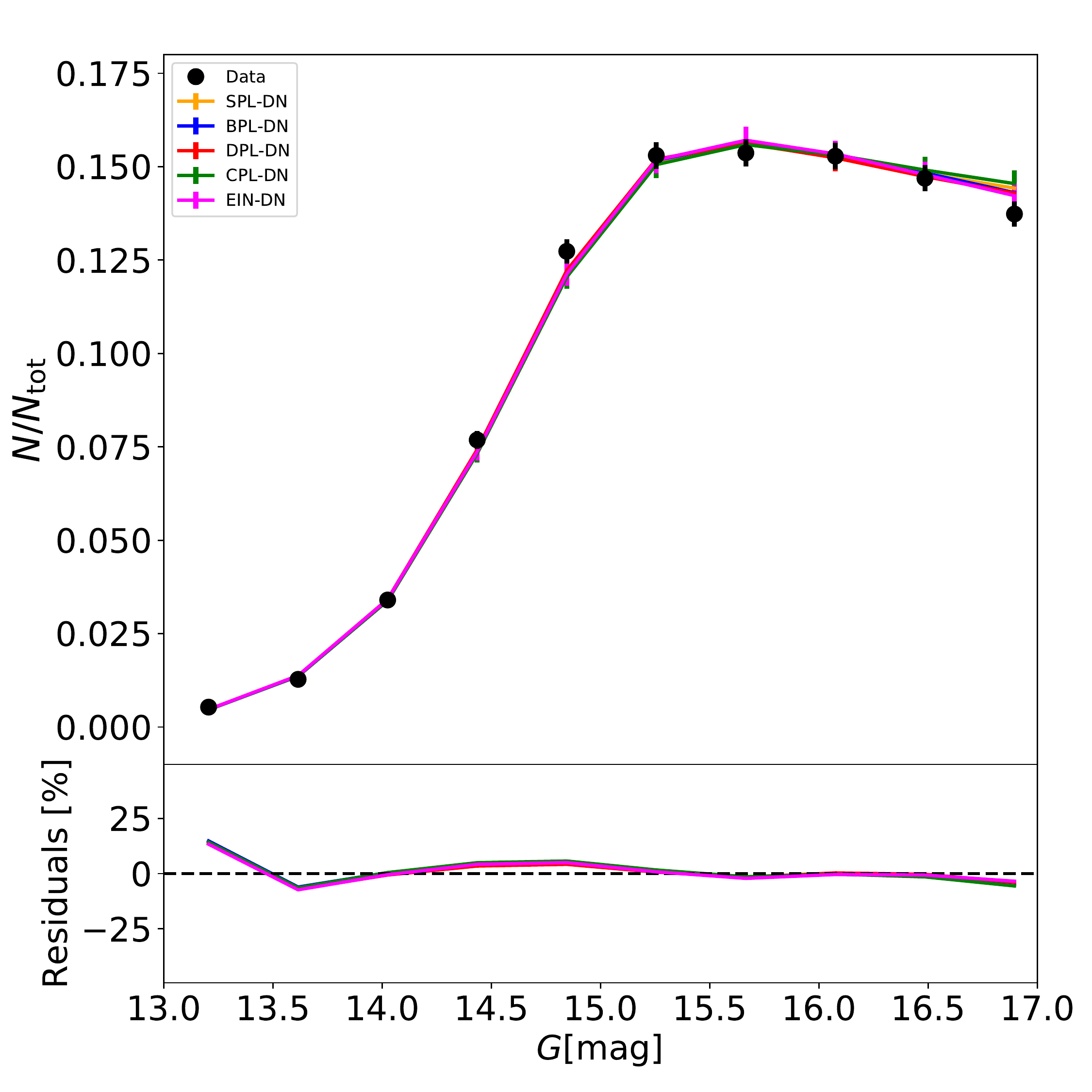}
\caption{ 
Top panel: the black points show the  fraction of stars 
in magnitude bins, while the curves show the same fraction expected  for halo models with  different density laws: SPL (blue), BPL (orange), DPL (green), CPL (red)  and EIN (magenta). Error bars on the data points and on the model distributions indicate Poisson uncertainties.
Bottom panel: relative residual (Data-Model/Model).
In all the cases  we  assumed  a DN geometrical halo model (see Tab.\ \ref{tab:fit}).
}
\label{fig:densg}
\end{figure}

\begin{figure}
\centering
\includegraphics[width=1.0\columnwidth]{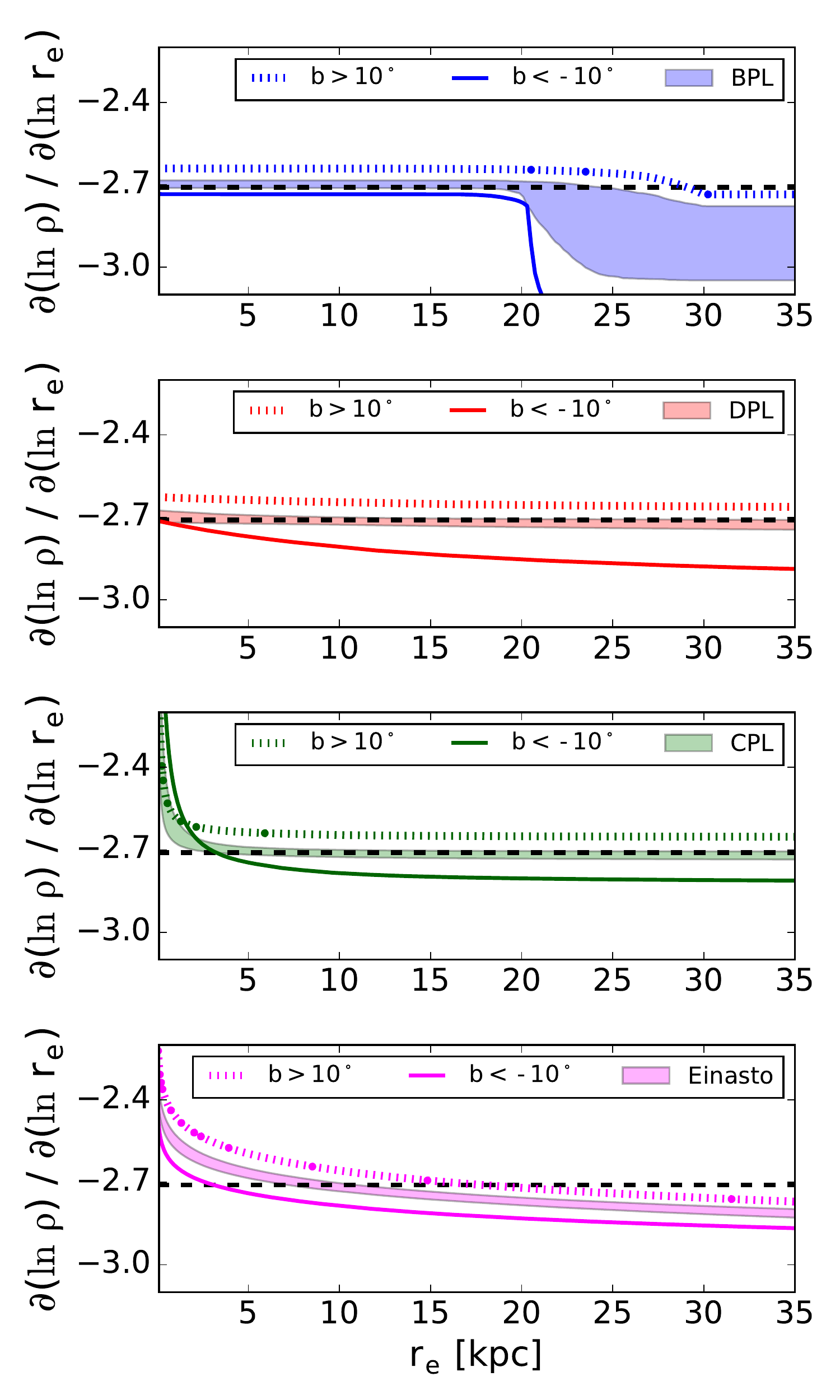}
\caption{Radial profile of the density slope  ($\partial(\ln \rho)/\partial(\ln \te{r}_\te{e}$)) for different density law models: BPL (1th-row), DPL (2th-row), CPL (3th-row) and EIN (4th-row).  The bands represent the posterior distribution between  16th and 84th  percentiles. The  coloured lines represent the  median of  the posterior obtained fitting  only the  stars  above  ($\te{b}>10^\circ$, dotted line) and below  ($\te{b}<-10^\circ$, solid line) the Galactic plane.
The dashed-black line shows the best-fit slope obtained for the SPL model. In all the cases  we  assumed  a DN geometrical halo model (Tab. \ref{tab:fit}).}
\label{fig:densfunc}
\end{figure}

We tested different density profiles assuming a DN geometrical model: the SPL (Eq. \ref{eq:plgeneral} with $\alpha_\te{inn}=\alpha_\te{out}$), the  DPL (Eq. \ref{eq:plgeneral}), the BPL (Eq. \ref{eq:bpl}), the CPL (Eq. \ref{eq:plgeneral} with $\alpha_\te{inn}=0$) and the EIN profile (Eq. \ref{eq:einasto}). 

The  logarithmic maximum likelihoods and the BICs obtained for these different models are very similar  as shown in Tab. \ref{tab:fit}, moreover Fig.\ \ref{fig:densg} shows  that the predicted  fraction of RRLs in bins of magnitude, for the  different density models, are practically coincident.
Therefore, the DPL, BPL, CPL and EIN models do not offer any significant improvement in the description of the RRLs distribution with respect to the simpler SPL.   
Fig. \ref{fig:densfunc} shows the comparison between the (elliptical) radial profiles of the density slope for different halo models. The slopes have been calculated as the logarithmic (elliptical) radial derivative of the logarithmic density, therefore  the result for a SPL is just a constant (as shown by the black-dashed line in the four panels).
The best-fit broken radius for the BPL  is $\te{r}_\te{eb}\approx22$ kpc: in the inner part the  best-fit slope  is practically the same as the SPL model then it decreases to  a slightly larger slope of about 2.9.

Concerning the DPL, the posterior  distributions of the slopes are compatible  with  $\alpha_\te{out}=\alpha_\te{inn}$, such that the final density profile is again an SPL with  a slope compatible with the best-fit SPL model (2th-row panel in Fig. \ref{fig:densfunc}), moreover the posterior distribution of $\te{r}_\te{e}$ is almost uniform in the prior range interval (Tab. \ref{tab:prior}). Similarly, the core radius  of the CPL model is very small and  the final CPL  model is compatible  with  no-core  and the outer slope is   the same of  the  SPL model (3th-row panel in Fig. \ref{fig:densfunc}). 

Finally, for the EIN  the  fit favours very large  values both for $\te{n}$ ($\approx38$) and for $\te{r}_\te{eb}$ ($\approx390$ kpc), so that  in the analysed radial range the variation of the slope is minimal  and the density profiles effectively mimic an SPL. However, since the BIC of the EIN model is higher than the BIC of the SPL models, there is no evidence in favour of an EIN profile for the inner part of the stellar halo.

The break radius  of  the BPL model  is  compatible with the one obtained in  \cite{xuehalo} after the subtraction of halo substructures, although the change of slope is  much more  significant  in their work with respect to our result (see Tab. \ref{tab:fitother}). 
In principle, the over-density of stars described in Sec.  \ref{sec:vasy} might make it difficult to detect the break, so  we repeated the maximum likelihood analysis  using   the RRLs only above and below the  Galactic  plane. 
The  results   are shown in  Fig. \ref{fig:densfunc} by the   solid   ($\te{b}<-10^\circ$) and dashed  
($\te{b}>10^\circ$)  coloured curve: in general the  stars above the Galactic plane  prefer a mean slope of about 2.65    while the  ones below have a mean slope of about  2.78.
Concerning  the change of slope  above the Galactic plane, the posterior distribution of the parameters of the BPL, DPL and CPL  models  are such that these models are similar to the SPL profile.
Below the Galactic plane both the BPL and  the  DPL  favour  a change of  slope, however  the one of the BPL model  is much more significant  going from about  2.73 in the inner part to  about  3.3 beyond $\te{r}_\te{e}\approx20$ kpc. It is evident  that  the stars  below the Galactic plane  have a   steeper decrease of the density with respect to the stars above the Galactic plane. This result could be   due to the  excess of  stars at high latitude, however in both the sub-samples the minimum BIC is still obtained for the SPL model. Therefore,  the  change of slope  is not  highly significant and the fact  that the  DPL  and  the BPL show  a  different behaviour could indicate  that the change of slope could be  due to some local artefact  (such as a local decrease of the completeness).

The comparisons discussed above refer only to DN halo  models, however we obtained similar results also  for DP and TR halo models. In particular,  we found that the SPL density model always provides the lowest BIC value, independent of the assumed halo geometry.

In conclusion, there is no significant evidence of deviation  from a SPL density law:
the RRLs follows  an SPL  with  an exponent  that ranges between  2.6, assuming a SH model, to  3  assuming a $\text{TR}^{\it qv}$ halo. 
These  results agree  with the  estimate of the density profiles obtained in Sec. \ref{sec:dprof} counting stars in ellipsoidal shells (Fig. \ref{fig:mdens}).

\subsubsection{Iso-density surfaces} \label{sec:isosurf}

\begin{figure}
\centering
\includegraphics[width=1.0\columnwidth]{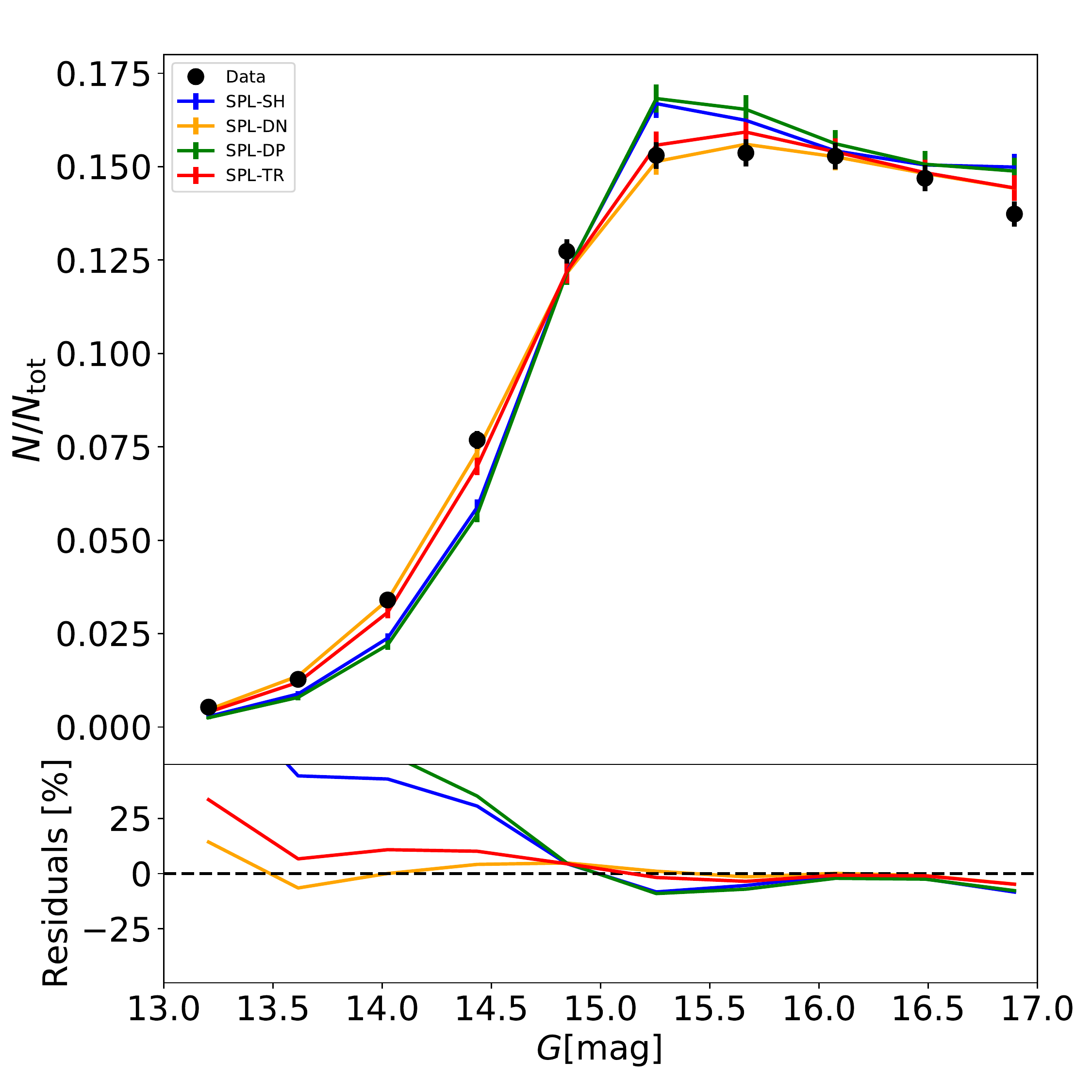}
\caption{
As in Fig. \ref{fig:densfunc} but for halo models SH (blue, $\te{q}=1$ and $\te{p}=1$), DN (orange, $\te{q}=0.58$ and $\te{p}=1$), DP (green, $\te{q}=1$ and $\te{p}=1.59$) and TR (red, $\te{q}=0.65$ and $\te{p}=1.27$). 
In all cases  we  assumed  a SPL for the density  profile  of the RRLs (with parameters given in Tab. \ref{tab:fit}).}
\label{fig:gdist}
\end{figure}

As shown  by the  BIC comparison in Tab. \ref{tab:fit}, the SH and DP  models  give significantly  worse fits with respect to  the DN and TR models. 
This result is confirmed by a comparison of the   properties of the RRLs in our sample with the ones expected  for the best-fit models.
For example,  Fig. \ref{fig:gdist} shows  the   fraction of  RRLs in our sample in different observed magnitude  bins (black points) compared to the predicted distribution  for  different halo models  (assuming a SPL, see Tab. \ref{tab:fit}).  The SH  and the  SP models predict too low a fraction of stars  at lower $G$ and  a significant excess around $G=15$, while  the  DN and the TR models  give  a good match  to the data.

\begin{figure}
\centering
\includegraphics[width=1.0\columnwidth]{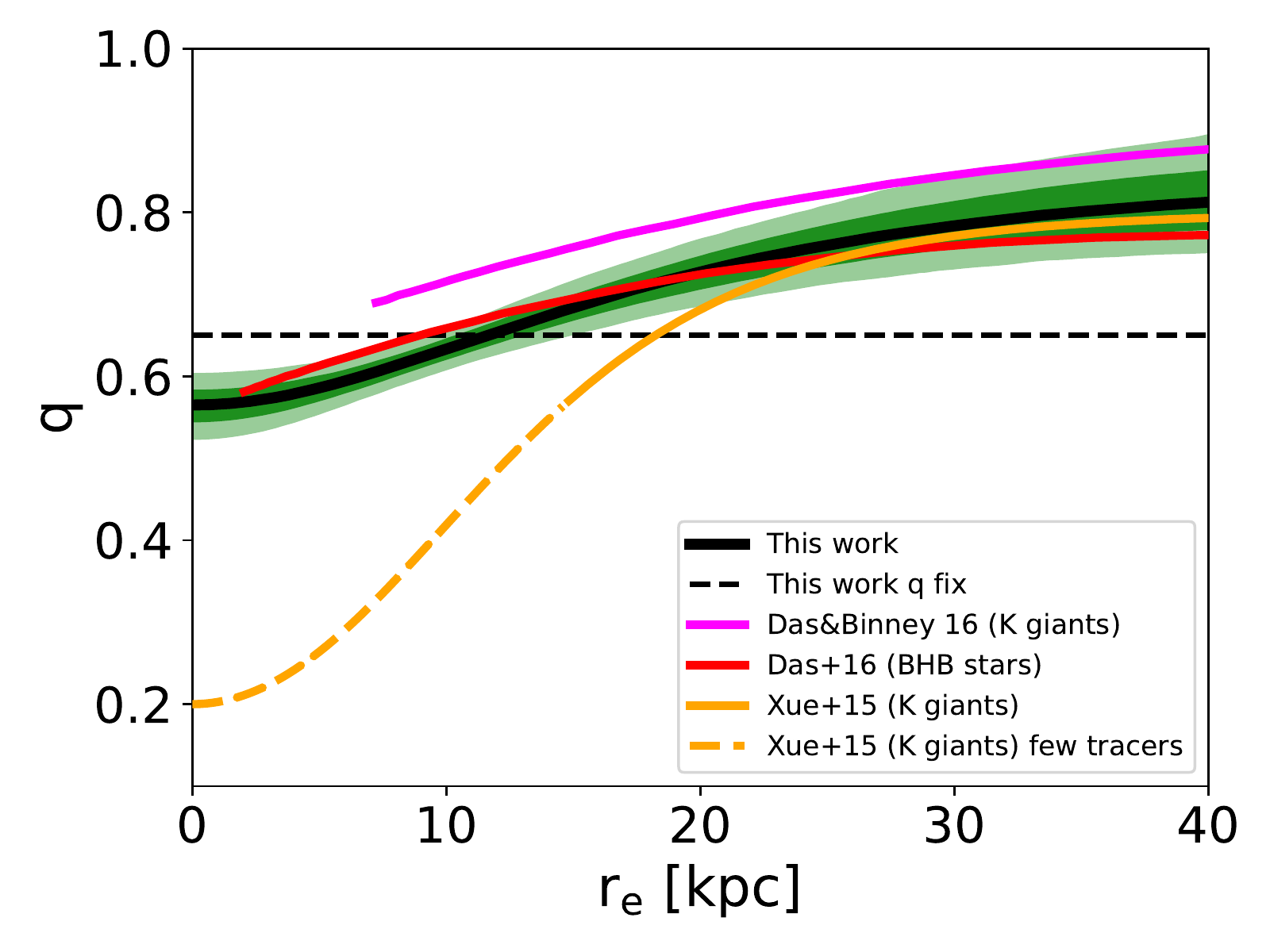}
\caption{
Halo flattening, q, as a function of the elliptical radius $\te{r}_\te{e}$ (Eq. \ref{eq:ellrad}). The black solid-line  shows the median of the functional-form distribution of q obtained  for our best  halo model  (SPL-TR$^{\it qv}$; last row in Tab. \ref{tab:fit}) while the dark and light green bands indicate the 68\% and 95\% confidence intervals.  The dashed  line  indicates the best-fit q  obtained   for the   SPL-TR halo model (8th row in Tab. \ref{tab:fit}). 
The orange curve shows the best-fit functional form of q found in  \protect\cite{xuehalo} using a sample of K giants. The other curves show the non-parametric estimate of q from \protect\cite{DasK} (magenta line) using a sample of K giants stars and from \protect\cite{das} (red line) using a sample of BHB stars.    
Concerning the best-fit parametric functional forms of \protect\cite{das}, we note that they found a profile that is very similar to the one of \protect\cite{xuehalo}(practically coincident, except at the very inner radii).
In the work of \protect\cite{xuehalo}, q is a function of the spherical radius $\te{D}_\te{g}$ rather than the elliptical radius $\te{r}_\te{e}$. In order to compare their and our results the orange line has been calculated along the Galactic vertical direction where $\te{X}_\te{g}=\te{Y}_\te{g}=0$, $\te{D}_\te{g}=\te{Z}_\te{g}$ (Eq. \ref{eq:frefDg}) and $\te{r}_\te{e} \te{q}(\te{r}_\te{e})=\te{Z}_\te{g}$ (Eq. \ref{eq:ellrad}). In this case our estimate of the flattening $\te{q}$ at $\te{r}_\te{e}$ should be compared with their estimates calculated at $\te{D}_\te{g} = \te{q} \te{r}_\te{e}$.
}
\label{fig:qvar}
\end{figure}

As a further refinement we tested the $\text{DN}^{\it qv}$ and $\text{TR}^{\it qv}$ models. 
Allowing  q to vary, we obtained a better description of the  distribution  of RRLs as shown by the  decreases of the  BICs in Tab. \ref{tab:fit}.  The  variation of the flattening is  quite significant as it is shown in Fig. \ref{fig:qvar}. 
The halo is largely flattened  ($\te{q}\approx0.57$) in the very inner part and then  it becomes more spherical reaching a flattening of about  0.75 at the border  of the Galactic  volume  analysed in this work  ($\te{r}_\te{e}\approx30$ kpc). 
These results are in agreement  with what we have seen directly  in the
distribution of stars in  the $\te{r}_\te{e}$-$\theta$ plane in Sec. \ref{sec:hflat} and  Fig. \ref{fig:mb10}.
Essentially the same result is obtained by considering the stars only above or only below the Galactic plane. 
Fig. \ref{fig:gdist2} shows the comparison between the magnitude distribution of the stars in our sample and the one expected for the varying q model.

In the (non-tilted) TR and the DP models the elongated axis makes an (anticlockwise) angle $\gamma\approx-20^\circ$  with respect to the positive part of the Galactic Y-axis. 
It is interesting to compare the orientation of the halo principal axes to the one of the Galactic bar. The orientation of the Galactic bar $\Phi_\te{bar}$\footnote{$\Phi_\te{bar}$ is defined as the anticlockwise angle between the direction of the bar and the Galactic X-axis.} is uncertain \citep{antoja_bar}, depending on the works it ranges  from about $-10^\circ$ (e.g. \citealt{robin_bar}) to about $-45^\circ$ (e.g. \citealt{benjamin_bar}). 
Given these large uncertainties, the angle $\gamma$ can be considered compatible with $\Phi_\te{bar}$. As a consequence, the  elongated axis (Y-axis) of the halo is almost perpendicular to the orientation of the Galactic bar. It is unclear if this correlation  is a coincidence, due to the large errors, or if it reflects a real link between the two structures.
For the purpose of this work, we note that the fact that the  elongated axis is almost perpendicular to the Galactic bar  rules out the hypothesis that the fit has been influenced by some contaminants coming from the inner Galactic bar.

\subsubsection{Tilt and offset} \label{sec:tilt}

%\begin{figure}
%\centering
%\includegraphics[width=0.9\columnwidth]{image/model3d.pdf}
%\caption{Iso-density surface at $\te{r}_\te{e}=30$ kpc for the best SPL-triaxial + tilt model. The  solid lines indicate the principal axes of the ellipsoid, while the dashed lines show the Cartesian Galactocentric frame of reference.}
%\label{fig:mod3d}
%\end{figure}

As a final analysis, we considered halo models  that can be  tilted with respect to the Galactic plane and offset  from the Galactic centre (see Sec. \ref{sec:ellsurf}).
We  added these  refinements  to  the  SPL-TR model.

For the offset model, the  best-fit parameters are 
\begin{itemize}
\item $\te{X}_\te{off}=0.39\pm0.05$ kpc,
\item $\te{Y}_\te{off}=-0.17\pm0.06$ kpc,
\item $\te{Z}_\te{off}=-0.01\pm0.01$ kpc,
\end{itemize}
and the size of the displacement vector with respect to the Galactic centre is $\te{D}_\te{off}=0.43\pm0.07$. 
The  offset we found is small and quite insignificant compared to the radial extent of the halo. Indeed these small values could also be due to the over-fitting of some local substructure at small elliptical radii.
Moreover, both the  posterior distribution of $\te{Y}_\te{off}$ and $\te{Z}_\te{off}$ are compatible (within $3-\sigma$ and $1-\sigma$, respectively) with 0 and most of the displacement is along the Galactic X-axis. This axis  connects the Sun to the Galactic centre, therefore the offset in this direction can be also interpreted as changing the distance of the Sun from the Galactic centre. Considering our results, the estimated Sun position is  $\te{X}_\odot=7.61\pm0.05$ kpc, which is totally compatible with the Sun-distance estimates in literature (e.g. \citealt{McMillan}).

For the tilt model, we take the triaxial model and allow the axes of symmetry to be arbitrarily oriented.  The best-fit results for the (anticlockwise) angles describing the orientation of the halo (see Sec. \ref{sec:ellsurf})  are
\begin{itemize}
\item $\gamma=-20.1^\circ\pm2.7^\circ$ (rotation around the Z-axis),
\item $\beta=5.7^\circ\pm0.8^\circ$ (rotation around the new Y-axis),
\item $\eta=3.13^\circ\pm0.5^\circ$ (rotation around the final X-axis).
\end{itemize}
The angle $\gamma$ represents the tilt of the elongated axis of the halo with respect to the Galactic Y-axis and it is compatible with the orientation found for the  non-tilted DN and TR disc-plane models (Tab. \ref{tab:fit}). The angles $\beta$ and $\eta$ measure the tilt of the halo meridional plane with respect to the plane of the Galactic disc. Even though the  BIC indicates a mild improvement  with respect to  the  non-tilted triaxial model ($\Delta\te{BIC}\approx-40$), the tilt is quite small.

\begin{figure}
\centering
\includegraphics[width=1.0\columnwidth]{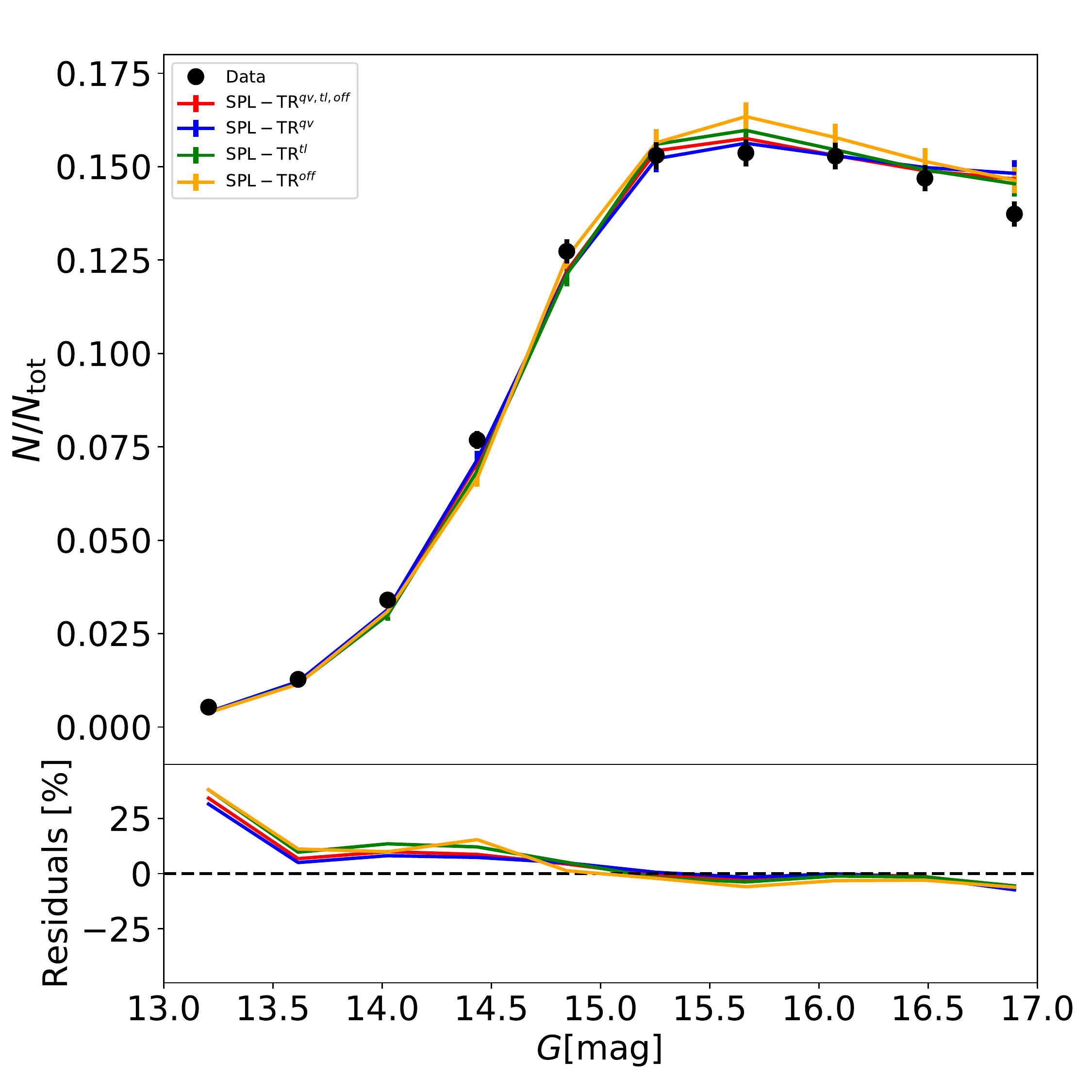}
\caption{
As in Fig. \ref{fig:densg} but for triaxial halo models with varying q (blue), halo tilt (green), centre offset (yellow) and  a combination of the previous models (red), see Sec. \ref{sec:tilt} for further details.}
\label{fig:gdist2}
\end{figure}

Finally, we  compared the ability of tilted and non-tilted models to reproduce the observed distribution of stars. Fig.  \ref{fig:gdist2} shows the comparison between data and different models for the magnitude distribution of the stars. The models are $TR^{\it qv}$, $TR^{\it tl}$, $TR^{\it off}$ and $TR^{\it qv,tl,off}$.
Note that all of them  are capable to  give a good match to the data.
                        
In conclusion, we  conservatively assume  that there  is  no need for  a significant tilt to describe the RRL distribution in the inner halo. A  more detailed study about this interesting properties  is  postponed   to future works  in which we  will extend the coverage of the Galactic volume (see Sec. \ref{sec:future}).

\section{Discussion}  \label{sec:ref}

\subsection{Best halo model} \label{sec:best}

\begin{figure*}
\centering
\includegraphics[width=0.8\textwidth]{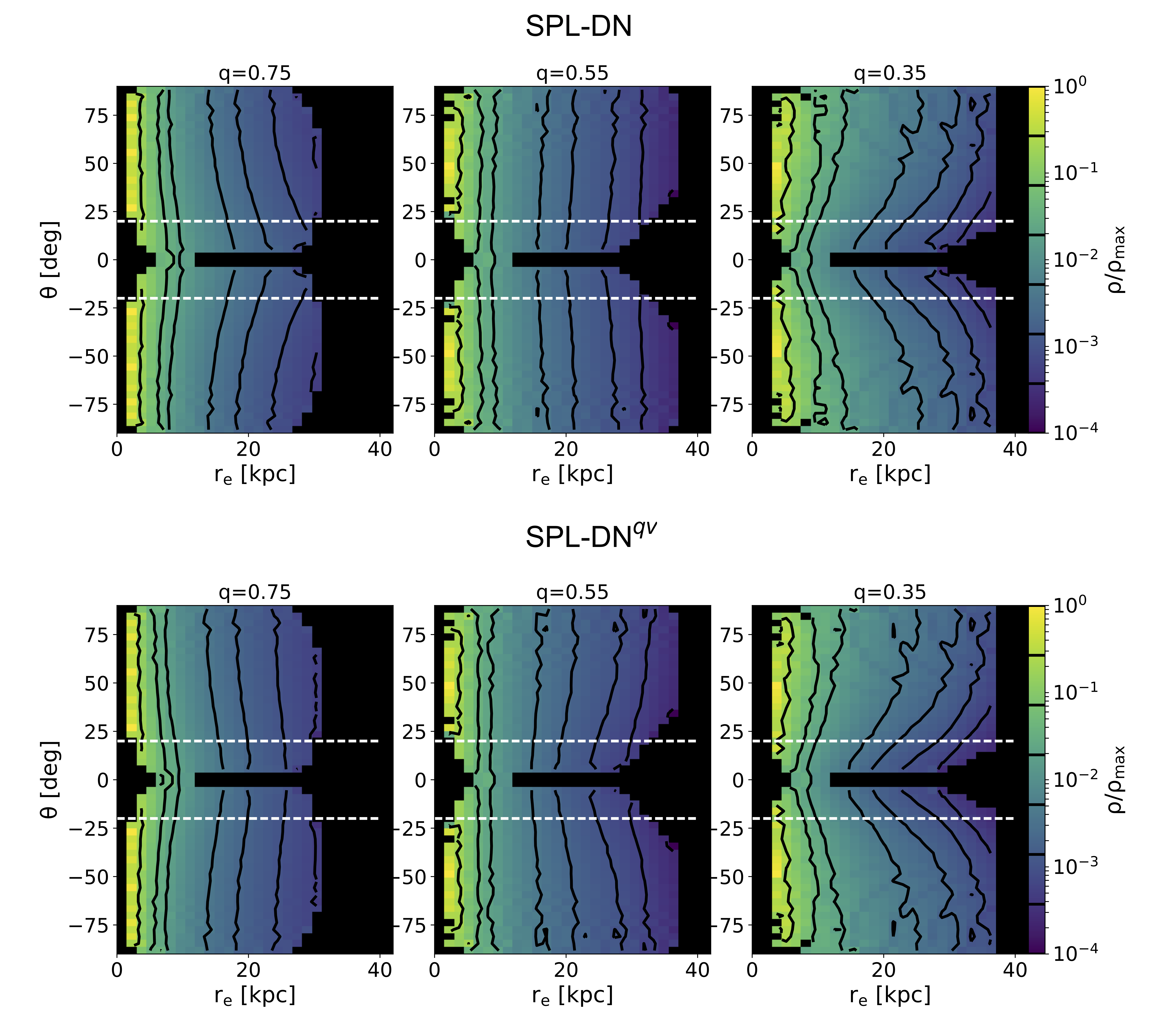}
\caption{Same as in Fig. \ref{fig:mb10} but for mock stellar
  halos. {\it Top:} density maps for the best-fit SPL-DN model. {\it Bottom:} density maps for the best-fit
  SPL-DN$^{\it qv}$ (see Tab. \ref{tab:fit}). The maps have been obtained by averaging
  over 1000 halo mock realisations. Note that for a halo with a fixed
  flattening, for the (almost) correct value of q (middle panel) the
  iso-density contours remain vertical across the whole range of
  elliptical radii. This is at odds with the RRL distribution in the
  Milky Way. As the middle panel of Fig.~\ref{fig:mb10}
  demonstrates, the contours indeed start vertical for $\te{r}_\te{e}
  < 20$ kpc, but beyond that, there is a noticeable bending away from
  the centre. Our best-fit model with varying q displays exactly the
  same behaviour (see middle panel of bottom row).}
\label{fig:mb_res_model}
\end{figure*}

\begin{figure*}
\centering
\includegraphics[width=0.8\textwidth]{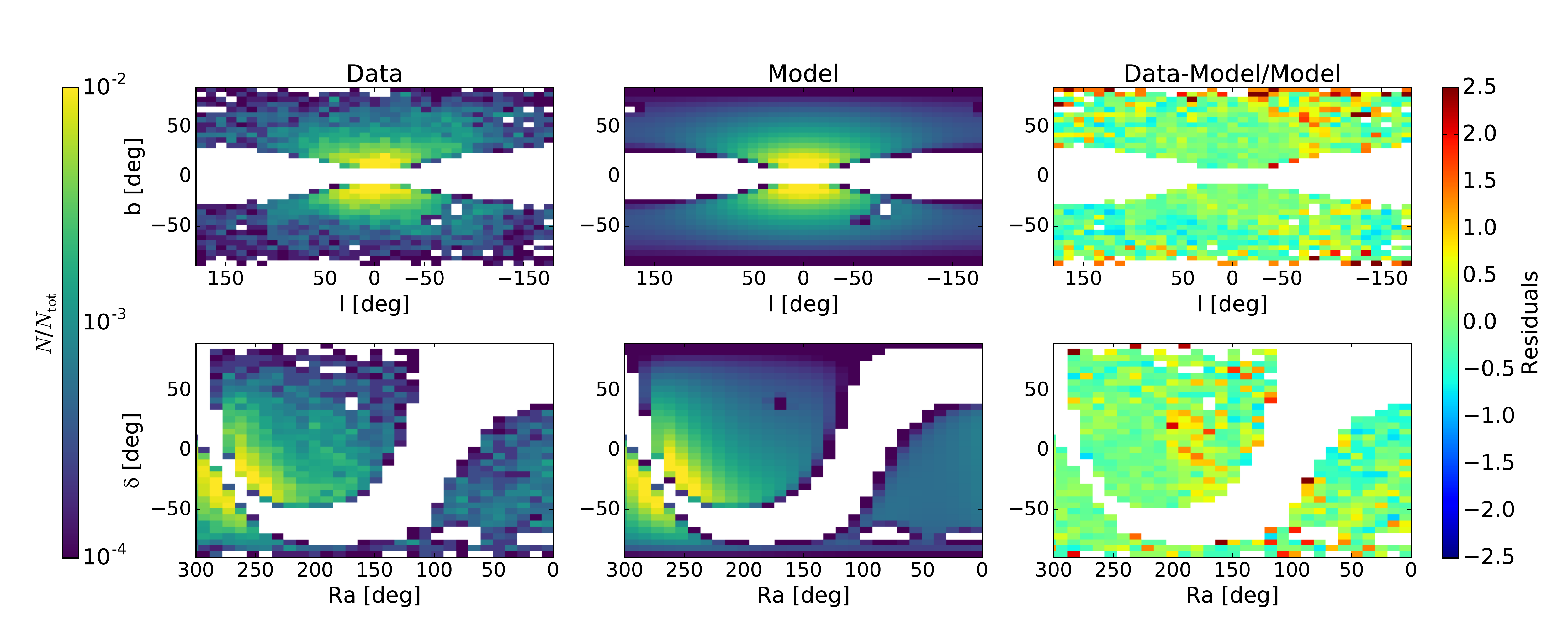}
\caption{Top panels: all-sky star number density maps in Galactic
  coordinates. Bottom panels: sky-maps density plots in equatorial
  coordinates. The left-hand panels show the data, while the middle
  panels show our best model (SPL-TR$^{\it qv}$, see
  Sec. \ref{sec:best} and Tab. \ref{tab:fit}), the right-hand panels
  show the relative data-model residuals. The model maps have been
  obtained integrating the star density distribution
  (Eq. \ref{eq:nurho2}) through the magnitude interval $10<G<17.1$
  assuming a constant absolute magnitude $M=M_\te{RRL}=0.525$.}
\label{fig:bm_res}
\end{figure*}
\begin{figure*}
\centering
\includegraphics[width=0.8\textwidth]{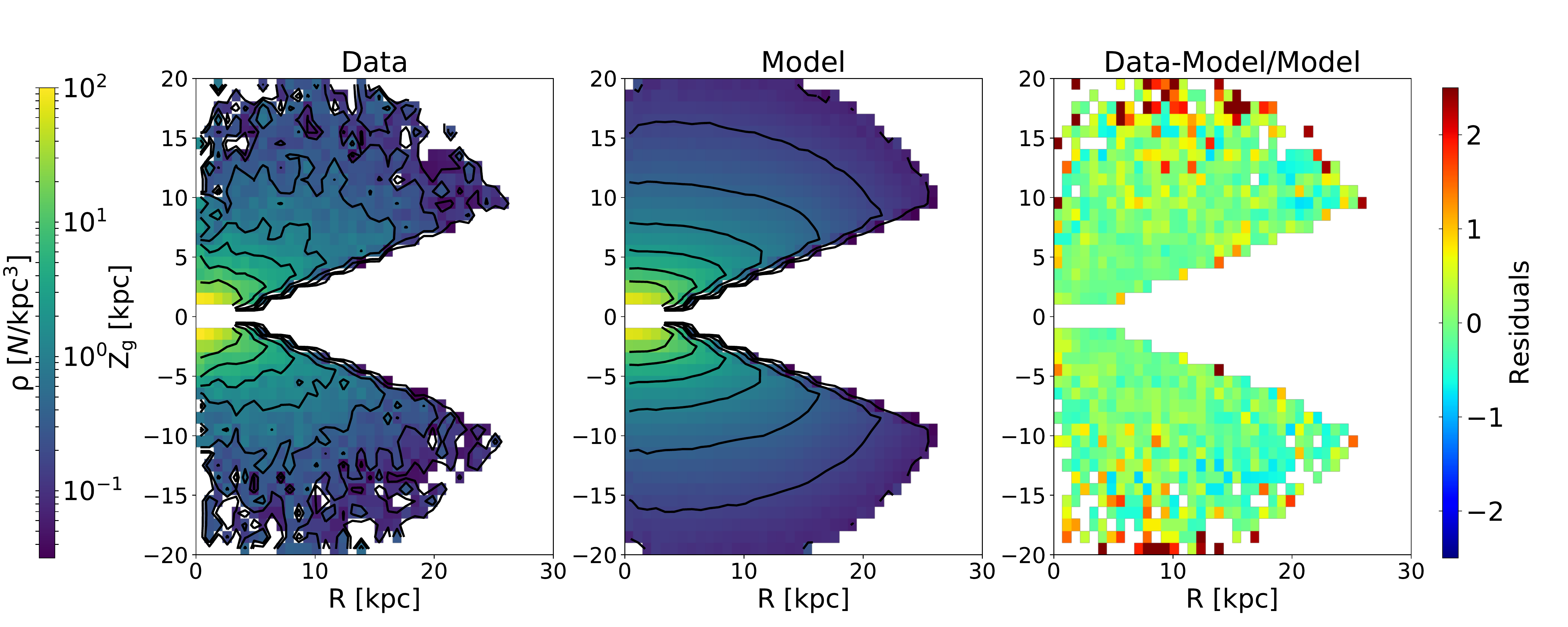}
\caption{Number density in the Galactic $\te{R}-\te{Z}_\te{g}$
  plane. Left-hand panel: density map for the RRLs in our clean sample
  (Sec. \ref{sec:csample}); middle panel: density map for our best
  model (SPL-TR$^{\it qv}$, see Sec. \ref{sec:best});
  right-hand panel: relative data-model residuals.  The black
  iso-density contours are plotted with interval of $\Delta \log
  \rho=0.4$ from $\log \rho=-3$ to $\log \rho=4$, where $\rho$ has the
  dimension $\te{kpc}^{-3}$.  The model map has been obtained
  averaging the maps of 1000 different mock catalogues made assuming
  the best-fit parameters obtained for the SPL-TR$^{\it qv}$
  halo model (see Tab.\ref{tab:fit}).}
\label{fig:rz_res}
\end{figure*}
\begin{figure*}
\centering
\includegraphics[width=0.8\textwidth]{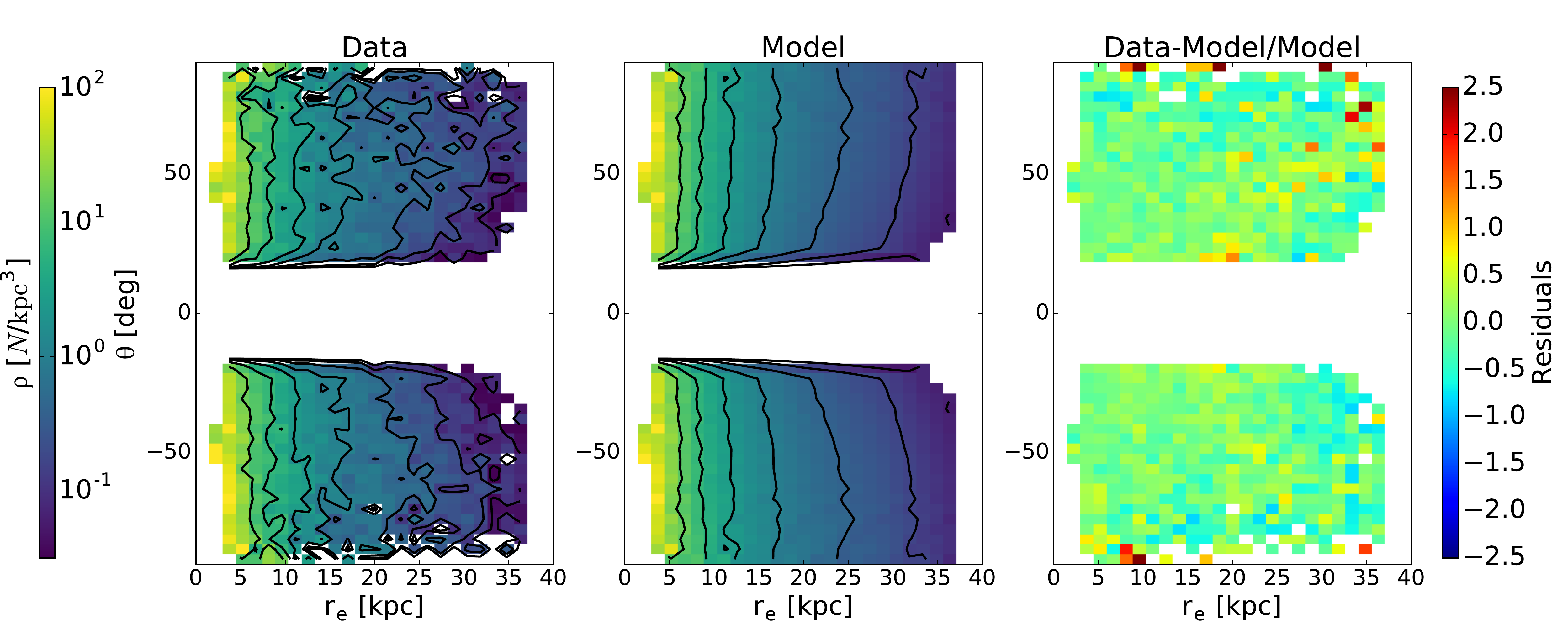}
\caption{Same as in Fig. \ref{fig:rz_res} but for the number density
  in the Galactic $\te{r}_\te{e}-\theta$ plane.  The $\te{r}_\te{e}$
  in the plot have been obtained assuming $\te{q}=0.6$ and $\te{p}=1$
  (see Sec. \ref{sec:hflat}).}
\label{fig:mb_res}
\end{figure*}

We have explored the properties of the Galactic stellar halo using two
independent methods. First, we analysed the density of the RR Lyrae
candidates binned along the $\te{Z}$ and $\te{R}$ dimensions. This direct method
was followed by an approach where the probability of observing each
individual star in the space spanned by celestial coordinates and the
apparent magnitude was described by a 3-D model, whose parameters were
optimized using mcmc likelihood sampling. Additionally, we
augmented each method with a model comparison exercise. The results of
the direct analysis (Sec. \ref{sec:halodist}) and the maximum
likelihood fit (Sec. \ref{sec:bay}) are highly compatible. Namely, i)
the density of the RRL stars in the inner 25 kpc of the halo appears
to be stratified on ellipsoids with a pronounced flattening along the
Galactic Z-axis (Fig. \ref{fig:Rzdens}), and ii) the radial density
profile can be described by a single power-law (SPL,
Fig. \ref{fig:mdens}). While - as evidenced by Tab. \ref{tab:fit} -
this simple model gives an adequate description of the RRLs density
distribution, it can be further refined by adding a mild elongation in
the Galactic plane in combination with a radially varying flattening
(along the Z-axis).  Based on the BIC analysis
(Sec. \ref{sec:compare}) as well as the direct comparison between the
properties of the data and the best-fit models, we chose to highlight
this SPL-TR$^{\it qv}$ (last row in Tab. \ref{tab:fit}) as
the best-performing model amongst the family examined.

The distribution of the RRLs in the inner halo as described by the
above best-fit model has the following properties: the radial density
is an SPL power-law with an exponent $\alpha=2.96\pm0.05$ and the
iso-density surfaces of triaxial shape. The minor axis of the density
ellipsoids is aligned with the Galactic Z. We find strong evidence for
an evolution of the vertical density flattening: the inner parts are
squashed with ($\te{q}\approx0.57$), however the flattening becomes
less pronounced ($\te{q}\approx0.75$) at the border of the Galactic
volume analysed in this work ($\te{r}_\te{e}\approx30$ kpc). Note that
the signs of the evolution of the halo flattening can be observed in
the data directly as demonstrated in Fig. ~\ref{fig:mb10}. In the
plane of Galactocentric latitude and elliptical radius, the iso-density
contours should appear vertical for the correct (and constant) value
of q. Clearly, the contours are slanted in the left and the right
panels of the Figure, where the halo flattening is over- and
under-estimated correspondingly. However, in the middle panel
(corresponding to our best-fit fixed q model), for a large range of
radii, the iso-density contours do look vertical. Beyond 20 kpc, the
contours start to bend away from the Galactic centre, signaling a
change in the halo shape. Fig.~\ref{fig:mb_res_model} illustrates
the effect of the radial halo shape evolution with the
$\theta-\te{r}_\te{e}$ views of the mock RRL realisations for two of
our best-fit (in each class) models. Note that the middle panel in the
bottom row of the Figure displays the same trend as the middle panel
of Fig.~\ref{fig:mb10}, albeit with a lot less noise. With regards
to the halo's major axis, in the Galactic plane, we measure a mild
elongation ($\te{p}=1.27\pm0.03$) in the direction which is rotated
with respect to the Galactic Y-axis by
$\gamma=-21.3^\circ\pm2.6^\circ$ (anticlockwise).

The measured RRL (as represented by the clean sample, see
Sec. \ref{sec:csample}) distribution can be compared to the best-fit
model in Figs. \ref{fig:bm_res}, \ref{fig:rz_res} and
\ref{fig:mb_res}. Fig. \ref{fig:bm_res} presents all-sky maps in both
Galactic (top panels) and equatorial (bottom)
coordinates. Fig. \ref{fig:rz_res} gives the number density in the
meridional plane, and, finally, Fig. \ref{fig:mb_res} shows the number
density in the $\te{r}_\te{e}-\theta$ space. In each Figure, the
right-most column displays the map of residuals in the corresponding
projection.

Reassuringly, the residual maps indicate that the model gives a
reasonable description of the RRLs distribution in the G2M
sample. Overall, the Milky Way's inner halo appears smooth, but some
systematic model-data mismatches are visible. The most prominent (with
residuals in some pixels corresponding to a density excess of order of
200\%) is the overdensity of stars visible in Fig. \ref{fig:bm_res} at
Galactic longitudes around -70$^\circ$ or RA of about
190$^\circ$. This structure is also discernible in
Fig. \ref{fig:rz_res} for $\te{Z}_\te{g}>15 \ \te{kpc}$ and at about
$\te{r}_\te{e}=35$ kpc and above $\theta=50^\circ$ in
Fig. \ref{fig:mb_res}. Most likely, this large stellar cloud is
related to the Virgo over-density, as discussed in
Sec. \ref{sec:vasy}.

Amongst other differences, is a mild deficiency of stars visible at
the edge of the Galactic volume (available to us) in the anti-centre
direction, i.e. around l=$180^\circ$. This mild depletion is
especially noticeable below the Galactic plane, however the amplitude
of these outer Galaxy residuals is much smaller as compared to the
mismatch caused by the Virgo over-density. We argue that this
deficiency of stars may be a hint of the steepening of the radial
density profile, but as discussed in Sec. \ref{sec:bay}, it does not
obey the smooth parametric model considered here. A more robust study
of the outer halo behaviour ought to be possible in the near future,
when the coverage of the Galactic volume is extended well beyond the
current radial limit (see Sec. \ref{sec:future}).

\subsection{Possible sources of bias in the maximum likelihood analysis} \label{sec:bias}

\subsubsection{Assumption on the absolute magnitude} \label{sec:biasamag}

Our halo fitting method (Sec. \ref{sec:results}) relies on the
assumption that all RRLs in the sample have the same absolute
magnitude $M_\te{RRL}=0.525$. However, as Fig. \ref{fig:Mg}
demonstrates, this is not strictly true as indicated by the
Gaussian-like wings in the $M_G$ distribution. To understand the
possible bias introduced by this simplification, we assumed that the
pdf of the absolute magnitude $P(M)$ in Eq. \ref{eq:denslamb_bn} is
represented by the best-fit double Gaussian (blue line in
Fig. \ref{fig:Mg}). With this addition to the model, we repeated the
fit of our best halo model (Sec. \ref{sec:mcmc}), this time
marginalising the likelihood over $M$. Perhaps unsurprisingly, the
final posterior distributions of the model parameters are totally
compatible with the ones obtained assuming a constant absolute
magnitude.

\subsubsection{Dust} \label{sec:dust}

Regions with high dust extinction add severe uncertainties in the
study of distribution of stars in the Galaxy.  In these regions it is
difficult to recover the faintest stars and the completeness could be
much lower compared to the rest of the sky.  Note however, that
applying the cut on $\theta$ (Sec. \ref{sec:csample}), we eliminated
most of the regions with high Galactic dust extinction: the 85\% of
the stars in our \vir{clean} sample has $E(B-V)<0.25$ and only the 1\%
are located in regions with $E(B-V)>0.8$.  To be sure that our results
are not influenced by the regions with high reddening, we repeated the
fit of our best halo model (Sec. \ref{sec:best}) masking the regions
with $E(B-V)>0.4$.
%that correspond to a sky window with $30^\circ<\te{l}<30^\circ$ and
%$0^\circ<\te{b}<30^\circ$.
We did not find any appreciable differences with respect to the
un-masked analysis.

\subsubsection{Contamination from the Galactic disc} \label{sec:biasdisc}

The final results of the fitting analysis (Sec. \ref{sec:results})
have been obtained considering the presence of only one smooth stellar
component, however disc RRLs (if any) and other disc variables,
including artefacts (see Sec. \ref{sec:comcont}) could \te{pollute}
our sample. Although the cut on Galactic b (Sec. \ref{sec:selection})
and $\theta$ (Sec. \ref{sec:csample}) is employed to limit such
contamination, we repeated the modelling with the addition of an extra
disc component. Thus, this new model has contribution from an
ellipsoidal halo with normalised density $\tilde{\rho}_h$
(Sec. \ref{sec:hmodel}) as well as a double-exponential disc with
normalised density $\tilde{\rho}_d \propto
\te{exp}[-\te{R}/\te{R}_\te{d}]
\te{exp}[-|\te{Z}|/\te{Z}_\te{d}]$. Here $\te{R}_\te{d}$ and
$\te{Z}_\te{d}$ are the radial and the vertical disc scale-lengths.
Considering the two components, the global pdf of the stellar
distribution is
\begin{equation}
\tilde{\lambda}(m,\te{l},\te{b}|f,\vec{\mu_\te{h}},\vec{\mu_\te{d}})=(1-f)\tilde{\lambda}_\te{h}(m,\te{l},\te{b}|\vec{\mu}_\te{h}) + f\tilde{\lambda}_{d}(m,\te{l},\te{b}|\vec{\mu}_\te{d}),
\label{eq:lamglobn}
\end{equation}
where $\tilde{\lambda}_\te{h}$ and $\tilde{\lambda}_{d}$ represent the
pdf of the halo and the disc stars, while $f$ is the the disc-to-total
stellar number ratio.  The pdfs that appear in Eq. \ref{eq:lamglobn}
above have been already marginalised over the absolute magnitude
distribution of the halo and disc stars. For the halo we used a Dirac
delta centred on $M=M_\te{RRL}=0.525$ (Sec. \ref{sec:amag}), while for
the disc a uniform distribution between $M=-2$ and $M=5$.
The absolute magnitude distribution of the stars in the disc could be
much more complicated with respect to a uniform
distribution. However, making use  of mock catalogues  we found that our assumption is a good choice to minimise the bias also in the presence of a truly complicated multi-peak distribution. 

The RRL sample was modelled (Sec. \ref{sec:mcmc}) using
Eq. \ref{eq:lamglobn} in the logarithmic likelihood in
Eq. \ref{eq:lnnp}: we considered $f$ as a free parameter, but fixed
the disc's stellar density profile assuming values for $\te{R}_\te{d}$
and $\te{Z}_\te{d}$. We tested a wide range of thin and thick disc
models using $\te{R}_\te{d}$ values between 2 and 4 kpc and
$\te{Z}_\te{d}$ values between 0.1 and 1 kpc.  In general, we did not
find significant differences with respect to the results shown in
Tab. \ref{tab:fit}, except for the SH and DP models. 
In these two cases, the models try to reproduce
the vertical flattening observed in the data with an high-percentage
of disc stars, but the final fits are poor as evidenced by the
presence of large residuals.  For all the other models considered we
found that the final posterior distribution for $f$ is lower than 3\%
and is consistent with 0. Therefore, we conclude that our \vir{clean}
sample (Sec. \ref{sec:csample}) does not harbor stars coming from the
Galactic disc and the observed flattening along the vertical direction
is a genuine property of the distribution of RRLs in the halo.

\subsection{Halo substructures}   \label{sec:struct} 

%
\begin{comment}
\end{}
\begin{figure*}
\centering
\includegraphics[width=0.8\textwidth]{image/sky_res_gbin.pdf}
\caption{Relative data-model residuals in Galactic coordinates in two
  magnitude bins: $10<G<15.5$ in the left-hand panel and $15.5<G<17.1$
  in the right-hand panel. The model is the same of
  Fig. \ref{fig:bm_res}.}
\label{fig:sky_res_gbin}
\end{figure*}
\end{comment}
%

\begin{figure}
\centering
\includegraphics[width=1.0\columnwidth]{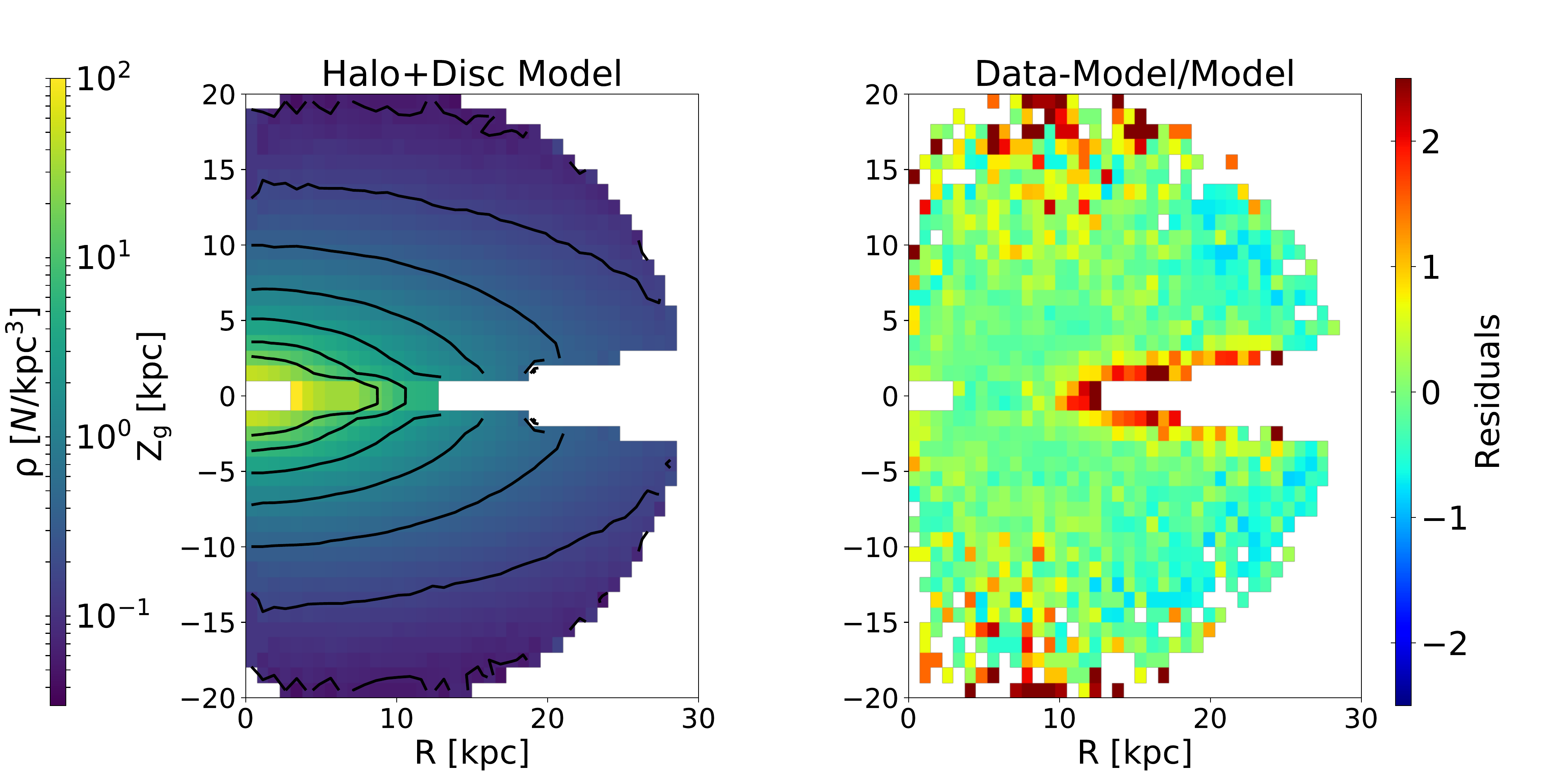}
\caption{ Left-hand panel: number density of stars in the Galactic
  $\te{R}-\te{Z}_\te{g}$ plane for a multi-component model with an
  halo and a disc; the black iso-density contours are plotted with
  interval of $\Delta \log \rho=0.4$ from $\log \rho=-3$ to $\log
  \rho=4$, where $\rho$ are in unit of $\te{kpc}^{-3}$.  Right-hand
  panel: relative residuals (Data-Model/Model) between the density in
  the $\te{R}-\te{Z}_\te{g}$ plane for our complete sample of RRLs as
  shown in Fig. \ref{fig:Rzdens} and the model shown in the left-hand
  panel.  The model map has been obtained averaging the maps of 1000
  different mock catalogues made assuming the best-fit parameters
  obtained for the disc + SPL-DN halo model (See
  \ref{sec:struct}).  The stars in the disc follow a double
  exponential distribution with a radial scale length
  $\te{R}_\te{d}=2.7$ kpc and a vertical scale height
  $\te{Z}_\te{d}=0.2$ kpc.}
\label{fig:disc_res}
\end{figure}

The main focus of this work is the study of the overall distribution
of RRLs in the inner halo. Reinforcing the principal assumption behind
the modelling, the distribution of RRLs in our sample appears smooth.
Nonetheless, several substructures are present, of which the
following two are the most obvious: i) a highly flattened distribution
of stars at low Galactic latitudes and ii) an excess of stars at high
$\te{Z}_\te{g}$. The first substructure is clearly evident in
Fig. \ref{fig:Rzdens} and in Fig. \ref{fig:mb10} below
$\theta=20^\circ$. In our investigation of this substructure we have
attempted two approaches: first, we tried to exclude the regions most
affected, i.e. those close to the disc from influencing the halo fit,
and, second, we aimed to model the excess as a contamination coming
from the Galactic disc.  In particular, we repeated the fit using our
original catalogue of about 22600 RRLs (see Tab. \ref{tab:tab_cut})
and the halo+disc stellar pdf defined in Eq. \ref{eq:lamglobn}. The
disc in the model was a realistic double exponential distribution: we
fixed the radial ($\te{R}_\te{d}$) and vertical ($\te{Z}_\te{d}$) thin
and thick disc scale-lengths to the classical values found in
literature (e.g. \citealt{das}).  Fig. \ref{fig:disc_res} gives the
density distribution in the $\te{R}-\te{Z}_\te{g}$ plane for a
disc+halo model in which the disc has $\te{R}_\te{d}=2.7$ kpc and
$\te{Z}_\te{d}=0.2$ kpc.  As illustrated by the residual map (right
panel), the halo+disc model is able to explain the flattened
distribution of stars but only within about 10 kpc. As a consequence,
the posterior distribution of the halo parameters favours a more
flattened density ($\te{q}\approx0.4$), thus concealing the evidence
of the radial variation of q, as measured and discussed above
(Tab.\ref{tab:fit}) as part of the analysis of the \vir{clean} sample
of RRLs (Sec. \ref{sec:csample}). The residuals suggest that the
two-component model gives a poor fit with a significant mismatch both
at low $\te{Z}_\te{g}$ for $\te{R}>10$ kpc and at high $\te{Z}_\te{g}$
for $\te{R}<10$ kpc. Note that these results do not change if
different values for the disc scale lengths are assumed
($\te{R}_\te{d}$ between 2 and 4 kpc and $\te{Z}_\te{d}$ between 0.1
and 1 kpc).  As a further test, we repeated the fit leaving the scale
lengths of the disc model as free parameters. This test yielded
unrealistically large values of $\te{R}_\te{d}\approx25$ kpc and
$\te{Z}_\te{d}\approx3$ kpc. This can be compared to the results of
the study of the thick disc's RR Lyrae by \citet{mateu_thick_disc} who
find the disc to be ``anti-truncated'', i.e. having a flatter density
profile in the outer parts.  In conclusion, it seems that this flat
and elongated structure can not be modelled simply by adding a double
exponential disc with the properties expected for the Milky Way.
Instead, we conjecture that the observed excess could possibly be
related to the Monoceros Ring \citep{monoceros}. Notice, however that
it is still debated \citep[see][and references
  therein]{monoceros_g2s} whether Monoceros represents a portion of
the accreted halo (e.g. \citealt{sollimaMono}) or a structure related
to the Galactic disc such as a warp and/or a flare in the outer disc
(e.g.  \citealt{lopezMono}).

Another, a more prosaic, explanation could simply mean that the excess
of stars at low latitude is due to \gaia artefacts such as the
cross-match failures (discussed in \citealt{bel16}). Such conclusion
is supported by the fact, that no survey so far has reported a clear
detection of RRL in the Monoceros Ring. 
Additionally, this structure
is not observed in a sample of metal poor K-Giants $[\te{Fe/h}]<-1$ from  the LAMOST survey \citep{lamost}. 
Of course, Monoceros Ring is indeed a prominent feature of the stellar density
maps in the direction of the Galactic anti-centre. However, so far it
has only been traced by typical disc stellar populations such as
metal-rich MS stars \citep[][]{yanny_monoceros,juric,morMono} or M
giants \citep[see e.g.][]{rocha_monoceros}.

The second substructure is a significant over-density of stars above
the Galactic plane: it is visible in the density map in
Fig. \ref{fig:Rzdens} for $\te{Z}_\te{g}>15$ kpc and in
Fig. \ref{fig:mb10} for $\theta>50^\circ$ and large elliptical radii;
it is the cause of the mismatch between the star counts above and
below the Galactic plane for $|\te{Z}_\te{g}|>10$ kpc (right-hand
panels in Fig. \ref{fig:zslab}).  Moreover, it is also traced by the
significant increases of the residuals in the data-model comparison in
Fig. \ref{fig:bm_res} and Fig. \ref{fig:rz_res}.  The spatial location
and shape of this structure indicate that it is most likely related to
the Virgo over-density \citep{juric}, a large and diffuse halo
substructure detected with a variety of tracers, but most notably with
horizontal branch stars, such as BHBs and RRL
\citep[see][]{duffau_virgo,Vivas,deasonhalo,Vivas2}. As far as the
quality of the halo models is concerned, by repeating the fit using
the stars only below the Galactic plane, we find that the Virgo Cloud
does not strongly influence our final results (see
Sec. \ref{sec:results}).

\subsection{Comparison with other works} \label{sec:comparison}

\begin{table*}
\centering
\tabcolsep=0.11cm
\begin{tabular}{cccccc}
\hline
 \multicolumn{1}{c}{Authors} & \begin{tabular}[c]{@{}c@{}}Tracer\\ (1)\end{tabular} & \begin{tabular}[c]{@{}c@{}}Range {[}kpc{]}\\ (2)\end{tabular} & \begin{tabular}[c]{@{}c@{}}Slope ($\alpha$)\\ (3)\end{tabular} & \begin{tabular}[c]{@{}c@{}}Axial ratios\\ (4)\end{tabular} & \begin{tabular}[c]{@{}c@{}}Catalogue\\ (5)\end{tabular} \\ \hline
\rowcolor[HTML]{C0C0C0} 
\cite{das} & BHB & 10-70 &  $\approx-4.7\pm0.3$ &$\te{q}=0.39\pm0.09\xrightarrow{}0.81\pm0.05 $ & SegueII \\
\cite{xuehalo} & K-Giants & 10-80 & $-4.2\pm0.1$ & $\te{q}=0.2\pm0.1\xrightarrow{}0.80\pm0.03 $ & SDSS \\
\rowcolor[HTML]{C0C0C0} 
\cite{sesarhalonew} & RRLs & 5-30 & $\approx-2.4$ & $\te{q}\approx0.65$ & LINEAR \\
\cite{deasonhalo} & BHB & 10-80 & $-2.3\pm0.1\xrightarrow{\te{R}_\te{b}\approx27 \ \te{kpc}}-4.6\pm0.2$ & $\te{q}=0.59\pm0.02$ & SDSS \\
\rowcolor[HTML]{C0C0C0} 
\cite{deprop} & BHB & 10-100 & $\approx-2.5$ & $\te{q}\approx1.0$ & 2dF QSO \\
\cite{watkins} & RRLs & 5-110 & $-2.4\xrightarrow{\te{R}_\te{b}\approx25 \ \te{kpc}}-4.5$& $\te{q=1}$ (assumed) & Stripe82 \\
\rowcolor[HTML]{C0C0C0} 
\cite{juric} & MSTO & 0-20 & $\approx-2.8$ & $\te{q}\approx0.6$ & SDSS \\
\cite{miceli} & RRLs & 3-30 &  $-3.15\pm0.07$ & $\te{q}=0.5\xrightarrow{}1.0$ (assumed)& SegueII \\
\rowcolor[HTML]{C0C0C0} 
\cite{Vivas} & RRLs & 4-60 & $\approx-2.8$ & $\te{q}\approx0.6$ & QUEST-1 \\ \hline
{ \bf This work} & RRLs & 1.5-28 & $-2.96\pm0.05$ & \begin{tabular}[c]{@{}c@{}}$\te{q}=0.57\pm0.02\xrightarrow{}0.84\pm0.06$\\ $\te{p}=1.27\pm0.03$\end{tabular} & \gaia DR1 + 2MASS \\ \hline
\end{tabular}
\caption{Galactic halo properties found in a sample of literature works. {\bf 1:} used halo tracers, Blue horizontal branch stars (BHB), K-Giants stars (K-Giants), RR Lyrae stars (RRLs), stars near the main-sequence turn-off (MSTO). {\bf 2:} range of sampled Galactocentric distances. {\bf 3:} exponent of the power-law describing the density profile of the stars in the halo; when two values are present they represent the exponents in the inner part and in the outer part of the halo separated by the transition cylindrical radius $\te{R}_\te{b}$. {\bf 4}: results on the shape of the halo iso-density surfaces, q represents the the Z-to-X axial ratio, p the Y-to-X axial ratio (if not shown $\te{p}=1$); when two values are present they represent the axial ratio in the inner part and in the outer part of the halo. {\bf 5}: Survey or Catalogue used, SDSS \citep{sdss}, Stripe82 \citep{S82}, SegueII \citep{segue}, 2dF QSO \citep{2df}, QUEST-1 \citep{quest}, LINEAR \citep{linear}, \gaia DR1 \citep{gaiaDR1}, 2MASS \citep{2mass}.}
\label{tab:fitother}
\end{table*}

Thanks to the all-sky view, stable completeness and substantial
purity, we have been able to test Galactic stellar halo models with a
degree of complexity previously unexplored. Indeed, our best halo
model (SPL-TR$^{\it qv}$, see Sec. \ref{sec:best}) can not be
directly compared with results from other works. In fact, to date,
only \cite{deasonhalo} attempted to fit a triaxial model to the
stellar halo. They found that the triaxial model does increase the
quality of the model of the distribution of A-colored stars, however
they considered this to be an overfit, mainly due to the presence of
substructures such as Monoceros and the Sagittarius stream. 
We have taken care to remove stars that may be related to Monoceros in our
modelling (Sec. \ref{sec:halodist} and \ref{sec:csample}).  With
regards to the Sagittarius stream, we have not found any worrying
levels of contamination by its members in the volume covered in this
analysis.  
Indeed, a recent work by \cite{lamost} shows that the residuals due to the Sagittarius stream are in regions not sampled in our work.
Comparing the two triaxial models, the best-fit p in
\cite{deasonhalo} is smaller than 1, while we infer
$\te{p}\approx1.3$. \cite{triaxialsim} found that, due to the gravitational effect of the bar/bulge,
a triaxial distribution of RRLs arises also from an initially oblate distribution of stars. This effect is restricted to a very small Galactic region ($\te{D}_\te{g}<5$ kpc), however gravitational effects at large scale (e.g. the influence of the LMC and/or of the spiral arms) could also cause halo triaxiality outside the innermost region of the Galaxy. In this context, in the future, it will be interesting to study the trend of p as a function of the elliptical radius as already done in this and previous works for the vertical flattening q (see below).

Perhaps not totally unexpectedly, we have found strong evidence for
the evolution of the shape of the stellar halo as a function of the
position in the Galaxy. Both parametric (Sec. \ref{sec:isosurf}) and
non-parametric (Sec. \ref{sec:hflat}) analysis require the flattening
of the halo along the Galactic Z-axis to decrease from the inner to
the outer regions, i.e. q to increase at larger distances. Most
recently, similar claims have been put forward in the works of
\cite{xuehalo}, \cite{das}, \cite{DasK} and \cite{lamost} (see Sec. \ref{sec:hflat}
and Fig. \ref{fig:qvar}). 
The results of \cite{xuehalo} have been obtained by employing the same functional form $\te{q}(\te{r}_{\te{e}})$  used in this work (Eq. \ref{eq:qvar}), while \cite{DasK} used a non parametric measure 
of $\te{q}(\te{r}_{\te{e}})$ estimated by the iso-density contours of their best-fit dynamical model.
Finally, \cite{das} estimated the flattening using both the methods described above.
Fig. \ref{fig:qvar} shows the results of these works compared to our best-fit flattening profile:
the non-parametric radial profile of \cite{das} is compatible with our results, while the results found in \cite{DasK} point towards a distribution of stars that is systematically more spherical.
While the best fit parametric functional  forms of both \cite{xuehalo} and \cite{das} are compatible with our
measurement of the flattening trend for $\te{r}_\te{e}>20$ kpc, there are several important differences. First of
all, these studies postulated the dependence of q on the
spherical radius. Here, to make our models self-consistent, we chose to
express $\te{q}$ as a function of the elliptical radius $\te{r}_{\te{e}}$ (see Eqs. \ref{eq:ellrad} and
\ref{eq:qvar}). Second, the largest discrepancies are at small radii ($\te{r}_{\te{e}} < 15$ kpc):
in our best-fitting model q is almost constant at
$\sim 0.6$, while in the other works the halo flattening keeps increasing
(q decreasing) towards the centre. For example, around the Galactic centre,
($\te{q}\approx0.4$ in \citealt{das} and $\te{q}\approx0.2$ in
\citealt{xuehalo}).  Note however, that the samples used in both these studies are actually
depleted in stars within 10 kpc from the Galactic centre, and thus the
reported behaviour of q there is an extrapolation rather than a
measurement. 
It is interesting to report that, in a recent work, \cite{xu} analysed the distribution of a sample of metal poor K giants with a novel non-parametric technique and found that the halo becomes almost spherical ($\te{q}>0.8$) already at small elliptical radii ($\te{r}_{\te{e}} \approx 20$ kpc).
Curiously, \cite{deasonhalo} found no evidence for the
variation of the halo flattening within 40 kpc from the Galactic
centre. The reason of these discrepancies is not clear, but we note that
both the sky coverage and the halo tracer population is different
compared to our study.
In conclusion, we are still a long way from a general convergence about the details of the flattening of the stellar halo, even for the same halo tracers. In this respect, we stress again that the superior sky coverage of our study (and of \gaia in general) should assure very robust results, especially in the study of the halo flattening.

Our best SPL-DN model (see Tab. \ref{tab:fit}) is
highly compatible with the previous results relying on RRLs (see
Tab. \ref{tab:fitother}) such as \cite{miceli} and \cite{Vivas}, but
partially in contrast with the result of \cite{sesarhalonew} who found
a similar slope outside of 16 kpc but a flatter radial density profile
($\alpha\approx-1$) in the very inner part of the halo. However, the
distribution of their RRLs in this region is very clumpy and can not
be properly fitted by a simple smooth power-law model.  Similarly,
\cite{watkins} found that the overall distribution of RRLs can not be
described by a smooth model, but we note that their results are based
on a drastically lower sky coverage with respect to this and other
RRL-based studies.

Comparing to inference with tracers other than RRL, our best
SPL-DN model is in good agreement with the work of
\cite{lamost} based on RGB stars and to \cite{juric} who used MS
main-sequence and MSTO main-sequence turn-off stars.  The studies
employing other tracers such as BHB and K-Giants
\citep{das,xuehalo,deasonhalo} derive similar flattening of the halo,
but usually report a steeper profile for the density law ($\alpha>-4$)
beyond 20-30 kpc. Note however that the work of \cite{deprop} found a
more shallow profile ($\alpha\approx-2.5$) up to 100 kpc. Some of the
differences can simply be due to different stars used to trace the
Galactic halo. In particular the RRLs tend to be more metal-rich as
compared to the BHBs - an important difference given that there is
evidence of a metallicity gradient in the halo \citep{carollo}.  While
we are not sensitive to the changes in the radial density profile
outside of 25 kpc, there have been claims that such ``break'' is an
artifact of not taking the shape evolution into account \citep[see
  e.g.][]{lamost,xuehalo}. Further discussion can be found in
Sec. \ref{sec:isosurf}). Table \ref{tab:fitother} summarises the properties of the Galactic
stellar halo as reported in the literature.

As a final note, it is important to mention that the distribution of our RRLs is nicely in agreement (in terms of both density profile and halo shape) with the one estimated by \cite{bulgeRRL} for a sample of genuine RRL stars in the Galactic bulge. This result seems to strengthen the hypothesis that the RRL population in the bulge is consistent with being the inward extension of the Galactic stellar halo, as also found in N-body simulations by \cite{triaxialsim}.

\subsection{Waiting for \gaia DR2}   \label{sec:future}

The \gaia collaboration announced the new data release (GDR2) for
April
2018\footnote{\url{https://www.cosmos.esa.int/web/gaia/release}}.  The
GDR2 will have a series of important improvements with respect to the
data we used in this work:
\begin{itemize}
\item the sky-coverage will be more uniform and the number of flux
  measurements ($N_\te{obs}$) for each star will increase, thus making
  our simple variability estimator AMP (Eq. \ref{eq:amp}) more precise
  and more robust;
\item the photometry for a larger sample of variable stars will be
  used to fine-tune the selection cuts of RRLs without relying only on
  external auxiliary datasets (Sec. \ref{sec:aux});
\item most importantly, photometric colours $G_\te{BP}$ and
  $G_\te{RP}$ will be used for RRL colour selection without
  cross-matching \gaia with other surveys (as 2MASS in this work).
\end{itemize}

We envisage applying the method presented here to the GDR2, shedding
the necessity to rely on shallow 2MASS data and thus extending our
analysis out to Galactocentric distances of about 100 kpc. This
unprecedented volume coverage, combined with stellar proper motions
will allow us to shed light on the substructures hidden in the
Galactic halo and thus decipher the Galaxy's formation history \citep[see
  e.g.][]{Helmi}.

\section{Summary} \label{sec:summary}

In this work, we presented the very first use of the \gaia DR1
photometric catalogue to study the properties of the Galactic stellar
halo as traced by RR Lyrae.  The key points of our study are listed
below.

\begin{enumerate}
\item We used the method proposed by \cite{bel16} to mine the \gaia
  DR1 for variable stars in advance of the official variable object
  release by the Collaboration. It is based on the estimate of the
  object's variability, AMP (Eq. \ref{eq:amp}) derived from the
  \gaiap's mean flux and its associated error.
\item Our principal selection cuts are those based on the AMP
  statistic and the colour index, $J-G$ obtained from the cross-match
  between \gaia and 2MASS. It is the depth of the latter survey which
  governs the reach of our most distant RR Lyrae. Our sample extends
  over the whole sky and contains 21643 stars within a sphere of 20
  kpc centred on the Sun, covering an unprecedented fraction of the
  volume ($\sim58\%$) of the inner halo ($\te{R}<28$ kpc). While the
  overall completeness of our sample can not compete with the levels
  attained by dedicated RRL surveys, it is stable across the sky and
  the magnitude range explored - the property most important for
  robust measurements of the stellar density behaviour in the
  halo. Additionally, we demonstrate that the sample's contamination
  is close to zero, and is not expected to exceed $\sim$10\%.
\item Assuming a constant absolute magnitude for all RRLs in our
  sample ($M_\te{RRL}=0.525$), we analysed their density distribution
  in the inner Galaxy. To that end, Figure \ref{fig:skymap} presents
  the first all-sky stellar halo maps. These heliocentric slices
  reveal a relatively smooth nearby halo, with prominent
  inter-quadrant asymmetries starting to become noticeable beyond 10
  kpc from the Sun.
\item We point out two main substructures superimposed on the
  otherwise smooth distribution of the Galactic RRLs. One is visible
  as a highly flattened distribution of stars close to the disc plane,
  at odds with the more ellipsoidal distribution observed at high
  $\te{Z}_\te{g}$. It contains a little less than half of the stars in
  our sample, but given its large radial extent it can not be
  explained as a contamination of the Galactic disc. It may be related
  to the Monoceros ring \citep{monoceros}, but most likely it is
  caused by artefacts in the \gaia data at low Galactic latitude
  \citep[see discussion in][]{bel16}. At high Galactic latitudes, a
  large over-density of stars is evident, clustered in a region with
  $\te{Z}_\te{g}\approx15$ kpc and between a Galactic radius of 5-10
  kpc, as illustrated in Figure~\ref{fig:zslab}.  We argue that this
  is likely a portion of the well-known Virgo over-density
  \citep{juric}. Its counterpart at negative Z is the Hercules-Aquila
  Cloud, also discernible in the Figure, on the other side of the
  Galaxy, beyond the bulge.
\item The RRL density distribution is stratified on ellipsoidal
  surfaces flattened along the direction perpendicular to the Galactic
  disc. The density decreases regularly following a power-law, without
  evidence of an inner-core or an change of slope within the radial
  range accessible, as demonstrated in Figure~\ref{fig:mdens}. For the
  first time, we are able to show the evolution of the halo flattening
  with radius directly in the data (see Figure~\ref{fig:mb10}). The
  Milky Way's stellar halo is rather flat in the inner parts but
  becomes more spherical at $\te{R}\sim20$ kpc. A similar property has
  been glimpsed in the sample of RGB halo stars considered by
  \cite{lamost}.
\item To mitigate possible biases associate with the low-latitude
  excess of the RRL candidates, we extracted a subset of 13713 stars
  from the original sample. We applied to this \vir{clean} sample a
  maximum likelihood fitting method, testing a large variety of halo
  models.  The parameters of our final best-model are in good
  agreement with the results obtained by the non-parametric analysis
  of the RRLs distribution. The halo is mildly triaxial with its major
  and intermediate axes in the Galactic pane ($\te{p}=1.27$) and a
  significant flattening along the Galactic Z-axis which varies from
  $\te{q}\approx0.57$ in the centre to $\te{q}\approx0.75$ at the edge
  of the radial range analysed. The halo's major axis is rotated in
  the anticlockwise direction by a angle $\gamma\approx-21^\circ$ with
  respect to the direction of the Galactic Y-axis.  The density slope
  is well approximated by a single power-law with exponent
  $\alpha=-2.96$.

%  We tested that these results are robust and not influenced by the
%  assumption of a constant absolute magnitude, by regions with high
%  dust extinction and by the possible contamination of the Galactic
%  disc.

%In this work we demonstrated that, concerning the study of the stellar
%halo, GDR1 is already competitive with previous surveys as the SDSS
%\citep{sdss} as well new surveys as LAMOST
%\citep{lamostpaper}. However, with the advent of the next \gaia data
%release it will be possible to further improve these kind of
%studies. In particular, using the information of the \gaia colours we
%will obtain an unprecedented deep all-sky view of the RRLs in the
%Galactic halo.

We have demonstrated the power of the \gaia data for the Galactic
stellar halo exploration. The wealth of the information available in
GDR1 is unexpected and signals that a paradigm shift in the halo
studies is imminent. One only has to wait for the GDR2 to be
unleashed.
  
\end{enumerate}

%%%%%%%%%%%%%%%%%%%%%%%%%%%%%%%%%%%%%%%%%%%%%%%%%%

\section*{Acknowledgments}

The authors wish to thank the members of the Cambridge Streams club
for stimulating discussions and the anonymous referee  for useful comments that improved this manuscript.

This work has made use of data from the European Space Agency (ESA)
mission {\it Gaia} (\url{http://www.cosmos.esa.int/gaia}), processed
by the {\it Gaia} Data Processing and Analysis Consortium (DPAC,
{\small
  \url{http://www.cosmos.esa.int/web/gaia/dpac/consortium}}). Funding
for the DPAC has been provided by national institutions, in particular
the institutions participating in the {\it Gaia} Multilateral
Agreement.

The research leading to these results has received funding from the
European Research Council under the European Union's Seventh Framework
Programme (FP/2007-2013) / ERC Grant Agreement n. 308024. V.B.,
D.E. and S.K. acknowledge financial support from the ERC. S.K. 
also acknowledges the support from the STFC (grant ST/N004493/1).



\bibliographystyle{mnras}
\bibliography{references}

%%%%%%%%%%%%%%%%%%%%%%%%%%%%%%%%%%%%%%%%%%%%%%%%%%
\clearpage

%%%%%%%%%%%%%%%%% APPENDICES %%%%%%%%%%%%%%%%%%%%%

\appendix
\section{Jacobian} \label{app:jaco}

The Jacobian needed to go from the set of Cartesian coordinate (X, Y, Z) to the observed one ($m$, $\te{l}$, $\te{b}$) is the determinant of the following matrix (see Sec. \ref{sec:dstars}):
\[ \text{J} = \renewcommand{\arraystretch}{1.55} \left[ \begin{array}{ccc}
 \frac{\partial \te{X}}{\partial m} & \frac{\partial \te{X}}{\partial \te{l}} & \frac{\partial \te{X}}{\partial \te{b}} \\
\frac{\partial \te{Y}}{\partial m}  & \frac{\partial \te{Y}}{\partial \te{l}}  & \frac{\partial \te{Y}}{\partial \te{b}}  \\
\frac{\partial \te{Z}}{\partial m}  & \frac{\partial \te{Z}}{\partial \te{l}} & \frac{\partial \te{Z}}{\partial \te{b}}  \end{array} \right].\] 
%We consider the  Cartesian  frame of reference as left-handed 
%and centered on the Sun: the X axis is directed toward the Sun and the Y axis in the direction of the Sun rotation. 
Therefore, the  intrinsic and observed  coordinates  are  related  as
\begin{equation}
\renewcommand{\arraystretch}{1.5}
\left\{ \begin{array}{lll}
         \te{X}=	\te{D}_\odot \cos \te{l} \cos \te{b}  \\
         \te{Y}=\te{D}_\odot \sin \te{l} \cos \te{b} \\
         \te{Z}=\te{D}_\odot \sin \te{b} \end{array} \right.
 \label{eq:sys}
 \end{equation}
where  $\te{D}_\odot = 10^{\frac{m-M}{5} -2} \ \text{kpc}$ is the distance of the object from the Sun. The Jacobian matrix becomes
\[ \text{J} = \renewcommand{\arraystretch}{1.5} \left[ \begin{array}{ccc}
 \te{dD}_\odot \cos \te{l} \cos \te{b}  & -\te{D}_\odot \sin \te{l} \cos \te{b} & -\te{D}_\odot \cos \te{l} \sin \te{b} \\
 \te{dD}_\odot \sin \te{l} \cos \te{b}   & \te{D}_\odot \cos \te{l} \cos \te{b}   & -\te{D}_\odot \sin \te{l}  \sin \te{b}   \\
\te{dD}_\odot \sin \te{b}  & 0 & \te{D}_\odot \cos \te{b} \end{array} \right].\] 
The determinant is
\begin{equation}
 |\text{J}|=\te{D}^2_\odot \te{dD}_\odot \cos \te{b}=\frac{\ln 10}{5}\te{D}^3_\odot \cos \te{b}.
 \label{eq:appA}
\end{equation}
The result of Eq. \ref{eq:appA} has been obtained  for a frame of reference  centered at the Sun, however it is also valid for a Galactocentric frame of reference. The two coordinate system differ only by an additive constant (the distance of the Sun from the Galactic center) in the first row of the system  \ref{eq:sys}., so the determinant is the same in the two cases.

\section{Exploration of the parameters space} 	\label{app:tec}

The exploration of the parameter space is performed  evaluating  the likelihood in  Eq. \ref{eq:likel3}, which is a very time-consuming step. The bottleneck of the process is the calculation of the  normalisation integral  $\te{V}_c$  in Eq. \ref{eq:vc2a}.
Given the presence of the selection function $W$, the integrand function is not continuous and shows abrupt decreases to 0 in some regions of the integration domain.
For this reason, the classical multi-dimensional quadrature methods are not able to give robust results, so we decided to make use of a Monte Carlo integration technique. In particular, we used the \textit{vegas} algorithm \citep{vegas} through its \texttt{Python} implementation\footnote{\url{https://github.com/gplepage/vegas}}.  
The final estimate of the integral comes from the average of $N_\te{it}$  \textit{vegas} runs with $N_\te{eval}$ integrand evaluations.
The values of $N_\te{it}$ and $N_\te{eval}$ have been chosen after comparing the results of \textit{vegas} and an adaptive quadrature technique\footnote{\url{https://github.com/saullocastro/cubature}} setting  $W$=1 in the integrand (Eq. \ref{eq:vc2a}). 
In particular we used two settings:
\begin{itemize}
\item{$fast$} with   $N_\te{eval}=10^5$ and  $N_\te{it}=20$;
\item{$robust$} with   $N_\te{eval}=5\cdot10^5$ and  $N_\te{it}=100$.
\end{itemize}
In principle, we can also divide our integration domain into multiple regions where $W=1$ and use the faster quadrature methods. However, given our cut in $\theta$, the integral extremes $G$, l and b become non-trivially interconnected and further time is expended on their calculation. Moreover, we also note that the time spent by a quadrature integral solver is roughly independent of the dimension of the sample space, therefore N different evaluations increase the
computational time by about N times. In conclusion, using this approach we do not obtain a significant decrease of the computational time.

In order to have the best compromise between computational time and good sampling of the parameter space we adopt a two-stage procedure.
\begin{enumerate}
\item  In the first \vir{explorative} stage we sample the parameter space using 200 walkers with 300 steps each after 100 \vir{burn-in} steps and  using the $fast$ estimate of the normalisation integral (see Sec. \ref{sec:lkdata}). The initial positions of the walkers are randomly extracted from their prior distributions (Tab. \ref{tab:prior}).
 Given the wide prior ranges (see Tab. \ref{tab:prior}), a large number of walkers is needed to have an unbiased sampling of the posterior distributions. 
\item  In the second stage we use 48 walkers with 300 steps each after 100 \vir{burn-in} steps using the $robust$ estimate of the normalisation integral (See sec. \ref{sec:lkdata}). The walkers are initially placed in a small region around the parameters for which we obtained the maximum value of the likelihood in the first stage.
\end{enumerate}

%\input{appendix/lnlike}
%\input{appendix/bias}

%%%%%%%%%%%%%%%%%%%%%%%%%%%%%%%%%%%%%%%%%%%%%%%%%%

% Don't change these lines
\bsp	% typesetting comment
\label{lastpage}
\end{document}